\documentclass[12pt]{article}%

\usepackage{bm}
\usepackage{appendix}
\usepackage{amsmath}
\usepackage{amssymb}
\usepackage[final]{graphicx}
\usepackage{float}
\usepackage{hyphenat}

\newcommand{\nn}{\noindent}

\newcommand{\be}{\begin{eqnarray}}
\newcommand{\ee}{\end{eqnarray}}
\newcommand{\bdm}{\begin{displaymath}}
\newcommand{\edm}{\end{displaymath}}
\newcommand{\bea}{\begin{eqnarray}}
\newcommand{\eea}{\end{eqnarray}}

\title{A Newtonian analogue of Kerr black hole}
\author{Areti Eleni \thanks{email: aeleni@phys.uoa.gr} \\
\footnotesize{Section of Astrophysics, Astronomy, and Mechanics, Department of Physics,} \\
\footnotesize{University of Athens, Panepistimiopolis Zografos GR15783, Athens, Greece} \and 
Theocharis A. Apostolatos \thanks{email: thapostol@phys.uoa.gr}\\
\footnotesize{Section of Astrophysics, Astronomy, and Mechanics, Department of Physics,} \\
\footnotesize{University of Athens, Panepistimiopolis Zografos GR15783, Athens, Greece}}

\date{\today}

\begin{document}
\maketitle

\begin{abstract}
A 250-year old Newtonian problem, first 
studied by Euler, turns out
to share a lot of similarities with 
the most extreme astrophysical 
relativistic object, the Kerr black hole. 
Although the framework behind the two 
fields is completely different, 
both problems
are related to gravitational fields that 
have quite intriguing analogies
with respect to orbital motions 
of a test-body 
in them. The fundamental reason responsible 
for their extraordinary similarity is the 
integrability of both problems, as well as
their common multipolar structure.
In this paper we demonstrate 
the existence of
a multitude of either qualitative, and sometimes 
quantitative, similarities between the
two problems. Based on this analogy, 
one could use the Newtonian 
problem to get insight 
in cases where the relativistic 
treatment of the 
field of a Kerr black hole becomes 
quite complicated. 
\end{abstract}


\section{Introduction}
\label{sec:1}
Struggling to solve the three-body problem in 
Newtonian gravity,  Euler studied in
1760 an easier version of the general problem: 
the problem of motion of a particle in the 
gravitational field of two fixed centers 
with masses $m_1$, $m_2$ at a distance
$2 a$ apart from each other \cite{Euler1760}. 
This problem, as it was much later shown by
Whittaker \cite{Whittaker}, is characterized by such an internal dynamical symmetry that leads 
to a new integral of motion (according to Noether's 
theorem for mechanical particle systems),
rendering this particular mechanical problem
fully integrable.

On the other hand Kerr black holes, that were 
first described and studied by Kerr \cite{Kerr}
as an exact solution of the vaccuum equations of Einstein,
proved to be the simplest macroscopic objects that 
Nature herself can create. They are fully described 
by only two parameters (the mass and the spin
of the black hole, if we assume that the net
electric charge  of 
astrophysical objects is negligible 
according to astrophysical
consensus); this is actually the physical 
context of the no-hair theorem \cite{nohair}. 
During the golden era of general relativity
(1970-1980) the physical
characteristics of Kerr black holes, including 
perturbations of this metric, were studied 
extensively \cite{Chandrabook}, and  
there  is still an ongoing 
research on the subject in the framework of  
gravitational waves emitted from compact binaries
\cite{LIGO}, that is binaries consisting 
of neutron stars and/or black holes. 

At least for binaries consisting of a massive 
rotating black hole and a neutron star, or a less 
massive black hole (known as extreme mass-ratio 
inspirals: EMRIs), could be studied perturbatively 
by considering  geodesic orbits of the less massive 
counterpart in the gravitational field of a Kerr 
black hole \cite{Emris}.
The geodesic orbits around a Kerr black hole are 
described by an integrable system of differential 
equations (exactly as with the 
Newtonian gravitational
field of the Euler problem), 
due to a Killing tensor 
field of the particular space-time 
that leads to a 
new integral of motion, the so called
Carter constant \cite{Carter,Rosquist}. 

One might think that the similarity between 
the Euler problem and the Kerr metric ends exactly
at this point, since the two gravitational fields 
do not look very similar from a physical point 
of view: Apart of the fundamental differences 
between the two corresponding physical frameworks 
(relativistic versus Newtonian gravity), 
the Kerr black hole describes an oblate 
gravitational field (due to its  spin related to
its axis of symmetry) of pure vaccuum,
while the Euler problem is by construction 
`prolate' in the sense that the mass
of the system is  distributed along its 
symmetry axis. However by introducing
a purely imaginary distance 
--instead of a real one-- 
between the two masses of  the 
Euler problem \cite{Arnold}, 
the corresponding gravitational potential remains
real (the only necessary additional condition to 
obtain a real potential then is that the two masses 
$m_1,m_2$ are equal), while the global multipolar 
structure of its potential field becomes oblate; 
thus it becomes analogous to the Kerr 
gravitational field. Moreover, as Will 
\cite{Will} has  shown, 
the only axisymmetrical and reflection symmetric 
Newtonian gravitational potential that leads to 
a third integral of motion that is quadratic 
in momenta (like the Carter constant) is the 
one that follows the same relation 
between successive multipole mass-moments
with that of Kerr.  
Oddly enough, the Newtonian multipole mass-moments of the Euler problem with an imaginary 
distance are exactly the same with the 
Geroch-Hansen relativistic multipole mass-moments of the Kerr metric itself (the prolate
version of the Euler field has the same ratio 
of moments but not the same moments).
This very similarity between the two problems 
is fundamental for all the similarities 
of the orbital characteristics arising in both 
gravitational fields.

In this paper we have studied thoroughly 
the fundamental frequencies of the Eulerian orbits
and have found numerous analogies with the
frequencies of bound orbits in Kerr. Apart of the
similarities in the expressions for the frequencies 
themselves, both problems have ISCOs. 
Moreover, one could find pairs of orbits 
--in both problems--
that are characterized by exactly the 
same triplet of frequencies.

Stimulated by all these aforementioned analogies 
between the two problems, we have performed 
an extensive comparison of properties between
the two problems and we found some new 
intricate similarities, that further persuaded 
us that the two problems could be considered 
 quite faithful analogues of each other in 
 Newtonian and relativistic frameworks,
respectively. We believe that one could use this
analogy to gain deeper insight into each one of 
them by studying its twin counterpart.
As an example we have applied this analogy 
to clarify and explain the rather subtle
argument of Kennefick and Ori \cite{KennOri} 
according to which the ``circular'' geodesic 
orbits in a Kerr background 
remain circular under their adiabatic 
evolution due to gravitational radiation. 
Not only the corresponding orbits
in the Euler problem have  analogous 
characteristics (they sweep over a surface 
of constant spheroidal radius that evolves
adiabatically under a weak dissipative force), 
but the formulation of the Newtonian problem itself 
allows for a much more transluscent explanation
than the one used by Kennefick and Ori for 
first time.  The Newtonian variation of this
proposition helps to better understand the 
foundations of that old 
but quite strong argument, and to explore it 
quantitatively in further detail. 

Furthermore, the integrability of Kerr 
metric has been used in 
\cite{AposGeraCont} to show that a hypothetical
non-Kerr object could in principle be recognizable 
by its characteristic gravitational wave signal,
due to its non-integrable character. 
During the crossing of an  orbit in phase space
through  a Birkhoff island 
(the existence of which is 
a direct theoretical consequence of a 
slightly perturbed integrable system according 
to KAM theorem \cite{KAM} and Poincar\'e-Birkhoff theorem \cite{PoincBirk}), 
the ratio of the corresponding 
frequencies in the signal 
spectrum should remain locked to a
constant rational value  for a while; 
this is the plateau effect which was
analyzed in \cite{AposGeraCont}. 
The initial computation of the relevant delay time  was 
based on the average energy loss and average 
angular momentum loss as computed for a generic 
orbit in Kerr, suitably adjusted to
account for the deviated lower multipole 
moments of the new space-time compared 
to the corresponding Kerr metric. 
In a foregoing paper, we plan to use the analogy 
between the Newtonian and the relativistic 
problem to 
check if this time interval could be systematically 
different if the evolution is computed by 
means of the 
instantaneous self-force acting on the particle,
instead of being based on averaging formulae. 
The Newtonian problem could be easily perturbed to 
construct a slightly non-integrable system
as a toy-model for a perturbed Kerr black hole.

The rest of the article is organized as follows: 
In Section \ref{sec:2}
an overall description of the Euler problem is given, along with its
oblate variant with an imaginary, 
instead of a real, distance. 
General characteristics of the orbits
in the oblate Euler gravitational field 
is presented in Section \ref{sec:3}. 
In Section \ref{sec:4}
a list of the  properties of the Euler problem,
that are similar to the properties of Kerr metric,
that are known up to now, are presented. 
In Section \ref{sec:5},
an extensive list of new properties 
that reveal the close analogy between the
two problems is presented. 
Finally, in Section \ref{sec:6}, the argument of
Kennefick and Ori from the perspective 
of the Euler problem is reformulated and explained. 
Furthermore, a quantitative result with respect 
to the evolution of the eccentricity of an orbit 
when the resonance condition is met --which actually
could happen in `circular' orbits in 
the Euler problem-- is constructed.
In Section \ref{sec:7} we summarize our findings, and suggest new
problems that the similarity between Kerr and Euler field could
further be used.

\section{The Euler problem}
\label{sec:2}

\subsection{The original problem}

The gravitational field of two point-like 
(or spherically distributed) 
masses located at fixed 
positions in an inertial (in the 
Newtonian sense) frame of reference
constitute the basis of the Euler problem, 
also known as the 
`{\it two-centre problem}'. Euler first 
studied the orbit of a test particle in 
such a gravitational field as an attempt 
to obtain analytical solutions in
special cases of the general three-body 
problem (the motion of
three particles of arbitrary masses 
under their mutual gravitational attraction).
The gravitational potential of such a system is 
\be
V^{(\rm E)}(r_1,r_2)=
-\frac{G m_1}{r_1}-\frac{G m_2}{r_2},
\ee
where $G$ is the gravitational constant, 
$m_1, m_2$ are the two point-like masses and 
$r_1, r_2$ are the distances of
the point, where the potential is computed, 
from the two fixed masses. If we use 
a  coordinate system such that the two 
masses lie on the $z$-axis
at equal distance $a$ from the origin, 
then the gravitational field assumes 
the following form
\be
V^{(\rm E)}({\bf r})=
-\frac{G m_1}{|{\bf r}-a{\bf \hat z}|}-
\frac{G m_2}{|{\bf r}+a{\bf \hat z}|}.
\ee
This potential is obviously conservative 
and axially symmetric; consequently
a test body orbiting this gravitational 
field will be described by a  constant energy
and a constant $z$-component of angular momentum.
The  potential is not
reflection-symmetric about the $x-y$ plane, 
except when the two masses are
equal. However, whatever the masses are, 
the problem is characterized by a hidden 
dynamical symmetry that leads to
an unexpected new integral of motion, 
quadratic in momenta.
This integral of motion --initially 
we shall call it `Euler's third integral'-- 
is derived by applying
the Hamilton-Jacobi method when we
perform separation of variables
in a suitable coordinate system \cite{Landau}. 
Although, this 3rd integral of motion is 
known for more than a century,
quite recently, Lynden-Bell  \cite{LyndBell}, 
trying to explain its physical meaning, 
offered a simple and straightforward
constructive method to build it. 
He proved that its kinetic part is the 
scalar  product of the angular momenta 
about the two centers of mass and defined it as:
\be
I_3=\frac{1}{2}({\bf r}_1\times{\bf v})
\cdot
({\bf r}_2\times{\bf v})
-G a{\bf {\hat z}} 
\cdot 
(m_1 {\bf \hat{r}}_1-
m_2 {\bf \hat{r}}_2) ,
\label{I3'}
\ee
where $\bf v$ is the particle's velocity, 
${\bf r}_{1,2}={\bf r}\mp a \hat{\bf z}$ are the 
vectors from either gravitating mass to 
the test particle, while 
${\bf {\hat r}}_{1,2}$ are
the unit vectors along the directions of 
${\bf r}_{1,2}$, respectively. 
$I_3$ does not depend on the mass of the test
particle orbiting the corresponding field.
The existence of three 
independent integrals of motion, 
render the motion in such a field 
describable by an integrable
set of equations. Only a few 
known physics problems are exactly 
integrable, and all of them are 
characterized by special common 
properties (for example the motion 
in phase space lies on a 3-torus 
and each such torus is characterized by  
a triplet of fundamental frequencies).

A more appropriate coordinate system to 
study the motion in the Euler gravitational 
field is that of prolate spheroidal coordinates,
where one of the two coordinates, $\xi$, is the 
sum of the distances from the two masses compared 
to the  distance between the masses (this is the
analogue of the  radius of spherical
coordinates, but endowed with an intrinsic length
scale), while the other one, $\eta$, is the
difference of the two distances divided 
again by   the  distance between the two fixed
masses (this is the analogue of the cosine
of the  polar angle in spherical coordinates). 
The third coordinate is the usual azimuthal angle
$\phi$ of spherical, or cylindrical coordinates.
Thus
\bea
\xi &= &\frac{r_2+r_1}{2 a}=
\frac{\sqrt{\rho^2+(z+a)^2}+\sqrt{\rho^2+(z-a)^2}}{2a},  \\
\eta &=& \frac{r_2-r_1}{2 a}=
\frac{\sqrt{\rho^2+(z+a)^2}-\sqrt{\rho^2+(z-a)^2}}{2a},
\eea
where $\rho$, $z$ are the usual cylindrical
coordinates. The spheroidal coordinates take values
within the intervals: $\xi \in [1,+\infty)$ 
and $\eta \in [-1,1]$. 
The inverse coordinate transformation yields
\bea
\rho &=& a \sqrt{(\xi^2-1)(1-\eta^2)}, \\
z &=& a \xi \eta.
\label{xitorho}
\ee
In terms of spheroidal coordinates $(\xi,\eta)$
the Euler potential assumes the following form
\be
V^{(E)}(\xi,\eta)=
-G \frac{(m_1+m_2) \xi + (m_1-m_2) \eta}{a (\xi^2-\eta^2)}.
\label{VEinspher}
\ee

The Lagrangian (per unit test-mass) of the 
Euler problem in spheroidal coordinates is then:
\bea
L=\frac{1}{2}a^2 \left[
 (\xi^2 - \eta^2)
 \left( \frac{{\dot\xi}^2}{\xi^2-1} +  \frac{{\dot\eta}^2}{1-\eta^2} \right)\right. \nonumber\\
+\left. \dot{\phi}^2 (\xi^2-1)(1-\eta^2) 
\right] -V^{(E)}(\xi,\eta).
\eea
The corresponding canonical momenta in these coordinates are
\bea
p_\xi &=& a^2 \frac{\dot{\xi}}{\xi^2-1} (\xi^2-\eta^2),  \nn \\
p_\eta &=& a^2 \frac{\dot{\eta}}{1-\eta^2} (\xi^2-\eta^2), \nn \\
p_\phi &=& a^2 (\xi^2-1)(1-\eta^2) \dot{\phi}.
\eea
and the Hamiltonian (per unit test-mass) assumes the following form
\bea
H &=&
\frac{p_\xi^2}{2a^2}\frac{\xi^2-1}{\xi^2-\eta^2} + \frac{p_\eta^2}{2a^2}\frac{1-\eta^2}{\xi^2-\eta^2} \nonumber \\
&&+ \frac{p_\phi^2}{2 a^2 (\xi^2-1)(1-\eta^2) }
+V^{(E)}(\xi,\eta).
\eea

According to Landau's analysis \cite{Landau} 
which is based on constructing the most general 
separable potential in such coordinates (called
elliptical in Landau's textbook), the 
separability of the particular
problem arises from
the very fact that the numerator in  equation
(\ref{VEinspher}) is a linear superposition
of a function of $\xi$ alone and a  function of 
$\eta$ alone. The integrability of this potential 
then arises as a direct consequence of the 
separability of Hamilton-Jacobi equation.
The third conserved quantity, $\beta$ in
\cite{Landau}, besides the energy $E$ 
and the $z$-angular momentum $L_z=p_z$,   
gets the following form in spheroidal coordinates:
\bea
\beta&=&(\xi^2-1)p_{\xi}^2+\frac{L_z^2}{\xi^2-1}
-2a^2 (\xi^2-1) E\nonumber \\
&&-2G(m_1+m_2) a \xi,  \label{theI3xi}\\
&=& - (1-\eta^2)p_{\eta}^2-\frac{L_z^2}{(1-\eta^2)}+2a^2 (1-\eta^2) E\nonumber \\
&&+2G(m_1-m_2) a \eta.
\label{theI3}
\eea
The two alternative expressions for $\beta$ in 
Eqs. (\ref{theI3xi},\ref{theI3}) 
are pure functions of $p_\xi$ and $\xi$, or 
$p_\eta$ and $\eta$, respectively, 
clearly demonstrating  the separability of the problem.
The constant $\beta$ is related with the expression 
for $I_3$ of Eq.~(\ref{I3'}) by $\beta=-2I_3$,
as one can verify by combining both expressions 
for $\beta$ (Eqs.~(\ref{theI3xi}, \ref{theI3})), expressed 
in cylindrical coordinates 
and performing a lengthy, but straightforward,
computation.

\subsection{The oblate version of the Euler problem}
\label{sec:2.2}

As mentioned in Section \ref{sec:1}, 
the gravitational field of the Euler problem
describes,  by construction, a prolate 
distribution of mass as a source 
(this will become more obvious later on, 
in Sec. \ref{sec:4.1}, when
we will present the multipolar structure 
of the Newtonian problem). 
Therefore it does not resemble
the gravitational field of a Kerr black hole, 
which is obviously oblate (its quadrupole
moment is negative).
However, it is easy to transform the 
original Euler problem into an oblate 
field by simply rotating
$a$ into a complex plane by $\pi/2$. 
Then $a$ will  transform into a purely
imaginary distance, but the gravitational 
field will still be real,
in the symmetric case where $m_1=m_2=M/2$. 
Only then the gravitational potential of 
each mass is given by the complex conjugate 
function of the potential of the other mass.
In order to avoid confusion we will keep 
considering the $a$ parameter real 
and simply replace  $a$ by $i a$ in the potential.
The corresponding gravitational field 
--henceforth called the oblate Euler
field-- assumes the following form:
\be
{V}^{(oE)}=
-\frac{G (M/2)}{|{\bf r}- i a {\bf \hat{z}}|}
-\frac{G (M/2)}{|{\bf r}+ i a {\bf \hat{z}}|}
\ee
where by $|{\bf k}|$ we mean 
$\sqrt{{\bf k}\cdot{\bf k}}$.
The latter vector product is  a 
complex number and in order to keep the
square root single-valued we should adopt 
a branch cut. We have chosen the negative real 
axis of the vector product as the branch-cut of
our potential function.  After some algebra 
the new potential 
(from now on we will use only this potential, so we will simply write it $V$) takes the following form
in usual spherical coordinates:
\be
V({\bf r})=-\frac{G M}{\sqrt{2}}
\frac{\sqrt{R^2+r^2-a^2}}{R^2},
\ee
where 
\bea
R^2 &=& \sqrt{(r^2-a^2)^2+(2 a {\bf r}\cdot 
{\bf \hat{z}})^2} \nonumber \\
&=&
\sqrt{(r^2-a^2)^2+4 a^2 r^2 \cos^2\theta }.
\ee
Although it is not obvious at this point that 
the new potential describes an
actually oblate field, its true character will 
be unequivocally revealed in 
Section \ref{sec:4.1}, where its multipole moments
are written.

It should be noted that the new field $V({\bf r})$ 
is defined everywhere
since $R^2 \geq r^2-a^2$, except of 
along the equatorial circle $(r=a, \theta=\pi/2)$
where the potential becomes indeterminate, 
since then  $R^2=r^2-a^2=0$. Also,
on the equatorial disk $(r<a, \theta=\pi/2)$ 
the  potential vanishes.
The oblate Euler field is 
reflection symmetric, as the original
prolate Euler field when the masses 
of its two gravitational centers are equal.

A more appropriate coordinate system to study 
the motion in this oblate field is that
of oblate spheroidal coordinates, 
$(\xi, \eta, \phi)$, which are defined as:
\bea
x &= & a \sqrt{(1+\xi^2)(1-\eta^2)}\cos{\phi},  \\
y &=& a \sqrt{(1+\xi^2)(1-\eta^2)}\sin{\phi},  \\
z &=& a \xi \eta,
\eea
where $\xi \in [0,+\infty)$, $\eta \in [-1,1]$ 
and $\phi \in [0,2\pi)$. The surfaces of constant 
$\xi$-coordinate are oblate ellipsoids of revolution
with focal circle $(r=a, \theta=\pi/2)$, while the
surfaces of constant $\eta$-coordinate  are one-sheet
half hyperboloids of revolution sharing the same focal
circle with the above ellipsoids.

In terms of oblate spheroidal coordinates
the Euler potential assumes the following simple form
\be \label{VoE}
V(\xi,\eta)=-
\frac{G M \xi }{a (\xi^2+\eta^2)}.
\ee
The Lagrangian (per unit test-particle mass) 
of the oblate Euler potential becomes:
\bea
L&=&\frac{1}{2}a^2 \left[(\xi^2 + \eta^2)
\left(  \frac{\dot{\xi}^2}{\xi^2+1} +
\frac{\dot\eta^2}{1-\eta^2} \right) \right.
\nonumber \\
&& \left.
+{\dot\phi}^2 (\xi^2+1)(1-\eta^2) 
\right] -V(\xi,\eta),
\eea
while the corresponding Hamiltonian is
\bea\label{Ham}
H&=&\frac{1}{2a^2} \left[
p_\xi^2 \frac{\xi^2+1}{\xi^2+\eta^2} + 
p_\eta^2 \frac{1-\eta^2}{\xi^2+\eta^2} \right. 
\nonumber \\
&&+\left.\frac{p_\phi^2}{(\xi^2+1)(1-\eta^2) }
\right] +V(\xi,\eta),
\eea
with the canonical momenta defined as:
\bea\label{momj}
p_\xi&=& a^2 \frac{\xi^2+\eta^2}{\xi^2+1} \dot{\xi}  
\label{momjr}\\
p_\eta&=&a^2 \frac{\xi^2+\eta^2}{1-\eta^2} \dot{\eta} 
\label{momjh}\\
p_\phi&=& a^2 (\xi^2+1)(1-\eta^2) \dot{\phi}. 
\eea
Repeating Landau's argument \cite{Landau}, for the
oblate field now, the very fact that the numerator in
Eq.~(\ref{VoE}) is again a linear superposition
of a function of $\xi$ and a  function of $\eta$ (no
presence of $\eta$ function here) lies behind the
separability of the given problem, and consequently, 
the integrability of this particular potential.
The third conserved quantity, $\beta$, besides 
the energy $E$ and the $z$-angular 
momentum $L_z=p_\phi$, takes the following form:
\begin{eqnarray}
\beta=&-(1-\eta^2) p_\eta^2
-\frac{L_z^2}{(1-\eta^2)}-2a^2 E (1-\eta^2)  \quad
 \\
=& (\xi^2+1) p_\xi^2-
\frac{L_z^2}{(\xi^2+1)}-2a^2E(\xi^2+1)- 2GMa\xi.
\label{theI3oc}
\end{eqnarray}
Once again, the separability of the problem is 
clearly manifested in these two expressions
since the 4D phase space of $\xi,p_\xi,\eta,
p_\eta$  breaks in two  independent 2D
phase spaces $\xi,p_\xi$ and $\eta,p_\eta$, 
and the motion evolves along a closed line
in each of these two phase planes.

\section{The orbital characteristics 
in oblate Euler}
\label{sec:3}

The gravitational potential of the  
Euler problem 
(henceforth we will only deal with 
the oblate version
of the Euler problem and we will 
omit any specific
notation mark) describes a 
conservative axisymmetric 
field that admits a constant of motion, 
as we have mentioned earlier, 
that is quadratic with respect to
momenta. This new constant could 
be considered 
as an analogue of the square 
of angular momentum 
of central fields. The new field 
is by construction not central 
though, but its dynamical 
structure is such 
that it renders the problem integrable.
Furthermore, the choice of imaginary distance 
between the two masses renders the field oblate 
with respect to its dynamics, 
instead of prolate.
This very fact make it more 
physical with respect to
qualitative resemblance with 
spinning astrophysical 
objects.

The Kerr metric is a relativistic 
object of extreme astrophysical 
interest, which shares a lot of
general properties with the Euler field as it will 
be shown in the following Sections of the article.
Both gravitational fields are (i) integrable 
(with respect to the description of geodesic orbits
of test particles orbiting around them), characterized by three, similar in context, constants of motion,
(ii)  have similar multipolar characteristics, 
and (iii) are fully described by only two physical
parameters, their total mass and the spin parameter 
(for the Kerr) or the imaginary part of the distance
between the two masses (for the Euler).

In the following section we will further 
study the orbital characteristics of the Euler problem
in order to demonstrate the extent of similarity 
between the two fields.
 
\subsection{Equations of motion }
\label{sec:3.1}

In order to compare the equations of motion 
in the  Euler potential with those of Kerr 
we define new coordinates:
\begin{eqnarray}
    r&=&a\xi,\\
    \theta&=&\cos^{-1}{\eta}.
\end{eqnarray}
The new $(r,\theta)$ coordinates play  the role of 
the radial and the longitudinal Boyer-Lindquist (BL)
coordinates of Kerr space-time, respectively. 
Although equivalent to the oblate spheroidal 
coordinates $\xi,\eta$, the new 
coordinates $r,\theta$ are better suited 
to reveal the analogies with the 
corresponding orbits of Kerr metric.
The 3rd coordinate, $\phi$, is the usual 
azimuthal angle that is common in both problems.
The comparison will be further simplified
by adopting geometrized units ($G=c=1$) in 
the Newtonian field as well.

The Euler potential in these new coordinates is given by:
\begin{equation}
    V(r,\theta)=-\frac{M r}{r^2+a^2 \cos{\theta}},
\end{equation}
while the corresponding Hamiltonian (\ref{Ham}) yields the following form:
\begin{equation}\label{Hamr}
H=\frac{(r^2+a^2)p_r^2+p_{\theta}^2}{2 \Sigma}
+\frac{p_{\phi}^2}{2 (r^2+a^2) \sin^2\theta} 
-\frac{M r}{\Sigma},
\end{equation}
where $\Sigma=r^2+a^2 \cos^2\theta$,  while $p_r$,
$p_{\theta}$, $p_{\phi}$ are the canonical momenta 
with respect to $r,\theta,\phi$, respectively. 
The momenta  $p_r$ and $p_{\theta}$ 
are related with the momenta $p_{\xi}$ and 
$p_{\eta}$ (c.f.,~Eqs.~(\ref{momjr}, \ref{momjh})), 
respectively, through the following relationships:
$p_r=p_{\xi}/a$ and $p_{\theta}=-\sin{\theta}\; p_{\eta}$.
Since $\phi$ coordinate is missing 
from the Hamiltonian, $p_\phi$ is conserved, 
and henceforth we will write it, instead, $L_z$.

Applying the Hamilton-Jacobi method in the 
above Hamiltonian (\ref{Hamr}), we obtain the 
following separated equations of motion 
for a test particle:
\begin{eqnarray}
\Sigma \left(\frac{dr}{dt}\right)
&=&\pm \sqrt{V_r(r)}, \label{quadr1} \\
\Sigma \left(\frac{d\theta}{dt}\right)
&=&\pm\sqrt{V_{\theta}(\theta)},\label{quadr2} \\
\frac{d\phi}{dt}&=&
\frac{L_z}{(r^2+a^2)\sin^2{\theta}}. \label{quadr3}
\end{eqnarray}
The radial potential $V_r(r)$ and 
the longitudinal potential $V_{\theta}(\theta)$ 
introduced in Eqs.~(\ref{quadr1}, \ref{quadr2}) 
are given by:
 \begin{align}
V_r(r)&=2Er^4+2 M r^3+\left( 2 a^2 E-Q-L_z^2 \right)r^2
+2 M a^2 r-Q a^2,
\label{Vr}\\
V_{\theta}(\theta)&=Q-\cos^2{\theta}\left(-2a^2E+
\frac{ L_z^2}{\sin^2{\theta}}\right).
\label{Vth}
\end{align}
where $E$ is the constant value of the Hamiltonian, $L_z$ is the conserved $z$-component of its angular momentum 
and  $Q$ is a third integral of motion
that naturally emerges from 
the above separation of variables, while  $t$ 
is the Newtonian time parameter. In the next 
section \ref{sec:3.2}, we define all these constants 
of motion in detail.

At this point, it should be noted that 
exactly the same equations of motion, but with slightly
different potentials $V_r, V_\theta$, show up 
in the description of the geodesics
in Kerr metric. However in Kerr case the proper time
$\tau$, instead of the coordinate time $t$, 
is the evolution parameter of the
spatial BL coordinates $r,\theta,\phi$. 

The Eulerian orbits are performing a radial 
oscillation and a precession while they revolve
around the axis of symmetry. The characteristics 
of $V_r$ are responsible for the radial 
oscillation, while those of $V_\theta$
are responsible for the oscillation of the
test particle about the equatorial plane.

\subsection{Bound orbits}
\label{sec:3.2}

As demonstrated in   the previous section
(c.f.~Section \ref{sec:3.1}), 
the Euler potential admits three constants of motion, 
which in terms of coordinates and momenta are given 
by the following expressions:
\begin{eqnarray}
E&=&\frac{p_r^2(r^2+a^2)+p_{\theta}^2}{2 \Sigma}
+\frac{p_{\phi}^2}{2(r^2+a^2)\sin^2{\theta}}
-\frac{ M r}{\Sigma},\label{constE} \\
L_z&=&p_{\phi},\label{consL}\\
Q&=&p_{\theta}^2+\cos^2{\theta}
\left(-2 E a^2+\frac{L_z^2}{\sin^2{\theta}} \right)
\label{consQth}\\
&=&-p_r^2(r^2+a^2)+2 E r^2+2 M r-\frac{L_z r^2}{r^2+a^2}. \label{consQr} 
\end{eqnarray}
The third integral of motion $Q$,
written in two alternative forms in Eqs.~(\ref{consQth}, \ref{consQr}), one with respect to
$r,p_r$, and one with respect to $\theta,p_\theta$,
is related to Lynden-Bell's $I_3$, and Landau's $\beta$,
through the relation: 
\begin{equation}
  \beta=-2 I_3 = -Q-L_z^2-2a^2 E.
\end{equation}
We have decided to use $Q$, instead of
$\beta$ and $I_3$, as the third integral 
of motion, because this form of $Q$ 
could be considered as the Newtonian 
analogue of Kerr's  Carter constant, 
as it will be shown later.

The two potentials $V_r,V_\theta$, presented in the
previous section, share a lot of similarities with 
the corresponding potentials of Kerr. 
More specifically, the later one, $V_\theta$, 
yields exactly the same form as $V_\theta$ of Kerr, 
if we simply replace $E$ by $(E^2-1)/2$ (see
Section \ref{sec:5.2}), while $V_r$ is of order four, 
 like that of Kerr, and most of the 
polynomial coefficients coincide with 
those of Kerr, if the previous reparametrization of 
$E$ is imposed here as well. 
Especially the fact that $V_r(r)$ is 
a quartic polynomial, leads to 
the possibility of two families of 
bound orbits: (a) one with lower radii: 
$0 \leq r_4 \leq r \leq r_3$, 
coexisting with another one with $r_3 
\leq r_2 \leq r \leq r_1$, where 
$r_1,r_2,r_3,r_4$ are real roots 
of the polynomial $V_r$, and (b) 
one with only a single range of radii
$r_2 \leq r \leq r_1$, while the other 
set of roots of $V_r$
are then complex conjugate to each other.
We will consider  bound orbits that
correspond to  the farthest family, 
if two of those exist.
The reason is the following: the bound geodesic 
orbits in Kerr are either those that remain
at the exterior of the event horizon, 
or plunging orbits that eventually cross the black 
hole horizon. The former ones are the ones 
at higher values of radii far from the horizon. 

\begin{figure}[b]
\includegraphics[width=\linewidth]{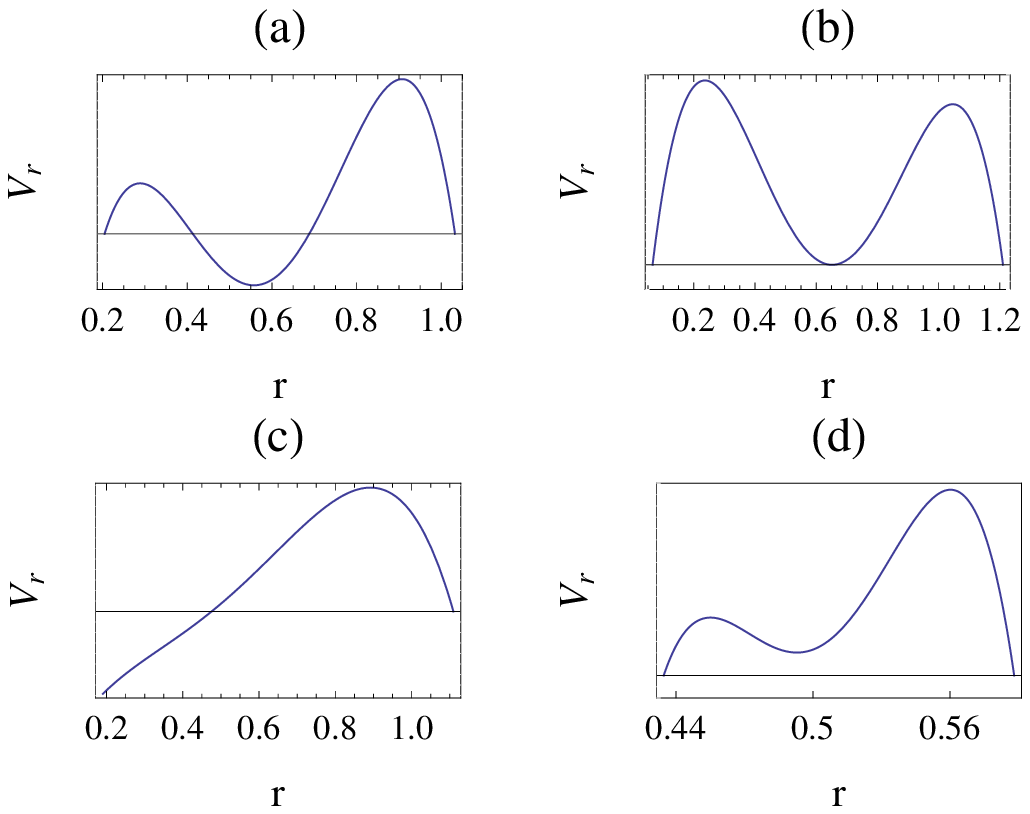}
    \caption{\label{fig:potentials}
    This is a collection of different 
    forms of potentials $V_r$ (by
    choosing different sets of
    parameters $E,L_z,Q$) leading 
    to different types of orbits. 
    The orbit evolves in-between 
    two successive roots of the 
    potential where $V_r>0$. 
    In plot (a) there are two distinct 
    regions of allowed $r$'s.
    The bound orbits we are mostly 
    interested in are orbits in the exterior 
    allowed region (the ones of the interior
    region are `plunging  orbits'. 
    In plot (b)     a marginal case 
    for the potential is depicted. Two of the
    roots of the potential coincide ($r_2=r_3$), 
    so the normal bound orbit spends infinite 
    time approaching $r_2$. This is 
    a separatrix case which is further 
    discussed in the next Section. 
    Plots (c) and (d) show two cases where
    two of the roots are complex. The former 
    one is  a normal bound orbit without 
    any plunging dual, while in the latter 
    one the two regions of plot (a) have been merged
    forming an `effectively plunging orbit'.
    We will not consider these in our analysis, since their analogue in Kerr refers to a geodesic orbit
    that eventually plunges under the 
    horizon of the black hole.}
    \end{figure}

The family of orbits in Euler corresponding to
lower radii, when
both families are present, will be considered
`plunging orbits' at close analogy to those of Kerr.
The second type (b) of bound orbits in Euler, 
with a single range of allowed radii, 
could either describe a normal bound orbit
(without any plunging pair), 
or an effectively `plunging orbit' in the sense 
that the two distinct regions of bound orbits 
of the first type have merged into a single region
through a potential neck that will eventually 
drive an orbit to a plunging one (see Figure 
\ref{fig:potentials}(d)). 
The latter type of orbits will not 
be treated as simple eccentric orbits
with a periastron and an apastron,
since it does not seem natural to describe 
them as orbits with a specific semi-latus 
rectum and eccentricity. 
In our description of possible bound orbits
in the Euler potential we will not 
consider such type of orbits.

Next we will follow the same procedure one uses to
study the bound orbits of Kerr: we will 
parametrize the roots of $V_r(r)$ as follows
\begin{equation}
    r_1= 
    \frac{p}{1-e} \quad,\quad r_2= \frac{p}{1+e},
\end{equation}
assuming they correspond to the outer   bound orbit 
(if there exists an inner region as well) 
described by the dimensional semi-latus rectum $p$ 
and the eccentricity $e$.
The rest of the roots of $V_r$, $r_3,r_4$ (either
real or complex)
could then be computed as functions of the orbital
parameters $p,e$ and the inclination angle of 
the orbit $\pi/2-\theta_{\min}$ 
(where $\theta_{\min}$ is the lowest polar angle of the orbit). 
The set of the three orbital parameters
$(p,e,\theta_{\min})$ could be used not only to compute the 
roots of $V_r$, but from them one could  compute the constants of motion, as well
(see Appendix \ref{App:2}).

Meanwhile, the oscillation of $\theta$ parameter around the equatorial plane ($\theta=\pi/2$) is governed by $V_\theta$ potential as mentioned previously.
The roots of $V_\theta$ are two real supplementary angles
which correspond to the turning points of orbital-plane oscillation and two complex imaginary
angles. The roots of $V_\theta$ are described in 
Appendix \ref{App:2}.

As mentioned above, the constants of motion 
$E$, $L_z$ and $Q$ 
are directly related to the orbital parameters
$p, e, \theta_{\min}$, but they  are not as easy to
handle as the orbital parameters. Although analytic
expressions for $p,e,\theta_{\min}$ as functions of
$E,L_z,Q$ could be written they are quite involved.
Furthermore, by fixing the constants of motion, one 
could get a set of two bound orbits, 
an interior one and an exterior one, 
but then one has to chose to which one a semi-latus
rectum and an eccentricity should be assigned.
In contrast, as long as one gets restricted in a
meaningful space of $p,e,\theta_{\min}$ the orbit is
unambiguously determined. This is actually the reason
we have chosen to use the orbital parameters 
in order to parametrize the orbits.


\subsection{The separatrix}
\label{sec:3.3}

In the 3-dimensional space of 
orbital parameters $(p,e,\theta_{\min})$
there is a special surface,  
which  corresponds 
to a pair of orbits: one normal bound orbit 
and a ``plunging'' one that share a common
turning point, that is $r_2=r_3$ (c.f.~Figure 
\ref{fig:potentials}(b)). 
This surface is the separatrix. 
The normal bound orbits of the separatrix 
are actually marginally stable orbits. 
Eventually these orbits will evolve 
into circular orbits
with radius $r(t \to \infty)=r_2=r_3$.
One expects that a slight variation 
of the physical
parameters of the orbit ($E,L_z,Q$), due to
any kind of dissipative self-force acting on 
the test particle,
could cause the two families of orbits 
to either communicate 
(by transforming the normal bound orbits 
into ``effectively plunging''
orbits), or move the two types of orbits
further apart.
Actually, the neighborhood of mostly 
the whole surface of the separatrix
corresponds to the latter case. Both sides
of the surface (but close to it) 
describe pairs of two distinct
separated families of orbits, one above the
separatrix with orbital parameters
($p_1,e_1,\theta_{\min}$) and one below the separatrix
with orbital parameters 
($p_2,e_2,\theta_{\min}$) with $p_1>p_s>p_2$,
such that both are described by the same $V_r$ potential with the same contants of motion.
The one with $p_2$ is actually the plunging one,
dual to the normal one with $p_1$; therefore we will 
deal only with orbits located `above' the
separatrix.

Near the edge of the separatrix (corresponding 
to the most inclined orbits of the separatrix)
there are orbits that are effectively
plunging ones like that of Figure
\ref{fig:potentials}(d) and as we mentioned
earlier, we will not study such orbits.

\begin{figure}[b]
    \includegraphics[width =\linewidth]{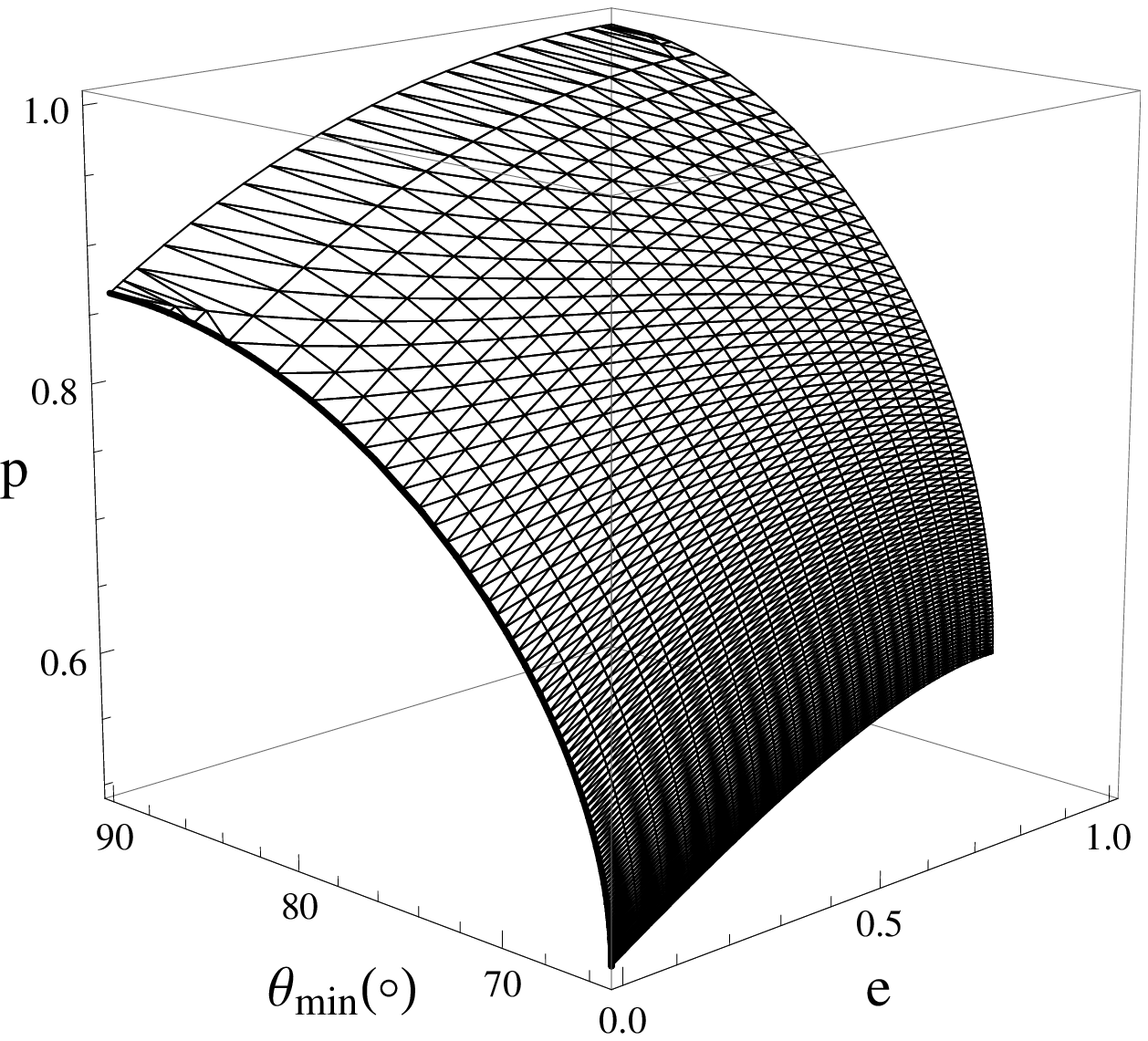}
    \label{fig:separa}
    \caption{The surface depicts the orbital 
    parameters of the  
    separatrix of the Euler problem in
    ($e,p,\theta_{\min}$)-space, for $a=0.5 M$. 
    The normal bound orbits we consider
    lie above this separatrix surface.
    The thick line (in front) corresponds 
    to spherical ($r={\rm const}=r_1=r_2=
    r_3$) orbits, that is, to orbits 
    with $e=0$. The upper left corner 
    of the surface at 
    $e=0,\theta_{\min}=\pi/2$
    corresponds to the ISCO. This plots shows 
    clearly that the separatrix extends up to 
    a minimum value of $\theta_{\min}$ which is a
    function of $e$ (the bottom boundary of the
    surface). It is obvious that there is no 
    separatrix of the Euler problem for orbits 
    that have a large inclination. The 
    semi-latus rectum is measured in units 
    of $M$.} 
    \end{figure}

The separatrix could be described as follows:
For a given pair of eccentricity, $e$, and
inclination, $\pi/2-\theta_{\min}$, there is 
a specific semi-latus rectum $p_s(e,\theta_{\min})$
that brings the two types of orbits (the exterior
normal orbit and the interior plunging one) in touch.
For equatorial orbits,
$\theta_{\min}=\pi/2$, one could easily obtain an
analytic expression for $p_s(e,\theta_{\min}=\pi/2)$
(by setting $Q=r_4=0$ and, $r_3=r_2=p/(1+e)$,
while $r_1=p/(1-e)$ in Eq.~(\ref{Vr}), and write it in 
terms of the roots of the polynomial).
For generic inclined orbits though it 
is a bit more difficult to obtain an
analytic expression for $p_s$ as a function 
of $e$, and $\theta_{\min}$. We found useful
to introduce an additional parameter $x := r_4/r_3$, 
in order to write an analytic expression 
for $p_s(e,x)$ and then plot the surface
$p_s(e,\theta_{\min})$ in parametric form, since
$\theta_{\min}$ itself could be directly expressed 
as a function of $e,x$ as well. In Figure
\ref{fig:separa} the separatrix surface has been
plotted for a specific value of $a$, namely $a=0.5 M$. All the above
analytic derivations are thoroughly analyzed
in Appendix \ref{App:3}.

The separatrix extends from $\theta_{\min}=\pi/2$
(equatorial orbits) to a minimum value of
$\theta_{\min}$ that depends --not very 
sensitively-- on the eccentricity, $e$. 
More specifically $\theta_{\min} \simeq 
65.5 ^\circ$, for $e=0$, and it increases 
monotonically to $\theta_{\min} \simeq 
70.5 ^\circ$, for $e=1$. 
The overall shape of the surface is the same 
for any value of $a$, while the value
of $p_s$ scales 
linearly with $a$, assuming its 
highest value, $p_{s,\max}=2 a/M$, for $e=1$, 
$x=0$ (which corresponds to 
$\theta_{\min}=\pi/2$). Beyond 
the lower $\theta_{\min}$ values
there is no separatrix; that is, there 
are no more four real roots of $V_r$. 
A complex pair of roots arise then.

Now let us  study in further detail  the region 
of parameter space outside the separatrix. 
On the region just above (but not far from) the separatrix ($p>p_s$) the two types of orbits (stable bound and ``plunging'') get separated ($r_2>r_3$). As we
mentioned previously, below the separatrix the
order of roots is alternated $r_2<r_3$; 
therefore there is lack of physical description 
of such orbits based on the
assumption that the order of roots
is $r_1 \geq r_2 \geq r_3 
\geq r_4$ and the bound orbit
oscillates radially between $r_1$ and $r_2$. 
Each such point ($p,e,\theta_{\min}$),
located below the separatrix, 
has its dual above the separatrix 
with a different set  $p',e'$
--but with the same inclination--
such that $r_1=p/(1-e)=
r'_1=p'/(1-e') 
\geq r'_2=p'/(1+e')=r_3
\geq r'_3=r_2=p/(1+e) \geq r'_4$,
therefore both these points correspond to a normal
bound orbit that oscillates radially between
$r'_1$ and $r'_2$.
Far from the separatrix surface
(either below or above it) the potential $V_r$
looses a pair of real roots, thus, then, 
there is only a single bound orbit
corresponding to it. Finally in a rather 
narrow space around the 
boundary of the separatrix, corresponding 
to the lower possible value of $\theta_{\min}$,
there is a bizarre type of bound orbits (the
effectively plunging orbits) arising from the
merging of a stable bound orbit with a 
``plunging'' one. The potential in
such cases has a local minimum between 
its two real roots (see Figure
\ref{fig:potentials}(d)). Although we will not deal 
with such orbits, there is 
a finite lower and a finite
higher allowed radius for those as well, so one could still use the analytic expressions
for the frequency of the radial oscillations, 
which we will introduce later on.

Especially the boundary of the 
separatrix with $e=0$
corresponds to margin-ally 
stable ``spherical''
orbits (or, as they are usually called in 
Kerr metric, ``circular'' orbits).  
It should be noted that
the boundary of the separatrix 
corresponding to equatorial orbits
($\theta_{\min}=\pi/2$) is given by a
monotonically increasing function $p_s(e)$,
like in Kerr, but it has the opposite
sign of curvature. In both 
problems the ``corner'' of the separatrix
at $e=0$ and $\theta_{\min}=\pi/2$ 
(equatorial orbit), which corresponds 
to the ISCO, represents the lowest
semi-latus rectum among all marginally 
stable spherical orbits.

\section{Known analogies}
\label{sec:4}

In this section we present a list of 
the analogies between the two problems, 
the Euler and Kerr, that have already been shown 
in the literature by various authors in the past.
Most of these analogies were  presented 
in different context from the one
followed in this article and in most
of them there is no clear connection
between Kerr and oblate Euler field.  

\subsection{Basic common characteristics 
and fundamental differences}
\label{sec:4.0}

During the golden era of black holes, when 
extensive mathematical studies had been performed,
Israel \cite{Israel70} ended up in the oblate 
Euler field (without recognizing it as such) as the 
Newtonian analogue of Kerr metric by means of 
the right source distribution of the
gravitational field. Actually the analogy between 
the two fields had been revealed even
earlier by Keres \cite{Keres1967}, but it was 
mainly focused on finding 
similar properties related to the ring 
singularity of the then recently discovered Kerr
metric and the corresponding avoidance 
of the ring singularity by geodesics. 
There was no demonstration of any connection 
between the two gravitational fields 
with respect to the orbital
characteristics in them.

The present study attempts to extend this
old found similarity between the
two fields, mostly in the direction of  
astrophysically oriented issues, like
geodesic orbits at the exterior of a 
Kerr black hole and their properties.

It should be pointed out though, that there are
fundamental differences between the 
two fields. At first glance there is a 
dimensional difference of the parameter $a$ 
showing up in the two fields. For the Kerr
metric the $a$ parameter is related 
to the spin of the Kerr black hole, thus
it has dimensions of length times velocity 
(it is actually the ratio of the angular momentum  of the black hole to its mass $a=S/M$).
In  Euler's oblate problem
the $a$ parameter is simply a length.
Although in geometrized units, both $a$ 
parameters have length dimensions, 
(or equivalently mass
dimensions) the two $a$'s still have completely
different physical origin. Despite that,
this parameter, (or its dimensionless counterpart
$a_\star=a/M$ in geometrized units where $G=c=1$),
seems to play the same role as 
an adjusting parameter
of the two fields and their correspondence.

However there is an essential difference 
in the use of $a$ in the two problems:
in Euler's problem $a_\star$ could  assume any
value, while in Kerr $a_{\star}$ cannot exceed 
the value of 1 (hyperextreme Kerr). However
the restriction for the $a_\star$ in Kerr is 
mainly related to the existence of
a horizon in Kerr, which has no analogue 
in Newtonian gravity in the first place.
Here we will restrict our study 
of the Euler field in the
range $a_\star \in (0,1]$.

Another technical difference arising 
between the two problems is the
asymmetry of Kerr metric under the 
transformation $\phi \to -\phi$ (one should
amend such a transformation with 
$t \to -t$, as well, to produce an isometry,
but then the Kerr spin parameter $a$ 
will change its sign). By contrast, 
Euler's problem (as any other type of 
axisymmetric Newtonian potential) is
completely symmetric under such a 
transformation $\phi \to -\phi$.
Consequently, the dragging of frames 
arising in stationary axisymmetric 
relativistic configurations, like in Kerr,
does not have its analogue in Newtonian gravity. 
Therefore we shall be very careful 
when we compare the quantitative characteristics 
of phenomena that are sensitive to the sense 
of rotation in Kerr, with the corresponding 
ones in Euler's field.

\subsection{Multipolar structure}
\label{sec:4.1}

More recently, Will \cite{Will} showed 
that the Newtonian axisymmetric
gravitational field that is characterized by a 
third conserved quantity analogous to
the Carter constant of a Kerr space-time 
(that is quadratic in momenta),
leading, consequently, to an integrable 
Newtonian potential, 
should have a multipolar exapnsion
that follows exactly the same relation as 
the mass multiple moments $M_{2l}$ of the
Kerr metric itself, that is
\be
M_{2 l}=M \left( \frac{M_2}{M} \right)^l,
\ee
while its odd-$l$ mass moments vanish 
(as one would expect for a reflection symmetric
gravitational field). It should be noted
that this multipolar structure
describes exactly the Euler's field 
(both the oblate and the prolate one)
with equal masses. More specifically,
if $M_2>0$ it corresponds to the original 
Euler's field (the prolate one),
while if $M_2<0$ it corresponds to the 
oblate version of the Euler's field.
Naturally, the case $M_2=0$ is  
the monopole, spherically symmetric, 
Keplerian field, for which the corresponding 
third integral of motion is simply the
square of the total angular momentum 
of an orbiting test particle,
instead of the constant $Q$, discussed 
in the previous section.

Of course a Newtonian stationary and 
axisymmetric potential cannot
reproduce any effect like the dragging 
of frames of the corresponding
relativistic field, and thus there 
are no current-mass multipole 
moments like that of Kerr. 
This is one of the basic differences 
between the two fields, and that renders 
the comparison between the corresponding orbits 
in the two fields more subtle. We will further
discuss this subtlety in Section \ref{sec:5.3}.

Later on Markakis \cite{Markakis} 
attempted to generalize
Will's result, looking for Newtonian gravitational 
fields that admit integrals of motion
of higher order (in particular quartic 
with respect to momenta). His analysis
yielded a null result. If the analogy 
between the Euler's problem
and the Kerr metric could be extended 
to problems with other type of
non-Carter-like integrals of motion, 
the negative result of
Markakis could be just a hint that 
probably there is 
no  integrable vaccuum stationary and 
axisymmetric solution describing an 
isolated object other than the Kerr in General Relativity.

 \subsection{The separability of the wave equation}
 \label{sec:4.2}

As it was shown in \cite{GlamApos12} the
gravitational field of the Euler's
problem, not only leads to a lot of 
similarities in the characteristics of 
test-particle orbits with those of the 
gravitational field of Kerr, 
but the separablity arising in the equations 
that determine the orbits (for both problems),
is also exhibited in the scalar wave equation 
in both gravitational fields. Therefore
the Newtonian type of wave equation
\be
\Box \Psi=-\kappa V \Psi,
\ee
where $V$ is  a Newtonian potential, and 
$\kappa$ is a constant with dimensions 
of frequency squared,
becomes separable in spheroidal coordinates  
if the potential is that of
the Euler's problem. Moreover, 
the wave solutions of the oblate 
Euler's problem can be written as a 
product of an angular part that has 
exactly the same form as the corresponding
angular part of the scalar perturbations 
in Kerr, while the radial part has
qualitatively very similar behavior, 
especially if $\kappa=4 \omega^2$,
where $\omega$ is the corresponding 
frequency of the wave solution.
To reveal this magnificent analogy 
one should consider the
following correspondence:
\begin{eqnarray}
\eta^{\rm(Euler)} &\leftrightarrow&
\cos\theta^{\rm (Kerr)}\quad,  \\
r^{\rm (Euler)}=a \xi^{\rm (Euler)} &\leftrightarrow& \tan\left(
a \int_{r_{\min}^{\rm (Kerr)}}^{r^{\rm (Kerr)}} \frac{dx}{\Delta(x)}
\right),
\end{eqnarray}
where $\Delta(x)=x^2-2 M x +a^2$ 
(for more details about the 
definition of $r_{\min}$ see 
\cite{GlamApos12}). The odd
correspondence between $\xi$ and Boyer-Lindquist
coordinate $r^{\rm (Kerr)}$ was chosen
in order to transform the radial part of the 
wave equation into a form that is as close 
to that of Kerr as possible, and  it 
is related to the fact that  
the radial coordinate in Kerr should not 
be interpreted as a spherical coordinate in flat 
space. For sufficiently large values of
$r^{\rm(Kerr)}$ though, $r^{\rm(Euler)}
\simeq r^{\rm(Kerr)}$, as it can be 
easily shown.

Therefore, not only the two problems 
lead to separable
wave equations, but the corresponding
eigenfunctions, on which any wave perturbation
can be decomposed, are quite similar. More
specifically the angular eigenfunctions
are exactly the same, while the radial 
ones, although not identical, they
have the same behavior at large radii.
 
\section{Revealing new analogies}
\label{sec:5}
 We devote this section in constructing 
 an extended list of new
 analogies that demonstrate the close 
 analogy between the relativistic 
 gravitational field of Kerr space-time 
 and the Newtonian gravitational
 field of the oblate Euler's problem.
 
 \subsection{Orbital precession}
 \label{sec:5.1}
 
It is well known that a bound geodesic 
orbit around a Kerr metric, generally
oscillates about the equatorial plane, 
while it revolves around the black hole. 
Thus an orbit with $L_z \neq 0$ never 
crosses  the symmetry axis; but
it oscillates within a maximum angular 
amplitude about the equatorial plane.
This angle is called the inclination 
of the orbit (see \cite{Chandrabook}),
and it could be easily obtained by computing 
the extreme  angles, $\theta_{\min},
\pi-\theta_{\min}$, constraining the
$\theta$-oscillation. This is directly 
regulated by the value of Carter constant $Q$. 
Especially, for $Q=0$ the orbit is 
strictly equatorial.
 
Since the connection between 
$\theta=\cos^{-1}\eta$, $p_\theta$ and $Q$ 
for the Euler's problem (\ref{QE1}) is exactly 
the same with that for Kerr 
(after adopting a redefinition of the constant $E\to (E^2-1)/2$, discussed 
in Section \ref{sec:3.2}),
the orbits  of a test particle in the oblate
Euler's problem will have similar
azimuthal properties with those of Kerr. 
As long as an Euler orbit is characterized 
by $L_z \neq 0$,
its  $\theta$ (related to the 
spheroidal coordinate $\eta$) oscillates  
back and forth around zero, while the
corresponding extreme values of $\theta$
are symmetrical to each other,  
determining the inclination 
of the orbit. Again when the Carter 
constant $Q$ of Euler 
vanishes, the orbit is equatorial.

Furthermore, in both problems, a bound orbit, 
while revolving around the axis of symmetry, and 
oscillating around the equatorial plane,
it also moves radially in and out between two 
extremal radii ($r_{\min} \leq r 
\leq r_{\max}$). If these two radii
are equal then we get a ``circular'' 
(or as we call it here ``spherical'')
orbit. In the Kerr case
circular orbits have been shown to be 
stable against gravitational wave perturbations.
Later on, we will demonstrate that this 
is not generally true for  bound orbits 
in Euler. Although this  is a qualitative
difference between Kerr and Euler,
this particular difference between 
the two problems 
enhances the qualitative similarity 
between the two problems, since
the resonance condition on which 
the whole argument about stability is based,
can be used for both problems.
It happens that in the Euler case 
there are physical parameters 
for which the resonance condition holds
true; consequently
the corresponding spherical orbits 
become unstable.
In Kerr case there is no such resonance for any 
bound orbit; thus spherical orbits are stable.
Section \ref{sec:6} is especially devoted
to demonstrate how this difference arises
when applying the same argument in the
two qualitatively similar problems.

Finally we should add that for both problems,
usually there is a pair of bound orbits, 
one of which corresponds to actually
plunging orbits in Kerr case 
(since such an orbit is partly buried beneath 
the horizon of the black hole). 
In contrast the Euler problem is not endowed 
with any horizons, thus this family of orbits 
are still regular orbits with lower 
extremal radii ($r_4,r_3$) 
than the corresponding extrema ($r_1,r_2$)
corresponding to its  normal counterpart 
orbit. To keep a close correspondence
with Kerr orbits though, we will baptize 
this new family of orbits  ``plunging 
orbits'' as well, and we will not study 
them furthermore.

As mentioned before, it should also be 
pointed out here that there is another
fundamental difference between
the two problems. In Kerr space-time 
the prograde orbits (orbiting at the
same sense as the spin of the black 
hole) and retrograde orbits (orbiting
at the opposite sense) are distinct; 
they have different characteristics.
This is due to the Lense-Thirring effect 
caused by the spin of
Kerr metric, and it arises due to 
the nonvanishing current-mass moments of the 
corresponding metric. 
The current-mass moments, though, 
are not present in a static 
Newtonian gravitational field; 
therefore the prograde and retrograde 
orbits  are completely equivalent
in the sense that they become identical 
under  the transformation $\phi \to  -\phi$. 
This qualitative difference between 
the two fields is emphasized in those cases
where the description of a phenomenon 
encompasses linear functions of $a$ in the Kerr metric, while only the square of $a$ 
shows up in the corresponding description
of the Euler problem. When some particular 
property of the Kerr field, that 
differentiates a prograde from a 
retrograde orbit, is compared to 
the corresponding one of the Euler field,
naturally, disagreements will appear.
We will handle this comparison with great care
by considering an average of a 
carefully chosen pair of orbits
(consisting of one prograde and 
one retrograde), both characterized 
by the same physically measurable quantity 
(see below at Section \ref{sec:5.3}). 

\subsection{The Carter-like constant}
\label{sec:5.2}
 
 The Carter constant is a conserved  quantity 
 along the geodesics in Kerr space-time.
 In Boyer-Lindquist coordinates, 
 when expressed in terms of $p_\theta$ 
 momenta, it takes the following form:
\be
Q =
 p_\theta^2 + \cos^2 \theta 
\left[ 
a^2 \left( 1-E^2 \right) +
\frac{L_z^2}{\sin^2\theta} \right]
\label{QQ1}
\ee
while, when expressed in terms of $p_r$ momenta,
it takes a quite different form:
\begin{eqnarray}
Q &=&\frac{\left[E
(r^2+a^2)-a L_z \right]^2}
{\Delta} -(L_z-a E)^2 
- r^2 - \Delta \; p_r^2,
\label{QQ2}
\end{eqnarray}
where $\Delta=r^2-2 M r +a^2$. 
In the expressions above,
$E$, $L_z$ 
are the conserved energy 
($-p_t$) and the  conserved  
$z$-component of the angular 
momentum ($p_\phi$) of an
orbiting test particle. Both momenta 
are proportional 
to the particle's rest mass $\mu$;
thus  $Q$ itself is 
proportional to $\mu^2$. 
Equivalently, one could construct 
the reduced quantities
$\tilde{Q}=Q/\mu^2$, 
$\tilde{E}=E/\mu$,
$\tilde{L_z}=L_z/\mu$,
$\tilde{p}_\theta=p_\theta/\mu$,
$\tilde{p}_r=p_r/\mu$, 
and rewrite the above expressions in terms of the
corresponding reduced quantities, 
that are not related to the test particle, 
but only on the particular geodesic.
For simplification we are going to use
Eqs.~(\ref{QQ1}, \ref{QQ2}) themselves, 
without the tilde signs, but referring 
to the reduced quantities.
This is equivalent to the quantities 
of a test particle with rest mass $\mu=1$.

Earlier (in Section \ref{sec:3.2}) we  
derived Euler's third integral 
of motion (c.f.~Eqs. (\ref{consQth}, \ref{consQr})). 
By rewriting the energy $E$ of the Euler
field
in terms of its relativistic
analogue at the non-relativistic limit:
\be
(E^{\rm (K)})^2 =
2 E^{\rm (E)}+1,
\ee
we get, after omitting the corresponding $^{\rm (K)}$
marks, the following expressions
for the Euler's Carter constant:
\be
Q=p_\theta^2+\cos^2\theta \left[ a^2(1-E^2)+\frac{L_z^2}{\sin^2\theta} \right] ,
\label{QE1}
\ee
in terms of $\theta,p_\theta$, 
and after some term rearrangement:
\be
Q&=&\frac{\left[E(r^2+a^2)-a L_z \right]^2}{r^2+a^2}
 -(L_z-a E)^2  -r^2 - (r^2+a^2) p_r^2+2Mr ,
\label{QE2}
\ee
in terms of $r, p_r$.
Although the second expression for 
$Q$ could be written in a simpler form where
the $a$ parameter shows up solely through 
$a^2$ (the terms linear in $a$ in Eq.~(\ref{QE2}) 
vanish if we expand it), we have chosen 
the above formulation in order to have a closer
comparison with the corresponding expression 
for Kerr (see Eq.~(\ref{QQ2}) above).
Note that the first formula (\ref{QE1}) 
has exactly the same form as the Carter constant 
of Kerr (\ref{QQ1}), while the second one (\ref{QE2}), 
although quite similar (both are rational 
functions built from polynomials of the 
same order with respect to $r$), 
apparently
it does not matches exactly the  
form of $Q$ for Kerr.  
However, it should be noted that 
the presence of the mass of the 
source, $M$, only at the last term 
of the expression (\ref{QE2}) 
for Euler's $Q$, although it seems 
to have no analogue term in the corresponding 
expression for $Q$ in Kerr, this is misleading. 
The presence of $M$ in $\Delta$ in 
Eq.~(\ref{QQ2}) will show up exactly 
as $2 M r$ when $\Delta$ in the 
denominator of the first term is expanded 
at lowest order with respect to $M/r$
and assume that the particle moves at the
weak gravitational field where $E \simeq 1$.
Of course the two problems are
not identical to each other, and 
this is the best analogy between the two 
expressions we could achieve. At the weak field
limit though the similarity is even better.

 \subsection{The ISCO}
 \label{sec:5.3}

The usual gravitational field of a Newtonian 
monopole  lacks the analogue of an innermost 
stable circular orbit (ISCO) which is 
present in the gravitational field of 
a Schwarzschild black hole. 
The reason is that the centrifugal 
potential in Newtonian 
gravitational fields acts always as a
repulsive potential which, compared to a suitably soft gravitational field
like that of a Newtonian monopole, it ensures that 
there is always a stable circular orbit at any
radius. This could be considered as a clear 
difference between the relativistic (Kerr) 
and the Newtonian (Euler) gravitational field 
at the limit $a=0$.

This fact is deceiving though. The Euler's problem 
has also an ISCO for any non-vanishing value of $a$.
If one considers a circular orbit on the equatorial 
plane ($\theta=\pi/2$) of the Euler's
oblate field, then the corresponding effective 
potential in cylindrical coordinates is
\be
V_{\rm eff}=\frac{ L_z^2}{2 \rho^2}+V(\rho,z=0)=
\frac{{L_z}^2}{2 \rho^2}-\frac{ M}{\sqrt{\rho^2-a^2}},
\ee
where $L_z$ is the reduced $z$-angular momentum 
of the test particle (or the angular 
momentum for a unit mass particle).
In terms of the oblate spheroidal coordinates 
instead, the above potential  assumes 
the following form:
\be
V_{\rm eff}(\xi,\eta=0)=
\frac{L_z^2}{2 a^2 (\xi^2+1)}-
\frac{M}{a \xi} .
\ee
By solving simultaneously the equations 
$V_{\rm eff,\xi}=V_{\rm eff,\xi\xi}=0$,
corresponding to the presence of an ISCO, 
leads to $\xi_{\rm ISCO}=\sqrt{3}$.
Thus the oblate Euler problem does 
have an ISCO, the actual radius of which
in cylindrical coordinates is
\be
\rho_{\rm ISCO}=a \sqrt{\xi_{\rm ISCO}^2+1}=
2 a, 
\ee
or in spheroidal radius
\be \label{riscoeuler}
r_{\rm ISCO}=a \xi_{\rm ISCO}=\sqrt{3} a.
\ee
Note that $r_{\rm ISCO}$ is not the real 
Euclidean distance from the origin to the 
test particle at ISCO (the Euclidean 
distance is $\rho_{\rm ISCO}$),
but it is the analogue of the Boyer-Lindquist 
radius of Kerr, which we use extensively 
in our paper in order to draw a  
faithful comparison of  Euler with Kerr.

Oddly enough, the existence of ISCO in the Euler 
problem is still present
even in the limit $a \to 0$; the corresponding
radius just tends to $r_{\rm ISCO}=0$ then. 
This is a new qualitative feature that the 
Newtonian monopole field ($a=0$) lacks, 
as mentioned previously. 
Therefore, the case $a \to 0$ 
(but $a \neq 0$) could be considered 
as the analogue  of a Schwarzschild black hole.
Note, also, that contrary to  oblate Euler problem, 
the original Euler problem (the prolate one) 
does {\it not} possess an 
ISCO. This is due to the fact that 
in the prolate Euler field
the attraction from the two point sources, 
located along the $z$-axis, is even  
softer than the Newtonian one from
a single point source; therefore 
the repulsive centrifugal potential 
rules out the existence of 
an ISCO in this case.

The critical value of $z$-angular momentum 
leading to the presence 
of ISCO in the oblate Euler field is
 \be
 L_z^2=\frac{16 M a}{3\sqrt{3}},
 \label{Nangmom}
 \ee
 while the corresponding expression for the Kerr
 field is quite involved and difficult to
 compare with the above simple formula,
 since the expression of the $z$-angular momentum for a prograde and a retrograde orbit as a function of $a$ is different, due to the Lense-Thirring effect.

\begin{figure}[b]
\includegraphics[width=\linewidth]{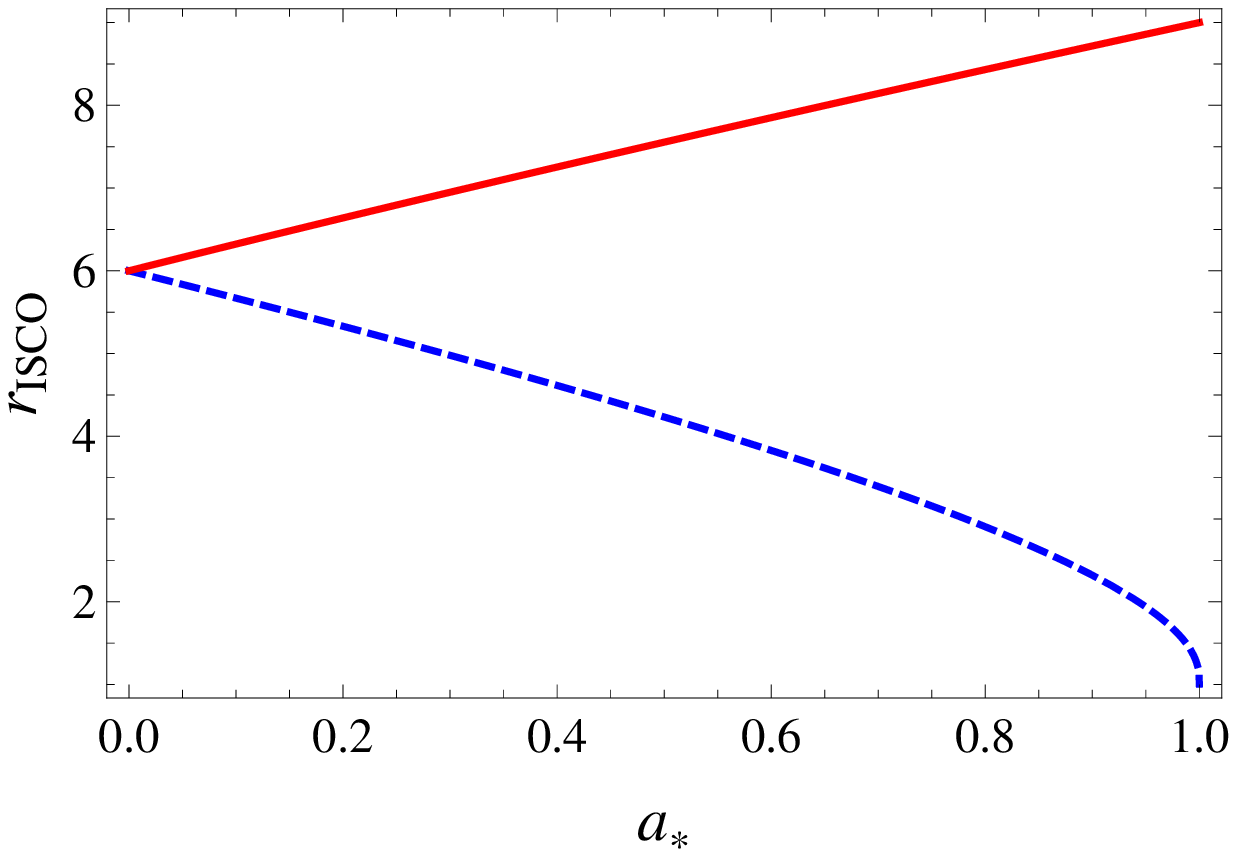}
\caption{This is the plot of $r_{\rm ISCO}$
as a function of $a_\star$ for prograde 
(blue dashed curve) and for 
retrograde (red curve) orbits in Kerr, 
according to Eq.~(\ref{ISCOKerr}). 
The radii are scaled with $M$.}
\label{fig:theKerrISCOs}
\end{figure}

Although the existence of an ISCO in the Newtonian problem is by itself 
a positive qualitative sign of the physical resemblance with the relativistic problem of a Kerr black hole, 
apparently it does not seem to share any quantitative
similarity with the dependence of  ISCO radius 
 in Kerr with  its spin parameter $a$ (see  \cite{BardPresTeuk72}):
 \be
 r_{\rm ISCO}=M \left[ 3+Z_2 \mp \sqrt{(3-Z_1)(3+Z_1+2 Z_2)} \right] ,
 \label{ISCOKerr}
 \ee
 where
\begin{equation}\label{theZs1}
 Z_1 = 1 +(1-a_\star^2)^{1/3} \left[ (1+a_\star)^{1/3} 
 + (1- a_\star)^{1/3} \right]  , 
 \end{equation}
 and 
 \begin{equation}
 Z_2 =\sqrt{3 a_\star^2+Z_1^2}, \label{theZs2}
\end{equation}
 with $a_\star=a/M \in [0,1]$ 
 while the two signs correspond to prograde
 (upper sign) and retrograde (lower sign) orbits,
 respectively. As mentioned previously  the  $a$'s showing up in the expressions for the
 ISCO radius in the two problems have 
 completely different physical origin:
 in  Euler field it's just a distance, while in  Kerr
 field it is related to  the angular momentum of the
 gravitational source itself; therefore apart of the
 existence of an ISCO
 radius in both problems, no qualitative similar behavior
 of $r_{\rm ISCO}$ with $a$ was  anticipated.
 In Figure \ref{fig:theKerrISCOs} 
 the  ISCO radii as a function of $a_\star$ for both 
 types of orbits in Kerr are plotted.
 The apparent non-linearity --especially for the 
 prograde orbit--, in contrast to the linearity of the
 oblate Euler problem, is clear.

\begin{figure}
      \includegraphics[width=\linewidth]{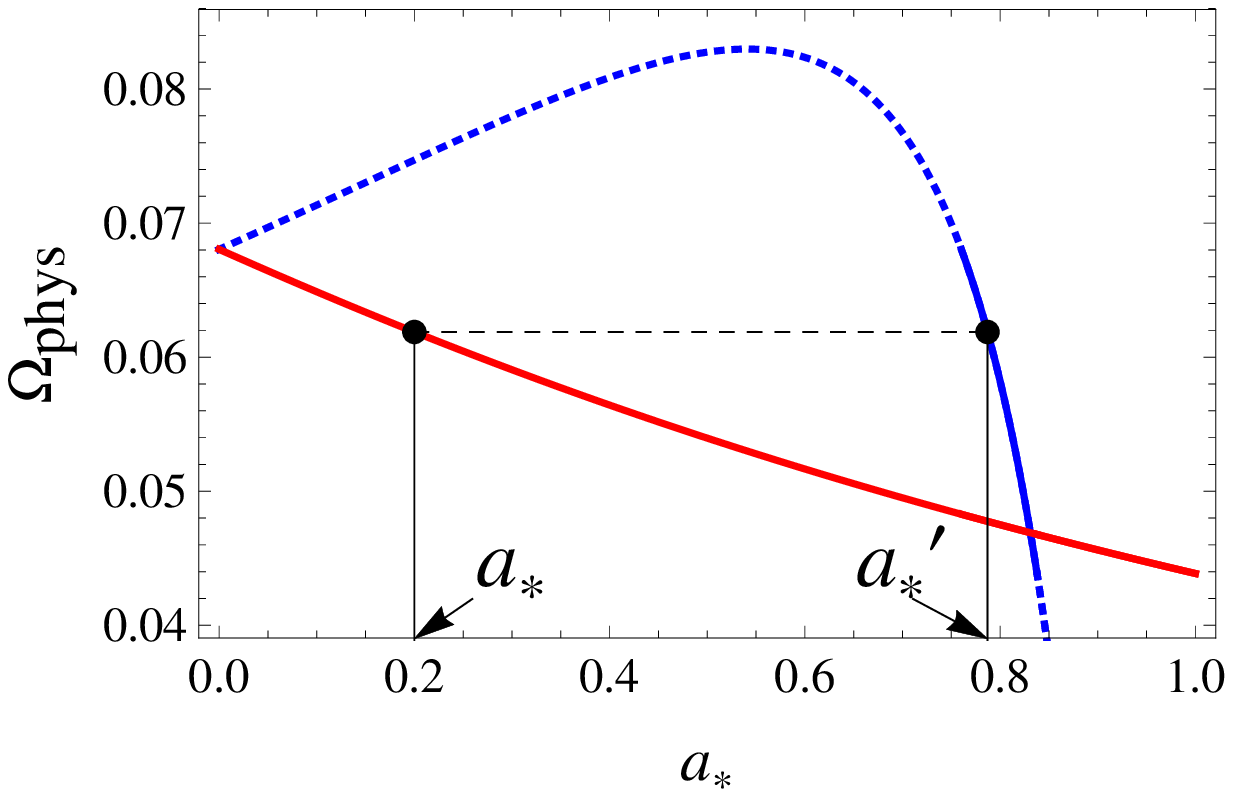}
     \caption{This is the plot of the absolute value of 
     $\Omega_{\rm phys}$, described in the text, 
     at $r_{\rm ISCO}$
     as a function of $a_\star$ for prograde 
     (blue dotted and solid curve) and for 
     retrograde (red curve) orbits. 
     In order to form pairs of ISCO
     counter-rotating orbits with equal 
     values of $\Omega_{\rm phys}$
     (like the pair of black dots) we assign the value 
     of $a_\star$ of the retrograde orbit to the 
     prograde orbit corresponding to a different
     actual
     $a'_\star$ but with equal $\Omega_{\rm phys}$. 
     The two parts of the blue curve (prograde orbits) 
     with no corresponding retrograde dual have been
     plotted as blue dotted curves. Only the 
     solid part of this curve have an 
     $\Omega_{\rm phys}$ that is equal
     to the corresponding value of a retrograde
     orbit. The average radius
     $\langle r_{\rm ISCO} \rangle$ of the two radii 
     for each such pair is considered the ISCO radius 
     corresponding to the particular $a_\star$ value.}
     \label{fig:ISCO1}
 \end{figure}

\begin{figure}
     \includegraphics[width=\linewidth]{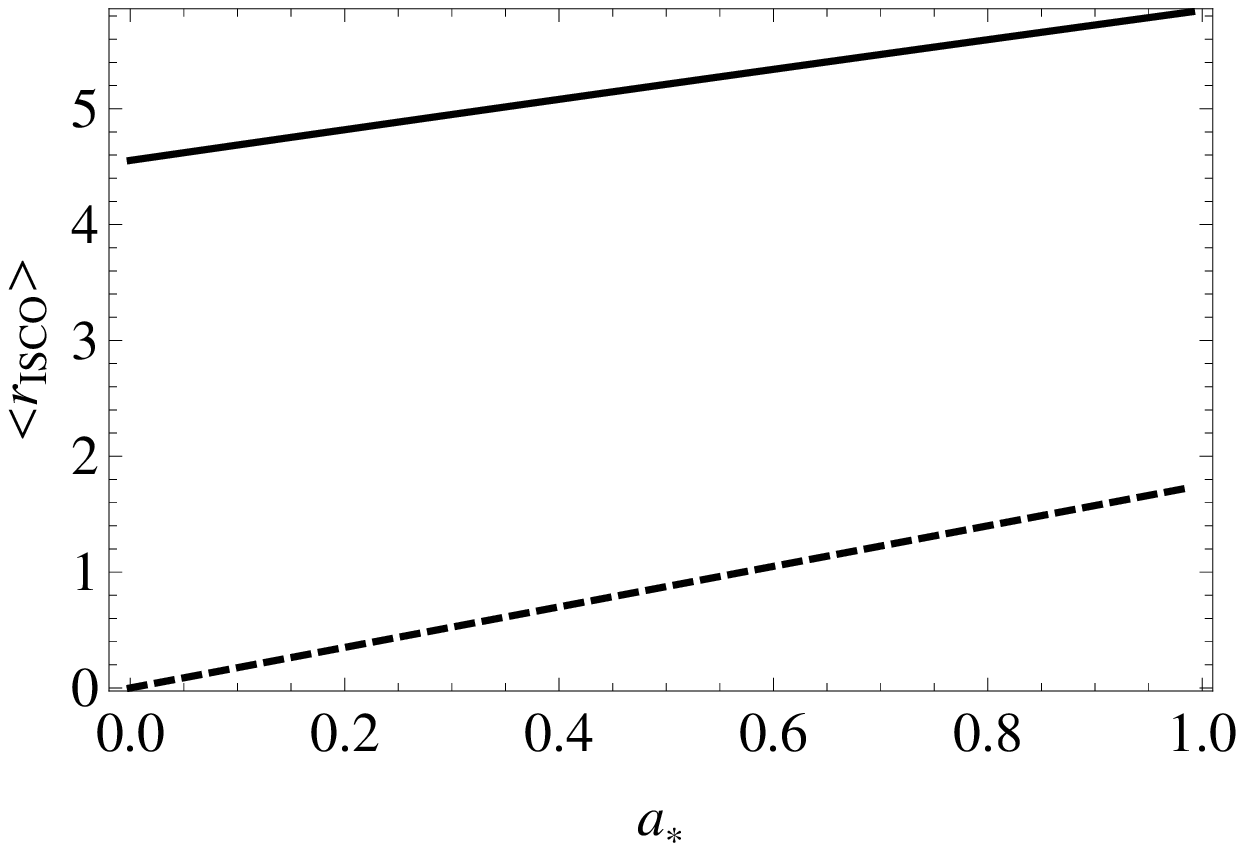}
     \caption{The solid curve shows the dependence of $\langle r_{\rm ISCO} \rangle$,  which is defined in Eq.~(\ref{rISCOaverage}), as
     a function of $a_\star$. Although it is apparently linear, this is not exactly true. The dashed curve
     shows the dependence of $r_{\rm ISCO}$ on $a_\star$ for the Euler problem. The oppositely directed orbits in the latter case have equal radii. The two lines have somewhat different slopes (the average slope of the almost linear curve corresponding to Kerr is $\sim 1.3$, while the 
     exactly straight line for Euler has slope $\sqrt{3}$).}
     \label{fig:KvsE_isco}
 \end{figure}

 On the other hand the comparison between 
 the two problems  is not exactly fair
 from a physical point of view. 
 The relativistic problem,
 although axisymmetric, strongly discriminates 
 between the two opposite senses of 
 rotation of an orbit (it is not symmetric 
 under the transformation $\phi \to -\phi$), 
 while
 the Newtonian one is absolutely symmetric under such a transformation. 
 Thus two circular equatorial orbits in Kerr  
 that rotate at opposite senses at the same radius 
 are not equivalent. It should be
more appropriate to compare two orbits 
(a prograde and a retrograde one) with 
opposite rotational frequencies, after taking into
account the frame-dragging of space-time itself,
as in the case for two oppositely directed 
 circular orbits in  Euler's problem  at the same radius.
 Of course such pairs of circular orbits in Kerr will 
 not have the same radii; thus one should assign some
 kind of an average value of radius for such a pair of oppositely, 
 equally rotating,
 orbits in Kerr. 
 To compute the actual rotational frequencies 
 in the Kerr metric, means that one should 
 firstly subtract  the rotational rate of  space-time itself,  that is the rotational rate of ZAMOs \cite{ZAMO}. Then one  should seek ISCO orbits
 with equal absolute values of rotational rates
 with respect to ZAMO observers,
in order to nullify the relativistic effect 
 of frame-dragging and put the oppositely directed
 orbits on equal footing. 
 In  order to accomplish such a comparison of
 ISCO radii between  oblate Euler's problem 
 and Kerr,  we   followed the  following process: 
 (i) For the Kerr case,
 first we  plotted  $\Omega_{\rm phys}=|\omega-\Omega_{ZAMO}|$
 at the radius of ISCO as a function of $a_\star$ 
 for both  prograde and  retrograde orbits (see
 Figure \ref{fig:ISCO1}). It is clear that the
 retrograde orbits have monotonically decreasing
 rotation rate with $a_\star$,
which qualitatively follows 
a similar behavior with the
rotation rate of circular 
Keplerian orbits $\Omega \propto r^{-3/2}$, while $r_{\rm ISCO}$ is increasing with $a_\star$.
 Instead, the rotation rate of
 prograde orbits have completely  different 
 dependence on $a_{\star}$.
 These orbits lie so deep in the 
 strong gravitational field of the
 black hole and the frame-dragging effect is then
 so dramatic (the region close to $a_{\star}=1$ is 
 actually buried inside the ergoregion which has no
 analogue in Newtonian gravity), that their 
 behavior with respect to rotational frequency 
 is completely different from those in the Newtonian
 problem. More specifically, there is no corresponding
 retrograde ISCO orbit that rotates at the same rate 
 (relative to ZAMO observers) for almost any 
 prograde ISCO orbit, except of
 the prograde ISCO orbits at large, but not extremal
 ($a_\star \simeq 1$), values.
 The range of $a'_\star$'s of prograde orbits
 with a corresponding dual retrograde orbit, sharing
 the same $\Omega_{\rm phys}$, is
 $0.760 \leq a'_\star  \leq 0.838$.
 On the other hand for any retrograde ISCO orbit
 (corresponding to any value of $a_\star \in [0,1]$) 
 there is a corresponding
 prograde one (one with the same 
 value of 
 $|\omega-\Omega_{ZAMO}|$).
 Therefore we could form pairs of retrograde-prograde 
 orbits with the
 same physical rotation rate $\Omega_{\rm phys}=|\omega-\Omega_{ZAMO}|$.
 On each such pair 
 we assigned the $a_\star$ of the retrograde one (since this is
 the one that has a counter-rotating dual with some $a'_\star$ value,
 for every value of $a_\star$).
 (ii) Then we numerically computed the 
 $r_{\rm ISCO}$'s for both 
 orbits (the retrograde (R) corresponding to 
 $a_{\star}$ and the prograde (P) corresponding to $a'_{\star}$, both characterized by the same
 $\Omega_{\rm phys}$). 
 (iii) For each such pair we computed the average
 value of ISCO radii, $\langle r_{\rm ISCO} \rangle$, according to  the relation
 \begin{equation}
  \langle r_{\rm ISCO} \rangle (a_\star)=\frac{1}{2}\left(
  r^{(R)}_{\rm ISCO}(a_\star) + r^{(P)}_{\rm ISCO}(a'_\star) \right)
 \label{rISCOaverage}
 \end{equation}
 that takes into account the two radii
 of the oppositely rotating orbits on equal footing.
 (iv) Finally we have plotted $\langle r_{\rm ISCO} \rangle$
 as a function of $a_\star$
 and the output was apparently almost linear
 (see Figure \ref{fig:KvsE_isco}), 
 like in the Euler case. 

Although it might seem rather artificial 
the way the  $\langle r_{\rm ISCO}\rangle$
was defined, we believe there is actually 
no other way to  define a radius of ISCO
that treats both the retrograde and the
prograde orbits on equal footing as 
it is always the case with axisymmetric 
Newtonian problems. 
The only alternative natural way we could  think of
in order to form  pairs of equivalent oppositely
directed orbits is by assuming orbits with equal
absolute values of angular momenta, instead of 
orbital  frequency. 
We tried that as well but there is 
no pair of retrograde-prograde
orbits with the same value of $|L_z|$. In contrast
to $\Omega_{\rm phys}$, $|L_z(a_\star)|$ is a monotonic function of $a_\star$, for $a_\star \in [-1,1]$, where the sign of $a_\star$ determines the rotational direction of the orbit. 
The deeper reason is that the frame-dragging 
effect is so enhanced for large values of 
$a_\star$ that 
one needs extreme value of angular momentum to keep
a circular orbit stable, when it lies 
deep 
in the gravitational potential.

For completeness,  we should 
note that the above method 
to construct equivalent pairs
of prograde and retrograde orbits, 
leads to a double solution only for the case of
$a_\star=0$, since then there are two 
$a'_\star$  values
for a prograde orbit with the same
$|\omega-\Omega_{ZAMO}|$
of a corresponding retrograde orbit; 
one of them being the
zero spin case. For continuity 
reasons though, we have ignored 
the  second root --that of zero spin.
 
As we mentioned earlier, 
the dependence of ISCO radius of the 
Euler's problem on  $a_\star$ parameter, 
is linear (c.f.~Eq.~\ref{riscoeuler})). Amazingly, this is approximately
the  behavior of $\langle
r_{\rm ISCO}\rangle$ in Kerr case with respect to the 
spin parameter $a_\star$, when  a retrograde and 
a prograde orbit are considered as a suitably 
equivalent pair. A last comment on this
similarity is that quantitatively 
the approximate
linear fashion of $\langle r_{ISCO}\rangle$ with $a_\star$ in Kerr, although of the same order, it
does not have the same slope with that of Euler,
which is $\sqrt{3}$.

\begin{figure}
      \includegraphics[width=\linewidth]{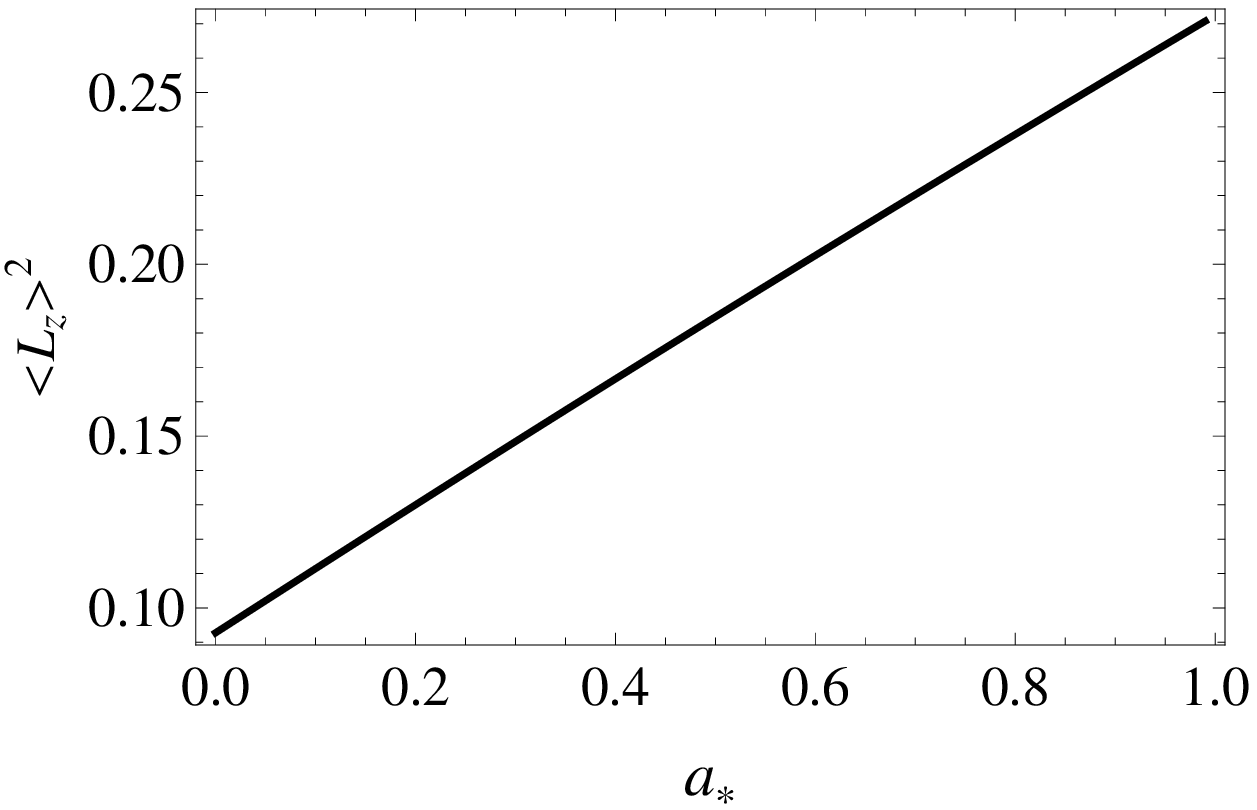}
     \caption{The dependence of $\langle L_z \rangle ^2$, which is formed as the average value of $L_z$'s for two oppositely directed ISCOs in Kerr that share the same physical rotation rate $\Omega_{\rm phys}$, on $a_\star$ is almost linear like in the Euler field. The slopes, though, are quite different.}
     \label{fig:averageL}
 \end{figure}

To further stress this peculiar equivalence
of oppositely rotating orbits in Kerr, we
have also plotted the square of the
average value of $L_z$'s 
for each such pair of equally rotating orbits
as a function of $a_\star$ (see Figure \ref{fig:averageL}).
Apparently this plot is also almost linear 
as in the Euler case, according 
to Eq.~(\ref{Nangmom}). The corresponding 
slopes though are remarkably different.

 \subsection{The fundamental frequencies}
 \label{sec:5.4}
 
The Euler potential is an integrable system with three independent and in involution integrals of motion. The conjugate momenta $p_r$ and $p_{\theta}$ (cf. Eqs.~(\ref{consQr}) and (\ref{consQth})) are functions only of $r$, and $\theta$, respectively.
For bound orbits  (see Section \ref{sec:3.2}), the space of trajectories is a compact and connected manifold. 
According to Arnold's theorem \cite{Arnold}, the phase space is diffeomorphic to a three-torus. 
Since the problem is integrable, even though the coordinates are not really
periodic, their oscillation (or libration for $\phi$) correspond to characteristic  frequencies, the so called fundamental frequencies.

Consequently, we could define a set of symplectic coordinates, the corresponding action-angle variables $(\boldsymbol{J},\boldsymbol{\omega})$, where the angle variables $\boldsymbol{\omega}$ are periodic
functions of time, and the action variable $\boldsymbol{J}$ corresponds to a fixed vector. 
The three frequencies related to the periodicities of ${\boldsymbol{\omega}}$'s are the fundamental frequencies that one obtains by Fourier analyzing the oscillating (or librating) time dependence of the coordinates $r,\theta$ and  $\phi$ coordinates of a bound orbit.
 
The generating function of the canonical 
transformation  $(q_i,p_i)\to(J_i,\omega_i)$ 
is the Hamilton's characteristic function
$W(\boldsymbol{q,F})$, which is the solution 
of Hamilton-Jacobi equation 
$\partial S/\partial t+H(q_i,\partial S/\partial q_i)=0$:
\begin{equation}
S(q_i,F_i,t)=-E t+ W(q_i,F_i),
\end{equation}
where $q_i$ are the old coordinates 
$(r,\theta,\phi)$ and their conjugate 
momenta $p_i$ are defined as: $p_r=\pm\sqrt{V_r(r)}/(r^2+a^2)$,
$p_{\theta}=\pm\sqrt{V_{\theta}(\theta)}$, and $p_{\phi}=L_z$,
(the corresponding potentials 
$V_r, V_\theta$ are 
given explicitly in Eqs. (\ref{Vr}, \ref{Vth})). 
The first integrals of motion are: 
$F_{i}=(H=E,L_z,Q)$. Due to the  
separability of the Hamilton-Jacobi 
equation its solution is of the form: 
\begin{equation}
    W(\boldsymbol{q,F})=L_z \phi\pm W_r(r)
    \pm 
    W_{\theta}(\theta),
\end{equation}
where: 

\begin{eqnarray}
W_r(r) &=& \int^r \frac{\sqrt{V_r}}{r^2+a^2}\;dr, \\
W_\theta(\theta) &=& \int^{\theta} \sqrt{V_\theta} \;d\theta.
\end{eqnarray}
The action variables are defined as (see \cite{Arnold}):
\begin{equation}
J_i=\frac{1}{2\pi}\oint p_id q_i,\label{J}
\end{equation}
where the integration is to be carried over 
a period of oscillation or rotation of $q_i$.

The action variables $J_{i}$ are constants 
of motion, as they depend only on the first 
integrals  $J_j=J_j(F_{i})$. Inverting 
them we can express the integrals $F_i$ 
as functions  of the action variables. 
In particular the Hamiltonian can be expressed as $H=F_1=H(\bf J)$ and it is cyclic 
with respect to $\omega_i$. 
The generating function also 
can be expressed in terms of 
the coordinates and the action variables: 
\begin{equation}
    W=W(\bf q,\bf J).
\end{equation}
The transformation equations   are:
\begin{eqnarray}
    p_i&=&\frac{\partial{W}}{\partial{q_i}}(\bf q,\bf J), \\
    \omega_i&=&\frac{\partial{W}}{\partial{J_i}}(\bf q,\bf J),
\end{eqnarray}
while the equations of motion in 
action-angle variables become:
\begin{eqnarray}
\dot{\omega}_{i}&=\frac{\partial{H}(\bf{J})}{\partial{J}_{i}}=\Omega_i,\\
\dot{J}_{i}&=-\frac{\partial{H}(\bf{J})}{\partial{\omega}_{i}}=0.
\end{eqnarray}
The corresponding angle variables are 
periodic and linear functions of time:
\begin{equation}
\omega_i(t) =\left(\Omega_{i}(\boldsymbol{J}) t+ \omega_i(0)\right)\mod{2 \pi}
\end{equation}
where $\Omega_{i}(\bf J)$ and $ \omega_i(0)$ 
are constants and  $\Omega_{i}(\boldsymbol{J})=
\partial{H}( {\bf J})/\partial{J}_{i}$  
describe the fundamental frequencies of the orbit. 
In Appendix \ref{App:4} we give analytic 
expressions of the above quantities, 
$J_i,\Omega_i$. Here we simply give 
the final expressions:
\begin{align}
\Omega_r&=\frac{\pi K(k)}{a^2z_{+}[K(k)-E(k)]X+YK(k)}\label{Wr}\\
\Omega_{\theta}&=
\frac{\pi\beta\sqrt{z_{+}}X/2}{a^2z_{+}
[K(k)-E(k)]X+YK(k)}
\label{Wth}\\
\Omega_{\phi}&=
\frac{ZK(k)+X L_z[\Pi(\frac{\pi}{2},z_{-},k)-K(k)]}
{a^2z_{+}[K(k)-E(k)]X+YK(k)}
\label{Wph}
\end{align}
where $K(k)$, $E(k)$ and 
$\Pi(z_{-},k)$ 
are the complete elliptic integrals 
of the first, second and third kind, 
respectively \cite{Abram}:
\begin{align}\label{ElK}
K(k)&=\int^{\frac{\pi}{2}}_0
\frac{d\theta}{\sqrt{1-k^2\sin^2{\theta}}},\\
\label{ElE}
E(k)&=\int^{\frac{\pi}{2}}_0
\sqrt{1-k^2\sin^2{\theta}}\;d\theta,\\
\label{ElP}
\Pi(z_{-},k)&=
\int^{\frac{\pi}{2}}_0
\frac{d{\theta}}{(1-z_{-}\sin^2{\theta})
\sqrt{1-k^2\sin^2{\theta}}},
\end{align}
with
$k=\sqrt{z_{-}/z_{+}}$ (where $z_{\pm}$ are the two roots of $V_\theta(\cos\theta)$, with $z_-<1<z_+$) and $\beta^2=-2a^2 E$.
The integrals $X$, $Y$ and $Z$ are related with the radial motion and are defined as:
 \begin{align}
X&=\int^{r_2}_{r_1}\frac{dr}{\sqrt{V_r}}  ,
\label{Xint}\\
Y&=\int^{r_2}_{r_1}\frac{r^2}{\sqrt{V_r}}dr   ,\label{Yint} \\
Z&=\int^{r_2}_{r_1}\frac{L_z r^2}{(r^2+a^2)\sqrt{V_r}}dr, \label{Zint}
\end{align} 
with $V_r$ being the radial potential $V_r(r)$ introduced in Eq.~(\ref{Vr}).

Although the orbit is not periodic, 
there are specific cases, where the 
motion is clearly periodic. A resonant orbit, 
where  the ratio $\Omega_r \div \Omega_{\theta} \div \Omega_\phi$ is a ratio of integers, is a more involved case of a purely periodic orbit since then an integer number of oscillations of $\theta$, $r$ and $\phi$ (not necessarily the same numbers)
are repeated in a finite time period. 

For comparison
the fundamental frequencies of a bound orbit in Kerr space-time have been derived by Schmidt \cite{Schmidt} and they are given by exactly the same expressions with that of Euler (\ref{Wr})-(\ref{Wph}), with $\beta^2=a^2(1-E^2)$ and $z_{\pm}$ the two roots of the Kerr polar potential (which is the same with the polar
potential of Euler).
The radial integrals for Kerr though are 
given by (see \cite{Schmidt}):
   \begin{align*}
X&=\int^{r_2}_{r_1}\frac{dr}{\sqrt{V_r}}  , \\
Y&=\int^{r_2}_{r_1}\frac{r^2}{\sqrt{V_r}}dr   , \\
Z&=\int^{r_2}_{r_1}\frac{L_z r^2 -2 M r (L_z - a E)}{(r^2-2 M r+a^2)\sqrt{V_r}}dr,
\end{align*}  
where the corresponding potential $V_r$ is
\begin{eqnarray*}
    V_r&=&(E^2-1) r^4 + 2 M r^3 + \left[(E^2-1)a^2  - Q - L_z^2 \right] r^2 \nonumber\\
    &&+ 
2 M \left[ (L_z-a E)^2+ Q \right] r - Q a^2.
\end{eqnarray*}
If we rewrite the Newtonian energy of the $V_r$
potential of Eq.~(\ref{Vr}) as previously:
that is by adopting the reparametrization $2E\to E^2-1$, we obtain a form of the radial potential of Euler 
which differs from that of Kerr only 
on the linear term. Therefore whatever
differences in frequencies between the Kerr
and the Euler field, arise from this 
difference in $V_r$ and the 
different expression of the 
$Z$ integral.


In Kerr space-time there is also a fourth constant 
$\Omega_t$ which is associated with the 
generalized time coordinate. However, 
the motion is not bounded in the timelike 
direction, so $\Omega_t$ cannot be 
interpreted as a physical fundamental 
frequency \cite{Schmidt,Hinde}.     

 \subsection{Pairs of isofrequencies}
 \label{sec:5.5}

Warburton et al \cite{Baracketal13} 
have shown that Kerr black holes 
(Schwarzschild black holes included) 
have an interesting property: there 
is no one-to-one correspondence between 
orbital characteristics and fundamental 
frequencies, that is, there are pairs 
of distinct bound geodesic orbits lying
in the strong field region, that are 
characterized by exactly the same  
triplets of frequencies (radial, 
azimuthal, and longitudinal). 
Motivated by the fact that the 
oblate Euler field has an ISCO, 
like the gravitational field of
a Kerr black hole, we looked for 
pairs of distinct, potentially 
synchronized, orbits in the Euler 
field as well.
  
\begin{figure}
\includegraphics[width =
\linewidth]{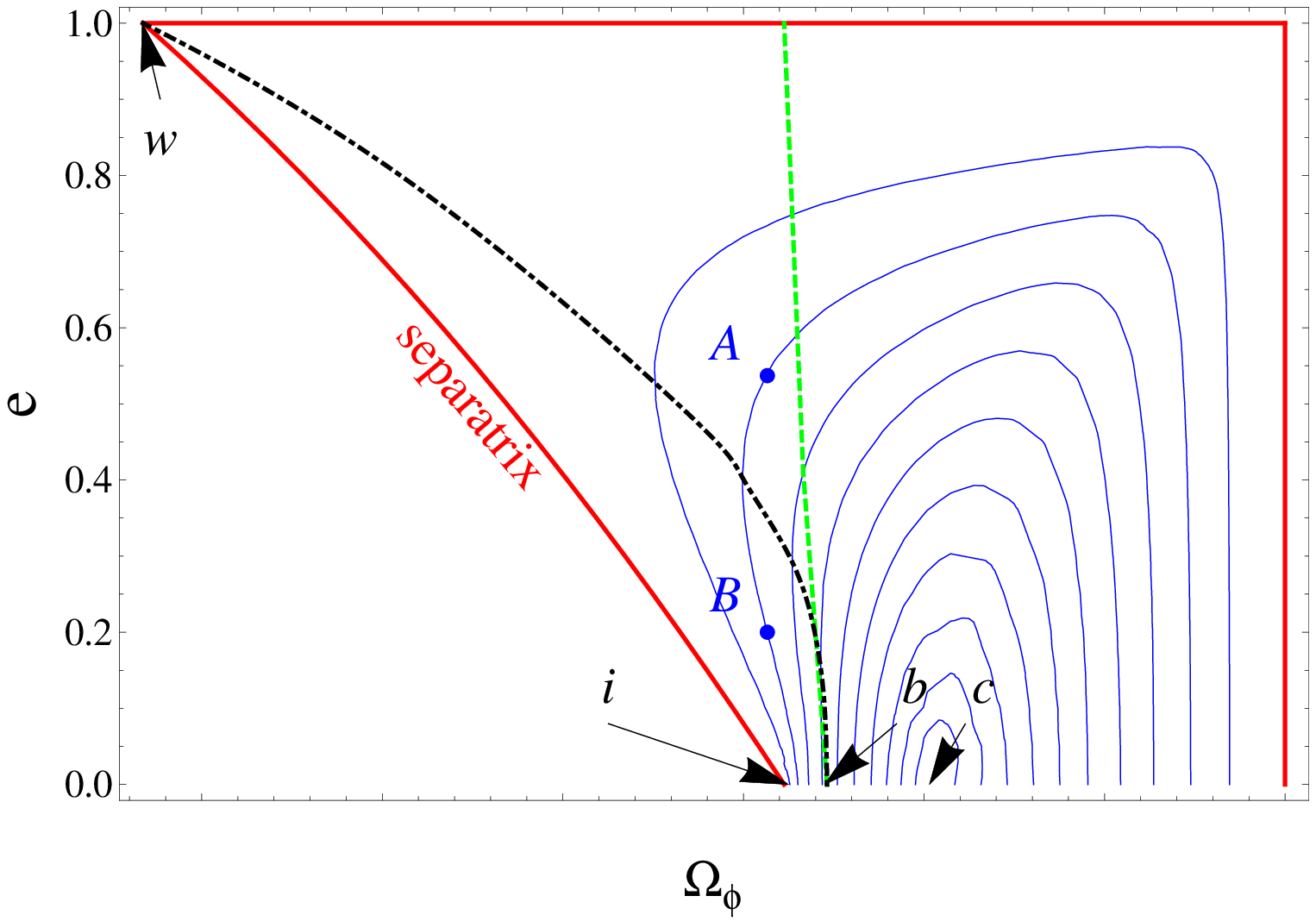}
\caption{Contour lines of $\Omega_r$  
(blue lines) for equatorial orbits 
in the oblate Euler potential in the 
$(e-\Omega_\phi)$ parameter-space 
($\Omega_\phi$ is increasing as you 
move towards the left in order to get a better 
comparison with Figure 4 of 
\cite{Baracketal13}
and there are no numbers marked
since the actual values of 
$\Omega_\phi$ depend on $\sqrt{M/a^3}$).
The boundary (red line) of all 
iso-$\Omega_r$ curves correspond to 
$\Omega_r=0$, while as one move 
towards the smaller contours, 
$\Omega_r$ increases. The point 
along the $\Omega_{\phi}$-axis, 
marked `$c$' correspond to the 
highest $\Omega_r$, (see Table 
\ref{tab.1}). The oblique, 
slightly curved part of the 
boundary is the separatrix, 
corresponding to orbits with 
$r_2=r_3$, for two of the 
four roots $r_1,r_2,r_3,r_4$  
of $V_r$ potential. The corresponding 
orbits evolve to eternally circular 
orbits with radius $r=r_2$. 
The highest value of $\Omega_{\phi}$, 
marked `$w$', corresponds to a 
marginally open orbit ($e\to 1$) 
along the separatrix. $\Omega_{\phi}$ 
is maximized for such an orbit 
because as you move higher 
on the separatrix the  double 
root $r_2=r_3$ is lowered, 
and the orbit moves closer 
to the strongest part 
of the field. The dashed green 
 line (the COD) is the
locus of all non-circular orbits 
with a circular dual that is characterized
by exactly the same set of frequencies $(\Omega_r,\Omega_{\phi})$. 
Finally the black dashed-dotted 
line is the singular curve along 
which all orbits are marginally 
different pairs of iso-frequency orbits. 
Among all pairs with the same 
set of frequencies that one could 
find on the left side of the 
green COD curve, we have isolated 
a single pair of orbits (two blue points
denoted $A$ and $B$) 
that we have thoroughly studied. 
Some of the contour lines, especially 
around point `$c$', are not very 
smooth, due to inaccurate 
interpolation of {\it Mathematica}
package.}
\label{fig:isofr_eq}
\end{figure}

Following the procedure of \cite{Baracketal13},
we first searched for pairs of equatorial orbits
with equal doublets of $(\Omega_r,\Omega_\phi)$
frequencies. While the third frequency, 
$\Omega_\theta$, could also be computed 
for such orbits, it does not show up in 
the orbital motion, since the orbit is
purely equatorial. In order to seek such 
double solutions in the frequency space, 
we have plotted the contours of $\Omega_r=
{\rm const}$ in the $(e,\Omega_\phi)$ plane.
Actually the very shape of the boundary of all 
possible equatorial orbits in the parameter 
space mentioned above, namely 
the contour-curve corresponding to $\Omega_r=0$,
is sufficient to ensure the existence of pairs 
of orbits with the same set of $(\Omega_r,
\Omega_\phi)$ frequencies. 
The boundary consists of orbits: (i) 
with infinite semi-latus rectum $p$, 
corresponding to $\Omega_\phi=0$, 
but with various eccentricities 
(infinitely distant bound orbits), 
(ii) with eccentricity $e=1$ 
(marginally closed orbits), 
corresponding to a range of 
$\Omega_\phi$ frequencies
depending on the semi-latus rectum, and
(iii) the separatrix, that is, orbits
corresponding to a potential $V_r(r)$
with a double root $r_2=r_3$ (and $V'_r(r_2)=0$), 
such that the orbit spends infinite 
time to complete an $r$-oscillation between 
$r_1$ and $r_2$.
The ISCO is simply the endpoint 
of the separatrix at $e=0$,
corresponding to a marginally stable 
circular orbit, due to a suitable 
tuning of the polynomial expression
for the potential $V_r$, to
obtain a triple root, $r_1=r_2=r_3=
\sqrt{3} a$ (see Section \ref{sec:5.3}). 
Along the separatrix of equatorial orbits, 
$\Omega_\phi$ is given by the simple expression
\begin{equation}
    \Omega_{\phi,s}^{\rm(eq)}=\sqrt{\frac{M}{r_2^3}},
\end{equation}
where $r_2=p/(1+e)$, since the particle
will eventually end up to radius $r_2$. 
Following Eq.~(\ref{pex}) 
of Appendix \ref{App:3}, the 
$\phi$-frequency of such orbits 
could be expressed as:
\begin{equation}
\Omega_{\phi,s}^{\rm (eq)}=   
\left( \frac{1+e}{3-e} \right)^{3/4}
\sqrt{\frac{M}{a^3}}.
\end{equation}
The separatrix has positive slope, 
$de/d\Omega_\phi|_s>0$. Consequently,
the boundary of the contour plot, 
$\Omega_r=0$, forms an inverted trapezoid 
(as in Kerr) in the parameter space 
$(e-\Omega_\phi)$.
Due to continuity of the function $\Omega_r(e,\Omega_\phi)$ 
for equatorial orbits, this  shape 
is conclusive for the existence of 
isofrequency pairs of orbits (see
Figure \ref{fig:isofr_eq}), as 
pointed out also in the case of Kerr 
\cite{Baracketal13}. 

In Appendix \ref{App:3} we give analytic 
formulae for the constants of motion 
at the separatrix and from them one 
could compute, based on the expressions 
of Appendix \ref{App:4}, 
the fundamental frequencies for these 
peculiar marginally whirl-zooming 
equatorial orbits. 
Apart of the separatrix, the bound equatorial 
orbits exhibit the following characteristics:
(i) There is a maximum value of $\Omega_r$, which corresponds to a circular orbit ($e=0$), 
marked as `$c$' in the contour
plot of $\Omega_r$,  Figure \ref{fig:isofr_eq}. 
The analytic expression for $\Omega_r$ at 
zero eccentricity is simply
\begin{equation}
\Omega_r(p,e=0)=
\frac{\sqrt{p^2-3a^2}}{p^{5/2}}
\sqrt{M}.
\end{equation}
Thus the maximum value of $\Omega_r(p,e)$ is
$\Omega_{r,c}=\sqrt{2/5^{5/2}}\sqrt{M/a^3}$,
for $p_c=\sqrt{5} a$ and $e=0$.
(ii) There is a line of non-circular orbits,
called `COD' curve (circular 
orbit duals) in
\cite{Baracketal13}, 
that have the same 
frequency set, $(\Omega_r,\Omega_\phi)$, 
with a single corresponding circular orbit. 
The range of $\Omega_\phi$'s that the
COD curve spans is
$[\Omega_{\phi,b},\Omega_{\phi,i}]$,
corresponding to specific circular orbits
marked as `$b$', `$i$' in the parameter 
space, respectively.  The `$i$' circular 
orbit is simply the ISCO orbit,
representing the maximum $\Omega_\phi$ 
value of a circular orbit (with $e=0$), 
the dual of which is a non-circular 
orbit with $e=1$
(the upper end-point of the green 
dashed curve of Figure \ref{fig:isofr_eq}). 
On the other hand `$b$' is the circular
equatorial orbit with a non-circular dual 
which has the lowest $\Omega_\phi$ and the
highest $\Omega_r$ that such an orbit could
yield. The `$b$' circular orbit is actually 
a singular case since the twin pair of the
corresponding circular orbit is exactly 
the same orbit, representing now a 
marginally non-circular orbit.
On the left side of the COD curve 
one could find
all possible iso-frequency pairs.
(iii) Finally, there is another special 
curve representing all iso-frequency 
pairs with marginally 
equal orbital parameters. 
Along this curve the Jacobian of 
the transformation between the frequency 
parameter space and the orbital parameter 
space $(\Omega_r,\Omega_\phi) 
\rightarrow (p,e)$  vanishes, 
which means that the transformation 
is singular: each point along
this line corresponds to a double
root of the system of equations 
\begin{eqnarray}
\Omega_r(p,e)&=&\Omega_{r 0}, \nonumber \\
\Omega_{\phi}(p,e)&=&\Omega_{\phi 0}.
\end{eqnarray}
This curve joins the  
points of the iso-$\Omega_r$ contour 
lines that represent the extremum 
values of $\Omega_\phi$ for each
$\Omega_r$. This singular curve 
spans all eccentricities 
from marginal bound orbits ($e=1$) 
to circular orbits ($e=0$)
meeting the COD line at point `$b$'.
As mentioned above, point `$b$' 
represents a singular circular 
orbit which is the dual of itself. 
The other end-point 
of the singular curve 
(the left-most corner of the 
plot in Figure \ref{fig:isofr_eq}) corresponds 
to the highest possible
$\Omega_{\phi}$ value for any bound orbit, 
and it is denoted as `$w$'
in the contour plot. Since this 
orbit is an orbit at the 
separatrix, its semi-latus rectum 
is $p_w=2 a$ (this is what one 
yields from the parametric Eq.~(\ref{pex})
when the values $e=1$ and $x=0$ 
are imposed), 
while its corresponding $\Omega_\phi$ 
frequency is $\Omega_{\phi,w}=\sqrt{M/r_2^3}=
 \sqrt{M/a^3}$.

In order to plot
the singular curve we have to solve the
equation
\begin{equation}
  J=\left| \frac{\partial(\Omega_r,\Omega_\phi)}
  {\partial(p, e)} \right|=0.  
\end{equation}
The Jacobian of the transformation
was computed numerically for arbitrary 
eccentricities. However the base of this line,
`$b$', corresponding to zero eccentricity
was derived analytically since then
the complicated functions $\Omega_\phi(p,e)$, 
and $\Omega_r(p,e)$, could be written as
simple analytical expressions, when expanded as 
Taylor series around $e=0$. Both
frequencies yield the form $\Omega_0(p)+
e^2 \Omega_2(p)$. Therefore the Jacobian  
of the transformation is linear with 
respect to $e$ near $e=0$. This 
explains why all iso-$\Omega_r$ 
contours are intersecting the 
$\Omega_\phi$-axis at right angles; 
that is, a slight eccentricity $e<<1$
does not alter both frequencies 
of the corresponding circular 
orbits at order $O(e)$. 
More specifically, the Jacobian 
determinant of the transformation 
near $e=0$ yields the following form
\begin{equation}
J|_{e \to 0}=\left|-
\frac{9 M a^2 (5a^4-15a^2 p^2+4 p^4) e}
{4 p^5 (p^2-3 a^2)^{3/2} (p^2+a^2) } 
+{\rm O}(e^2)\right|.
\end{equation}
Thus the starting point `$b$' of the 
COD line (at $e=0$), which coincides 
with the starting point of the singular 
line, is given by the solution
of the algebraic equation 
$5a^4-15a^2 p^2+4 p^4=0$, which is
\begin{equation*}
p_b=a \sqrt{(15+\sqrt{145})/8} \simeq 1.839 a,
\end{equation*}
(the second solution is lower than 
the ISCO radius so it has been omitted;
this corresponds to the semi-latus rectum of
the dual plunging orbit
that lies beyond the separatrix). 
Finally, from the semi-latus rectum one 
can compute the two frequencies 
$\Omega_{\phi,b}$, and $\Omega_{r,b}$. 
The numerical values
of these frequencies are 
$\Omega_{r,b}\simeq 0.135 \sqrt{M/a^3}$ and $\Omega_{\phi,b}=0.401 \sqrt{M/a^3}$.

\begin{table}[b!]
    \centering
    \begin{tabular}{|c||c|c|c|c|}\hline \hline
                       & $c$   & $b$   & $i$   & $w$ \\
                       \hline \hline
$\Omega^\star_{r,}$    & 0.189 & 0.135 & 0     & 0   \\
\hline
$\Omega^\star_{\phi}$  & 0.299 & 0.401 & 0.439 & 1   \\
\hline
$p^\star$              & 2.236 & 1.839 & 1.732 & 2   \\
\hline
$e$                    & 0     & 0     & 0     & 1   \\
\hline
\end{tabular}
\caption{The characteristic frequencies in the
($e-\Omega_\phi$) parameter-space for the 
equatorial Eulerian orbits. The  
$\Omega^\star$'s are simply the 
dimensionless frequencies that 
arise when frequencies are written
in terms of $\sqrt{M/a^3}$. These 
numerical values are independent 
of $a$. $p^\star$'s are the dimensionless 
semi-latera recta ($p/a$).}
\label{tab.1}
\end{table}

\begin{table}[b!]
    \centering
    \begin{tabular}{|c||c|c|}
    \hline     \hline 
                      & $A$         & $B$  \\
     \hline \hline
$\Omega^\star_{r}$    & \multicolumn{2}{|c|}{ 0.0790569} \\
\hline
$\Omega^\star_{\phi}$ & \multicolumn{2}{|c|}{ 
0.453379} \\
\hline
$e$                   & 0.5370696   &  0.2    \\\hline
$p^\star$             & 1.94569     &  1.83337\\\hline
\end{tabular}
\caption{The two orbits, marked as $A$ 
and $B$ in Figure \ref{fig:isofr_eq}, have been
isolated in the region where iso-frequency 
pairs exist. One of them (B) was fixed 
and the other one was tracked down by 
numerically solving the complicated
equation $\Omega_r(e,\Omega_{\phi}(p,e)
=\Omega_{\phi,B})=\Omega_r(e_B,
\Omega_{\phi,B})$ with respect 
to $e$ and $p$. All frequencies 
are scaled with the dimensional 
quantity $\sqrt{M/a^3}$, that is 
$\Omega_i=\Omega_i^\star \sqrt{M/a^3}$,
while the 
semi-latus rectum $p^\star$ is 
 $p/a$.}
\label{tab.2}
\end{table}

\begin{figure}[b]
\includegraphics[width=\linewidth]{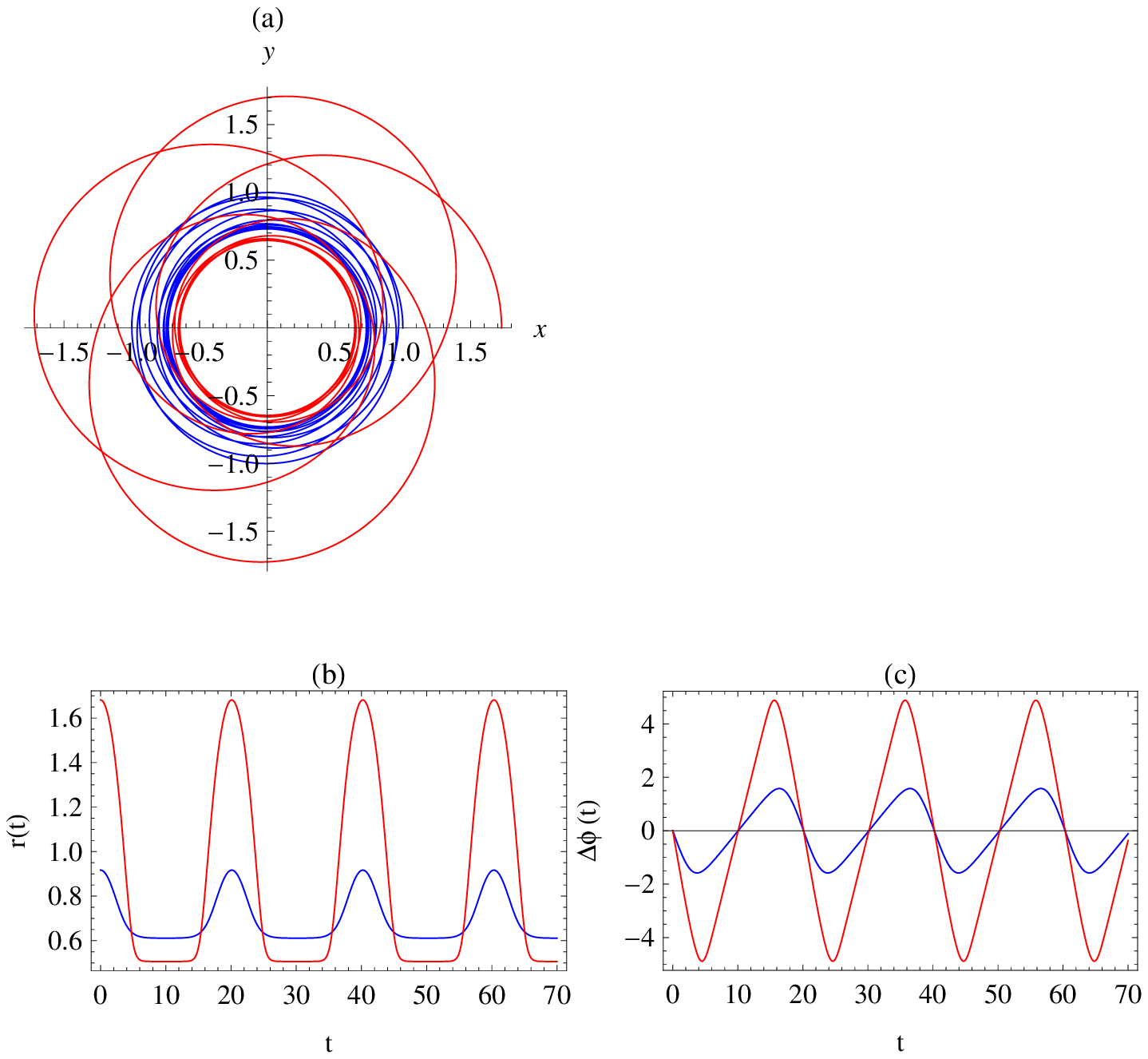}
    \caption{This plot shows the two distinct  
    equatorial orbits $A$ (the red more eccentric orbit) and 
    $B$ (the blue less eccentric one) of Figure \ref{fig:isofr_eq} (diagram a),
    as well as the synchronized time evolution of their 
    coordinates (diagrams b and c). 
    The last diagram (c) shows the plot of 
    $\Delta \phi(t)
    =\phi(t)-\Omega_\phi t$. The last two diagrams
    exhibit the common frequencies of the two orbits, despite the very different morphology of them.
    \label{fig:2orbits_eq}}
\end{figure}

In Table \ref{tab.1} we summarize 
the above results by presenting a 
complete list of the values of all 
characteristic frequencies
discussed in the previous paragraph
(all are simply multiples of the 
dimensional quantity $\sqrt{M/a^3}$), 
as well as the corresponding semi-latera 
recta, $p$, of these orbits. 
It should be emphasized that not only the 
characteristic frequencies, but the frequencies 
of any bound orbit in the Euler field, 
are all scaled with $\sqrt{M/a^3}$, 
independently of $a$ and $M$; they 
depend only on the characteristic 
orbital parameters $e$ and $p_\star=p/a$. 
Therefore the contour plot
of Figure \ref{fig:isofr_eq} does 
not represent a specific
$a$ value, as in the Kerr case. 
In contrast the exact form of the contour 
curves in Kerr does depend on $a$, 
due mainly to the dragging of frames in 
the corresponding relativistic problem.
The shape of the $\Omega_r$-contour lines on the left of the COD line ensures that one could find pairs of equatorial orbits with different orbital parameters but with the same set of frequencies. Two such orbits have been plotted in Figure \ref{fig:2orbits_eq}
and have been marked (as $A$ and $B$) on the contour plot of Figure \ref{fig:isofr_eq}.

There is also one more similarity
connected with the orbital frequencies.
The Schwarzschild gravitational field
is a specific case of a Kerr metric,
the orbits of which are not closed due to
different values of $\Omega_\phi$ and $\Omega_r$.
Also the Euler field,
even when $a \to 0$,  
has orbits that are not closed, as well, 
(the ratio $\Omega_r/\Omega_\phi$ is not 
identically equal to unity 
and depends only on $e$ and $p/a$ and not
on the actual value of $a$). Therefore this
$a \to 0$ Euler field has orbital characteristics 
that are closer to Schwarzschild than to Kepler.
The existence of ISCO in the, almost Newtonian, 
Euler field is a singular outcome of the
above diversity of frequencies.

\begin{figure}[b]
\includegraphics[width=.9\linewidth]{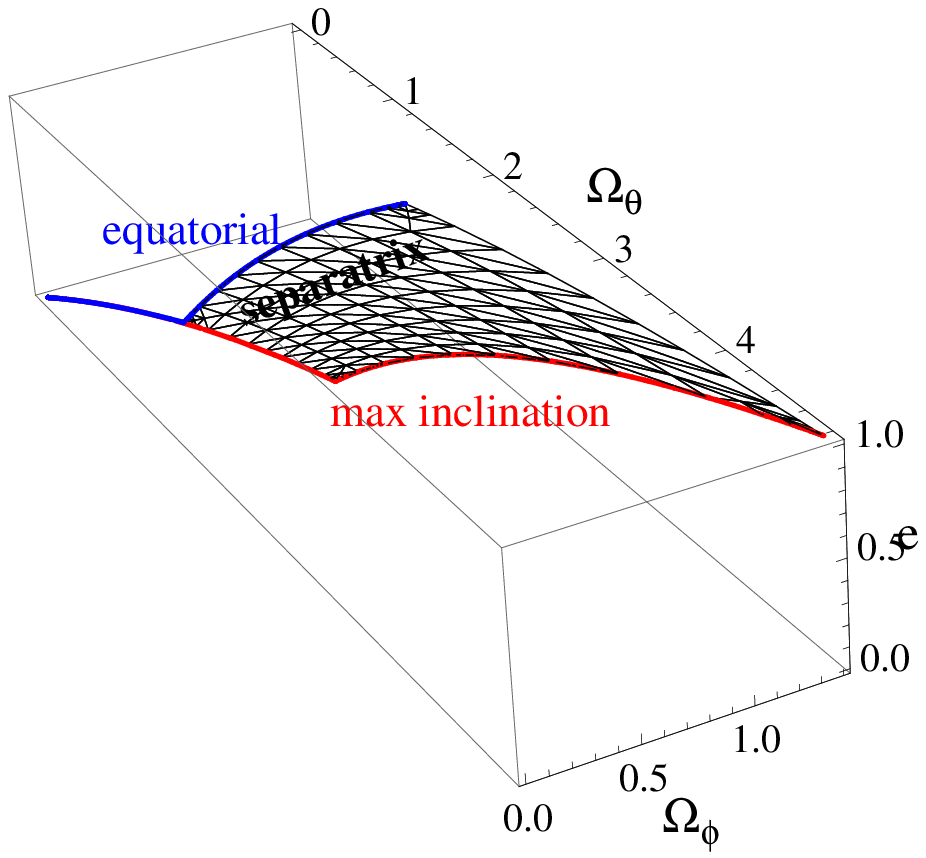}
    \caption{This plot shows two sides of  
    the 3-dimensional parameter space
    $(\Omega_\phi,\Omega_\theta,e)$ of 
    all normal bound orbits of the Euler gravitational
    field (one base and one of the lateral surfaces). 
    The plot  shows: (i) the separatrix 
    strip (the grid-surface which is the 
    locus of all orbits with $\Omega_r=0$, 
    apart of those with $e\to 1$
    and those of infinite semi-latera recta), 
    and (ii) the locus of spherical ($e=0$) orbits that are either equatorial 
    (blue solid line lying on the plane $e=0$), 
    or with maximum inclination (red 
    line on the plane $e=0$). 
    These two lines are the boundaries of 
    the base which are then extended along 
    the separatrix strip, defining 
    its boundaries (the equatorial orbits 
    at separatrix and the most inclined orbits 
    at separatrix). The two lines on 
    the plane $e=0$, are so close to
    each other that they look like a 
    single curve (both are starting from 
    the origin $(0,0,0)$, while the
    red one has a slightly greater $\Omega_\theta$
    frequency at a given $\Omega_\phi$). 
    Furthermore, the basis of the separatrix itself 
    is  almost parallel to the maximum 
    inclination curve on
    the plane $e=0$. Thus the base of 
    this 3-dimensional
    body of orbits is actually a 
    very thin curved triangle.
    The same is true also with all sections 
    of this body with the planes $e={\rm const}$
    (not shown in this diagram).
    The full body of the orbits forms, 
    in this parameter space, a very 
    thin curved and skewed 
    triangular prism, that looks more like a 
    2-dimensional surface. 
    The axis of $e$ correspond to all 
    infinitely distant orbits 
    that are characterized by $\Omega_r=\Omega_\theta=\Omega_\phi=0$. 
    A section of this body with a plane
    of constant $\Omega_\phi$, that intersects the separatrix strip has been drawn in Figure \ref{fig:isofr2_gen}, along with a 
    few iso-$\Omega_r$ contour lines.}
    \label{fig:isofr_gen}
\end{figure}

\begin{figure}[b!]
    \includegraphics[width=\linewidth]{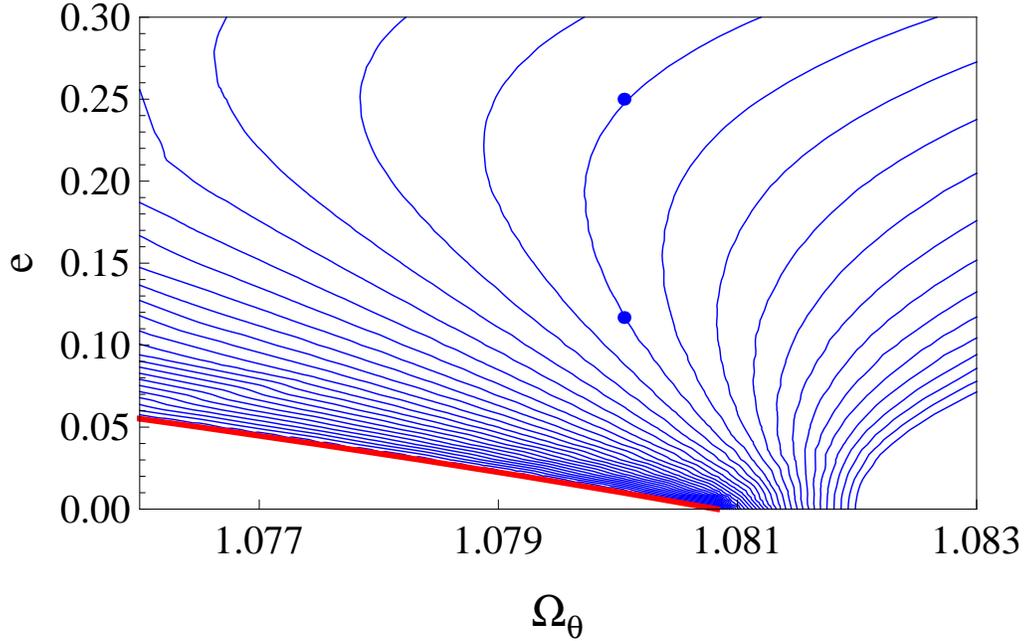}
    \caption{This is the contour plot of $\Omega_r$
    in the parameter space $(\Omega_\theta,e)$
    of the Eulerian generic orbits,
    while     $\Omega_\phi$ is held fixed at a constant value, such that the plane $\Omega_\phi={\rm const}$ intersects the separatrix surface. All frequencies are expressed as multiples of $\sqrt{M/a^3}$. The segment of
    the parameter space depicted here is very small especially along the axis of $\Omega_\phi$
    due to the tiny width of the 3-dimensional 
    body of orbits.  It should be noted that 
    the separatrix (the red line at the boundary of the contour plot) represents the lower
    values of $\Omega_\theta$. This can be easily
    explained from the 3-dimensional shape
    of the body of orbits depicted in Figure \ref{fig:isofr_gen}. In contrast, the 
    separatrix of generic orbits of Kerr 
    represents the higher values of $\Omega_\theta$
    along the section of $\Omega_\phi={\rm cons}$.
    This is due to the different orientation
    of the separatrix strip in the parameter space
    $(\Omega_\phi,\Omega_\theta,e)$ of Kerr.
    A specific iso-frequency pair has been marked with two dots on the diagram. The characteristics of these two synchronized orbits are written in Table \ref{tab:3}, while the corresponding orbits are depicted in Figure \ref{fig.iso3d_pair}. }
    \label{fig:isofr2_gen}
\end{figure}
 
\begin{figure}[h!]
\includegraphics[width = \linewidth]{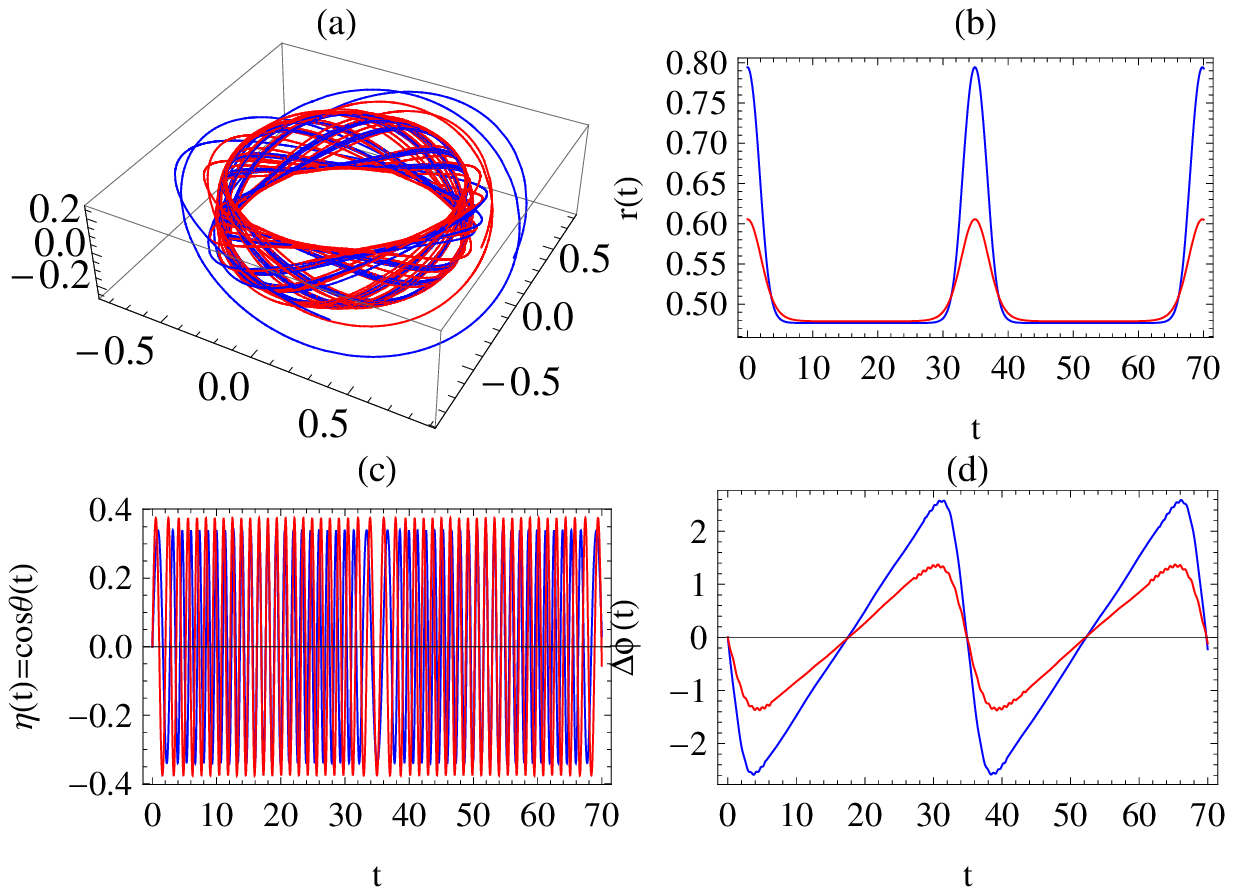}
    \caption{The specific iso-frequency pair
    mentioned in the previous paragraph has been plotted in this multiple figure. Diagram (a) shows the actual orbits regarding this pair (red is the orbit with the lower eccentricity, while blue is the most eccentric one). The rest three plots depict the evolution of  $r(t)$, $\eta(t)=\cos(\theta(t))$, and $\Delta \phi(t)=\phi(t)-\Omega_\phi t$,
    respectively, for both orbits. It is clear that all three oscillations
    of these particular orbits are synchronized; even
    $\eta(t)$ in diagram (c), where there is a periodic 
    shift between the two curves, there is no net shift
    when a large number of $\eta$ oscillations are
    taken into account. The 
    much higher value of $\Omega_\theta$ (see Table 
    \ref{tab:3}) with respect to $\Omega_r$ is 
    directly reflected to the corresponding
    plots (b and c). On the other hand the
    fundamental period of $\Omega_\phi$ (which
    is of the same order of magnitude with 
    that of $\Omega_\theta$) is not directly
    presented in plot (d), since it has been
    subtracted through the term $\Omega_\phi t$. 
    The apparent 
    periodicity of this diagram is simply a multiple
    of the corresponding fundamental period.
    \label{fig.iso3d_pair}}
\end{figure}

Next we investigate the existence of 
iso-frequency pairs in generic, non-equatorial orbits in the Euler gravitational field. Such orbits are characterized by a triplet of frequencies, $\Omega_r,\Omega_\theta,\Omega_\phi$, 
all of which are present in the evolution of the orbit.
In this case it is quite more complicated
to seek for a pair of orbits having the same triplet of frequencies $(\Omega_r,\Omega_\theta,\Omega_\phi)$,
since the 3-dimensional parameter space ($e-\Omega_\theta-\Omega_\phi$), spanned by all types of bounded orbits has the shape of a skewed 
triangular prism, which is so thin that it looks like a 2-dimensional slice in the parameter space
(see Figure \ref{fig:isofr_gen}).  
The separatrix of these orbits is now a strip 
parametrized by the functions $\Omega_\theta(\zeta,e)$, $\Omega_\phi(\zeta,e)$,
where $\zeta$ is a single parameter that 
varies the inclination of the orbit, while it
adjusts the semi-latus rectum $p$, so that $r_2=r_3$, at any given value of the eccentricity. This strip 
spans the whole range of eccentricities 
from $e=0$ to $e=1$. Even though the two frequencies
$\Omega_\theta$, $\Omega_\phi$ increase monotonically as one moves from $e=0$ to $e=1$, for any given inclination of the orbit (the strip has 
the right slope to allow for iso-frequency pairs), 
this is not sufficient to
ensure that there are iso-frequency pairs.
The contour surfaces of constant
$\Omega_r$ are also strips that fill, like onion-shells, the whole prism-like 3-D body 
of orbits in the new parameter space.
If we  intersect these iso-$\Omega_r$ contour 
surfaces,  with the plane of constant $\Omega_\phi$
(or one of constant $\Omega_\theta$),
the intersection will not necessarily
span the whole range
of eccentricities $0 \leq e \leq 1$,
due to the obliqueness of the body of orbits 
and its tiny thickness. This renders the
search for iso-frequency pairs unattainable
for a wide range of frequencies.
More specifically, if one starts from an orbit 
in the region close to the separatrix and 
moves upwards (towards higher $e$) or downwards 
(towards lower $e$) in order to find its 
potential iso-frequency pair, one
may end up at the oblique lateral boundary 
of the space of orbits, 
before reaching the initial $\Omega_r$ value.
This potential failure is strengthened by the fact that 
the thickness of the 3-dimensional body, 
describing all possible orbits in the parameter space,
is extremely tiny.
Therefore the starting point of this exploration might be quite essential. 
The search should only be restricted in 
a region close  to that part of the  separatrix strip
that lies near the surface which plays 
the role of the singular line of 
equatorial orbits. 
Then a numerical computation of the 
frequencies in the neighborhood 
of that initial point
follows. Our investigation ended up in 
the contour plot of Figure \ref{fig:isofr2_gen}, 
 which depicts a segment
of the iso-$\Omega_r$ contours on the 
intersection of the parameter space with the plane $\Omega_\phi=0.632456 \sqrt{M/a^3}$. Due to the
very narrow width of the parameter space, the
horizontal axis covers a very small range of $\Omega_\theta$ frequencies.
This contour plot diagram looks like
Figure 6 of \cite{Baracketal13}
referring to generic orbits in Kerr.
However, there is a small difference:
in the Euler field the 
iso-$\Omega_r$ contours lie at 
higher $\Omega_\theta$ values than those 
at the separatrix, while in Kerr 
the relative position is the opposite.
The reason is that the separatrix
surface is differently oriented with respect
to the rest orbits in the three-dimensional
parameter space $\Omega_\theta,\Omega_\phi,e$
in Kerr and in Euler;
thus the rest of the bound orbits lie on
opposite sides of the separatrix surface in
the two problems.

\begin{table}[b!]
    \centering
    \begin{tabular}{|c||c|c|}
   \hline     \hline 
   {}                   & A                    & B  \\
     \hline \hline
$\Omega^\star_{r}$      & \multicolumn{2}{|c|}{0.0455368 } \\
\hline
$\Omega^\star_{\phi}$   & \multicolumn{2}{|c|}{0.632456 } \\
\hline
$\Omega^\star_{\theta}$ & \multicolumn{2}{|c|}{1.08005 } \\
\hline
$e$                   & 0.25                  & 0.116833918947677 \\
\hline
$p^\star$             & 1.48957                 & 1.33702 \\
\hline
$\theta_{\min}(^{\circ})$             & 70.148692322964                 & 67.94240594 \\
\hline
    \end{tabular}
    \caption{These are the characteristics (orbital parameters and fundamental frequencies) of the pair of orbits
    shown up in Figures \ref{fig:isofr2_gen}
    and \ref{fig.iso3d_pair}. Frequencies are given
    in units of $\sqrt{M/a^3}$ and the semi-latera recta
    in $a$ units. The two orbits have different orbital
    characteristics, but their fundamental frequencies are exactly (up to a numerical accuracy of 80 significant figures) the same.}
    \label{tab:3}
\end{table}

 Following, then, the numerical scheme we
 used in the equatorial orbits, we 
 pinpointed  two distinct orbits
 with the same triplet of frequencies (the two
 points along the same iso-$\Omega_r$ contour line
 of Figure \ref{fig:isofr_gen}). The orbital
 characteristics of these two orbits have 
 been written in Table \ref{tab:3}, while the $r$-, $\theta$-, and 
 $(\phi-\Omega_\phi t)$-oscillations of these
 orbits, along with the orbits themselves, 
 have been depicted in Figure \ref{fig.iso3d_pair}.
 
Note that for non-equatorial orbits, as well as for equatorial orbits, all frequencies could be written in terms of the dimensional qauntity $\sqrt{M/a^3}$, 
that is the numerical value
 of all frequencies that multiplies $\sqrt{M/a^3}$, is independent of the actual value of $a$ and $M$. This is the reason why we have not assigned any specific value of $a$ in any of the plots of Figures
 \ref{fig:isofr_gen}, \ref{fig:isofr2_gen}.

 \section{ ``Circular'' orbits remain circular}
 \label{sec:6}

In this section we will exploit the
great similarity of the Euler's potential with the 
Kerr field in order to investigate the stability of spherical orbits (the ``circular orbits'' of Kerr
as they are mostly known) in both problems.
The initial argument in favor of this proposition, for the Kerr case,
was given by Ori and Kennefick \cite{KennOri}
back in the 90's. The argument was analytical,
but rather obscure, while the resonance case
$\Omega_r=2 \Omega_\theta$, which was
the condition for the argument not to hold, 
was not further studied. Later on, in the
late 90's Ryan \cite{Ryan} presented an elegant
argument for the stability, when the resonant 
condition is not met, based simply on the basic
symmetries of Kerr.

Here we will present an extensive analytic argument
to explain this stability, constructed in terms
of the Euler problem. The argument could be
recast, though, in the form of the Kerr case. 
In our study we have managed to translate
the problem in a driven harmonic oscillator
which has a continuously increasing amplitude
when the above resonance condition is met.
Especially the Euler case, in contrast to Kerr
case, could be set at such an initially spherical
condition that the resonance condition is met. 
We have shown
that such an orbit will eventually deviate from
sphericity when a generic dissipative self-force 
is taken into account. This is an example that 
strongly supports and further
explores  the ``spherical stability'' proposition.

Let us write down the equation of motion for the $r$ coordinate of a particle in
an Euler field in terms of Mino time 
$\lambda$ (c.f.~Eq.~(\ref{dtdMino}) of 
Appendix \ref{App:5}), amended by 
a tiny extra force that drives 
adiabatically the particle away from its 
geodesic orbit:
\begin{eqnarray}
\frac{d^2 r}{d\lambda^2} &=&
\frac{d}{d\lambda} \left( \pm \sqrt{V_r(r)} \right) + \epsilon F^{(SF)}_r
\nonumber \\
&=&  \frac{1}{2} V_r'(r) + \epsilon F^{(SF)}_r
\label{eqm0}
\end{eqnarray}
where $'$ denotes a derivative with respect to $r$, $\epsilon$ is a small parameter,  analogous to $\mu^2/M$ of an EMRI ($\mu$ is the test-particle's mass, while $M$
is the total mass of the Eulerian gravitational field). This 
is the usual scale of the
relativistic gravitational self-force at lowest order. This force plays the role 
of the  self-force of a relativistic 
test-particle orbitting around a  
Kerr black hole.

The potential $V_r$, related to orbital 
$r$-oscillation, 
is the quartic polynomial of
$r$ of Eq.~(\ref{Vr}), with coefficients 
that are given as functions of the three integrals 
of motion $E,L_z,Q$ (of the geodesic equation),
which are not constant anymore. Thus 
\be
V_r(r)=a_4 r^4 + a_3 r^3 + a_2 r^2 + a_1 r + a_0
\ee
with $a_i=a_i(E,L_z,Q)$;
thus
\be
\frac{d^2 r}{d \lambda^2}= \left( 
2 a_4 r^3 +\frac{3}{2} a_3 r^2 +a_2 r + \frac{1}{2} a_1 
\right)
+ \epsilon F^{(SF)}_r .
\label{eqm1}
\ee
Furthermore, the potential is characterized by a local
minimum $r_0$, around which the orbit evolves, at least initially. Thus $r(\lambda)\simeq r_0$, and $V_r'(r_0)=0$.
$r_0$ is the instantaneous center of $r$-oscillations,
the amplitude of which is directly related to the eccentricity of the orbit (which is assumed extremely
small at the beginning).   
Subtracting from the equation above the vanishing
derivative of the potential at $r_0$ we obtain the
following equation:

\be
\frac{d^2 r}{d \lambda^2} &=& 2 a_4 (r^3 - r_0^3) +\frac{3}{2} a_3 (r^2-r_0^2) +
a_2 (r-r_0)  + \epsilon F^{(SF)}_r \nonumber \\
&=& (r - r_0) \left(2 a_4 (r^2 + r r_0 + r_0^2) +
\frac{3}{2} a_3 (r+r_0) +
a_2 \right)  + \epsilon F^{(SF)}_r.
\label{eqm11}
\ee
At this point it should be emphasized that the new parameter $r_0$ showing up in the last expressions could also be considered a function of $E,L_z$, and $Q$, since it is simply the maximum root of the cubic equation $V'_r(r)=0$, which could be directly expressed in terms of $a_1,a_2,a_3,a_4$.

If  the corresponding geodesic orbit ($\epsilon=0$) 
is initially almost 
``spherical'', that is $r(\lambda) \simeq r_0$, 
 the equation above describes an approximate  harmonic oscillator
with $\omega_r^2=-6 a_4 r_0^2 - 3 a_3 r_0 - a_2$, which oscillates with very small amplitude. (Note that this $\omega_r$ is simply the $Y_r$
part of the fundamental frequency $\Omega_r$ mentioned in Appendix \ref{App:5}, since it is the frequency with respect to Mino-time $\lambda$). However,
when self-force is present, $a_i$'s (consequently $r_0$, as well),  will evolve
as mentioned previously; therefore
$r(\lambda)$ will adiabatically 
 deviate somehow from its
pure oscillatory fashion.

In order to study the new type of evolution
when any type of self-force is present, we will seek a solution in the form of
\be
r(\lambda)=r_0(0)+e_0 \Delta(\lambda),
\ee
assuming $r_0(0)$ is the initial value of $r_0$ (the instantaneous 
minimum of $V_r$) and $e_0$ is the  initial amplitude of
$r$-oscillations (which is proportional to the small initial eccentricity 
of the orbit), while $\Delta(\lambda)$ is a function 
of $\lambda$, of order zero (while eccentricity $e$ and 
magnitude of self-force $\epsilon$ are assumed to be of order one) that describes the 
overall evolution of $r$ (both oscillatory and secular evolution).
By direct replacement in Eq.~(\ref{eqm11}) we obtain the following equation of motion with respect to $\Delta$:
\be
e_0 \frac{d ^2 \Delta}{d \lambda^2}=-(r(\lambda)-r_0(\lambda))\omega_r^2(\lambda) +\epsilon F_r^{(SF)},
\label{eq0Delta}
\ee
where $\omega_r^2(\lambda)$ is the instantaneous 
value of $-6 a_4 r_0^2-3 a_3 r_0-a_2$, due to adiabatic changes of all these parameters.

Assuming $\epsilon,e$ are two comparable small 
quantities, as mentioned above, we will only keep  quantities of order 
${\rm O}(e)$ and ${\rm O}(\epsilon)$ in the equation above, and after using the full expression for $r(\lambda)$ from Eq.~(\ref{eq0Delta}) we get: 
\be
\frac{d ^2 \Delta}{d \lambda^2}=-\omega_r^2 \Delta- 
\frac{(r_0(0)-r_0)}{e_0} \omega_r^2  
+ \frac{\epsilon}{e_0} F_r^{(SF)},
\label{eqm01}
\ee
where we remind that the quantities $\Delta,\omega_r,r_0$
are functions of $\lambda$.

Next we will further analyze the drift of $r_0$, 
$\delta r_0 \equiv r_0-r_0(0)$, 
caused by the self-force.
Since, by the definition of $r_0$, $V_r'(r_0)=0$
(where $V_r$ is a function of $r$ and $\lambda$--through
the $\lambda$-dependence of its coefficients):
\be
0 &=& V_r'(r_0(\lambda),\lambda)\nonumber\\
&=& V_r'(r_0(0)+\delta r_0,\lambda)\nonumber \\
&\simeq& V_r'(r_0(0),\lambda)+V_r''(r_0(0),\lambda=0) \delta r_0,
\ee
the drift of $r_0$, $\delta r_0$, is approximately given by
\begin{equation}
    \delta r_0 \simeq
    - \frac{V_r'(r_0(0),\lambda)}{V_r''(r_0(0),\lambda=0)}= 
    \frac{0 + \delta V_r'(r_0(0),\lambda)}{2 \omega_r(0)^2},
\end{equation} 
where $\delta V_r$ denotes the shift of  $V_r$ due to  the evolution of the
coefficients of its polynomial expression, while the initial value $0$ at the numerator marks simply the
value of the derivative of the initial (at $\lambda=0$) $V_r$, at $r_0(0)$. The denominator $V_r''(r_0(0),\lambda=0)$ has been directly replaced by its value, $- 2 \omega_r(0)^2$ (see the paragraph after Eq.~(\ref{eqm11})),
calculated at the initial form of $V_r$.

By replacing $\delta r_0$ in Eq.~(\ref{eqm01}) 
with our final answer, 
and neglecting the drift of $\omega_r^2$ with $\lambda$ 
in the second term of the right hand side, as a higher order term, we obtain
\be
\frac{d ^2 \Delta}{d \lambda^2}=-\omega_r^2 \Delta +
\frac{\delta V_r'(r_0(0),\lambda)}{2 e_0}  
+ \frac{\epsilon}{e_0} F_r^{(SF)}.
\label{eqm11a}
\ee

Next, we will show that the value of the
evolved $V_r'$ at $r_0(0)$ due to secular change of 
the parameters $a_i$'s is simply
proportional to $\lambda$, at lowest order. All coefficients $a_4,a_3,a_2,a_1$ (the same also  
holds for $a_0$, but $a_0$ is not present in $V_r'$) are simple linear functions of $E,L_z^2$,
and $Q$ (c.f.~Eq.~(\ref{Vr})).
Therefore 
\be
\delta a_i &=& \int_0^{\lambda} \frac{d a_i}{d t} \frac{dt}{d\lambda'} d\lambda' \nn \\
&=& \int_0^{\lambda} \left( 
\frac{\partial a_i}{\partial E} \frac{d E}{dt} +
\frac{\partial a_i}{\partial L_z^2} \frac{d L_z^2}{dt} +
\frac{\partial a_i}{\partial Q} \frac{d Q}{dt} 
\right)
\frac{dt}{d\lambda'} d\lambda'.
\ee
On the other hand, each one of these time derivatives of the integrals of motion are exactly equal to 0
when there is no self-force (that is at the limit $\epsilon \to 0$). However, the derivatives $dE/dt,dL_z^2/dt,dQ/dt$ are not
 vanishing when a self-force is present. For example
\be
\frac{d E}{d t} &=& 
\frac{\partial E}{\partial r} \dot r +
\frac{\partial E}{\partial \eta} \dot\eta +
\frac{\partial E}{\partial \dot r} \ddot r +
\frac{\partial E}{\partial \dot\eta} \ddot\eta+
\frac{\partial E}{\partial \dot\phi} \ddot\phi,
\ee
since $E$ (as well as $Q$ and $L_z^2$) is a function of either all $r,\eta,\dot r,\dot\eta,\dot\phi$,
or a few of those. 
In the expression above, 
$\eta$ is simply an abbreviation for $\cos\theta$.
The dependence of the expression  above, for
$dE/dt$, on the self-force is hidden only in the 
double time derivatives $\ddot r,\ddot\eta,\ddot\phi$. All other terms, including that part of the double derivatives corresponding to no
self-force, have a vanishing net result, 
since $E$ is an integral of motion for
pure gravitational force (geodesic motion),
without any extra self-force. Thus
\be
\frac{d E}{dt} &=& \epsilon \left(
\frac{\partial E}{\partial \dot r} F_r^{(SF)} +
\frac{\partial E}{\partial \dot\eta} F_\eta^{(SF)}+
\frac{\partial E}{\partial \dot\phi} F_\phi^{(SF)}
\right).
\ee
Actually all integrals of motion $E,L_z^2,Q$ are bilinear functions of $\dot r,\dot\eta,\dot\phi$,
therefore $\partial E/\partial \dot r, \partial E/\partial \dot\eta$, and $\partial E/ \partial \dot\phi$ are simply
linear functions of
$\dot r,\dot\eta$, and $\dot \phi$, respectively.
Thus, collecting all these partial results we end up with a general expression for
all $a_i$'s:
\be
\delta a_i &=& \epsilon \sum_k
 \int_0^{\lambda}  F_k^{(SF)} \left( 
\frac{\partial a_i}{\partial E}G_{E,k} +
\frac{\partial a_i}{\partial L_z^2} G_{L_z^2,k}  +
\frac{\partial a_i}{\partial Q} G_{Q,k} 
\right)
\frac{dt}{d\lambda'} d\lambda',
\ee
where $x_k$ denote the coordinates $r,\eta,\phi$ for
$k=1,2,3$, respectively, while
$G_{E,k}=\partial E /\partial \dot x_k$ (and similarly for 
$G_{L_z^2,k}, G_{Q,k}$) which are linear with respect to $\dot{x}_k$. Note that all quantities inside the integral
should be computed along a geodesic orbit, since $\delta a_i$ itself is of
order $\epsilon$, whereas any deviation from geodesic will cause higher order corrections.
Furthermore, assuming that $F_k^{(SF)}$ is of the form $-\dot x_k f_k(r,\eta,\dot{x}_l^2)$ 
--that is, of purely dissipative character--, $\delta a_i$ will be given by
integrals of $(\dot x_k) ^{2}$ and other more complicated functions of 
coordinates $r,\eta$ and $\dot{x}_l^2$.
Finally, the term $dt/d\lambda'$ is also a quadratic function 
of $r$, and $\eta$:
\begin{equation}
    \frac{dt}{d\lambda'}=r^2+a^2 \cos\theta,
\end{equation}
(see Eq.~(\ref{dtdMino})).

Now, taking into account the almost constant
value of $r$ of spherical orbits, the integrand for each $\delta a_i$ will oscillate, mainly due to $\eta$-oscillations,
around its average value.
Thus all integrals related with $\delta a_i$'s (both $\omega_r^2$
and $\delta V_r'(r_0(0),\lambda)$) will consist of a part that scales linearly with $\lambda$,
due to the average value of the integral,
plus an oscillating part,
due to $\eta$-oscillations.
The total $\lambda$-time of integration to 
compute
$\delta a_i$ is assumed sufficiently short 
to be insensitive to the drift of $r$-coordinate caused
by the self-force, but sufficiently long to
span at least a few complete periods of $\eta$. 
Of course, for longer time periods,
higher order terms, than the linear terms
with respect to $\lambda$, will show up.
Combining all previous results, $\Delta$ will obey the following generic
equation:
\be
\frac{d^2 \Delta}{d \lambda^2}=
-\left[ \omega_0^2 + \epsilon (B \lambda + A_\eta(\lambda)) \right] \Delta
-\frac{\epsilon}{e_0} \left[ D \lambda + C_\eta(\lambda) + \dot r f_r \right],
\label{eqm2}
\ee 
where $\omega_0^2=\omega_r(0)^2$ and 
\be
B \lambda + A_\eta(\lambda)=\frac{\delta \omega_r^2}{\epsilon}, 
\ee
while
\be
D \lambda+ C_\eta(\lambda)=\frac{1}{\epsilon}
\left[
4 \delta a_4 r_0(0)^3 + 
3 \delta a_3 r_0(0)^2 + 
2 \delta a_2 r_0(0) +
\delta a_1 \right].
\ee
The terms $B \lambda, D \lambda$ denote the linear 
part of the integrals mentioned above, while  $A_\eta, C_\eta$ denote the oscillating part of the integrals due to $\eta$-oscillations of the orbit itself.
The last term of the second bracket of the right hand of Eq.~(\ref{eqm2}) is of higher order than the rest terms,
since it is  proportional to 
$e_0$ (the tiny amplitude of the oscillation of $r$) --the other terms $D \lambda$ and $C_\eta(\lambda)$
are of order unity--, thus it could be omitted. Eq.~(\ref{eqm2}) describes a harmonic oscillator
with a drifting frequency, that is driven by an external force which consists of a linear part with respect to time and an oscillating part caused by
$\eta$ oscillations. As a consequence, $\Delta$ will oscillate with frequency that varies continuously
in an adiabatic fashion, while it slightly oscillates 
at even harmonics of $\omega_\eta$ (since all functions of $\eta$ are 
quadratic with respect to $\eta$ due to the reflection-symmetry
of $V_\theta$). On the other hand, $\Delta$ adiabatically drifts  away with $\lambda$ 
(due to $D \lambda$ term) and is driven by the
$C_\eta(\lambda)$ term that oscillates again at
even harmonics of $\eta$.
Especially, if $\omega_r=2 m \omega_\eta$ (where $m$ is some integer), resonance will take place 
and  $\Delta$ oscillations will grow in amplitude, 
until the drift of frequencies (mainly 
due to the $B \lambda$ term)
will bring the system out of resonance. 
Therefore the mechanical model for the
time-dependence of $\Delta$ could be
described approximately by an equation of the form
\begin{equation}
    \frac{d^2\Delta}{dt^2}+[\omega_0^2+\epsilon_1 t     +\epsilon_2 \cos(k \omega_0 t+\phi_1)]\Delta+
    \epsilon_3 t + \epsilon_4 \cos(k \omega_0 t+\phi_2)]=0,
    \label{model}
\end{equation}
where $\epsilon_1,\epsilon_2,\epsilon_3,\epsilon_4$
are small numbers, $\phi_1,\phi_2$ are
random phases, while $k$ is a factor
that regulates resonance or non-resonance condition.
The difference in   sign of the driving force
(the last two terms in the equation above), compared to that of Eq.~(\ref{eqm2}) is deceiving; the $D$
parameter in (\ref{eqm2}) is negative
for a dissipative self force that makes
the orbit drift closer to the strong field region.
Notice also that the forced oscillator described by 
Eq.~(\ref{eqm2}) and its simplified model
(\ref{model}) is free of any dissipation, so its 
amplitude could grow indefinitely, as long as the 
resonance condition is met --actually the term $\dot{r} f_r$ in Eq.~(\ref{eqm2}), that we omitted
in Eq.~(\ref{model}), operates like a 
dissipative force but of tiny
strength.

Finally, we should mention that the oscillatory term $A_\eta(\lambda)$ in Eq.~(\ref{eqm2}) (represented
in our harmonic-oscillator model by the term $\epsilon_2$) could in principle lead to
parametric resonance as well (when $\Omega_r/(2 \Omega_\theta)=m/2$, where $m$ is an integer).
However this is quite difficult to work
since the parametric resonance is quite sensitive
to the resonance condition. 

The similarity between the Kerr problem and 
the Euler problem is such that the whole
process described above fits perfectly well in
the analysis of the evolution of spherical
orbits in both cases. The Kerr case though
does not meet the resonance condition for any
kind of spherical orbits as it was suggested 
in \cite{KennOri} (we did not find 
any such orbit in Kerr, as well). 
On the other hand the
Euler problem, having small but distinctive
quantitative differences
from Kerr (the corresponding quartic polynomials 
of $V_r$ are not identical), does actually admit
initial parameters that describe spherical orbits
with $\Omega_r=2 \Omega_\theta$.
These Eulerian orbits offer an ideal testbed to
check our analytic predictions for the 
evolution of small $r$-oscillations.

\begin{figure}[b]
    \includegraphics[width=\linewidth]{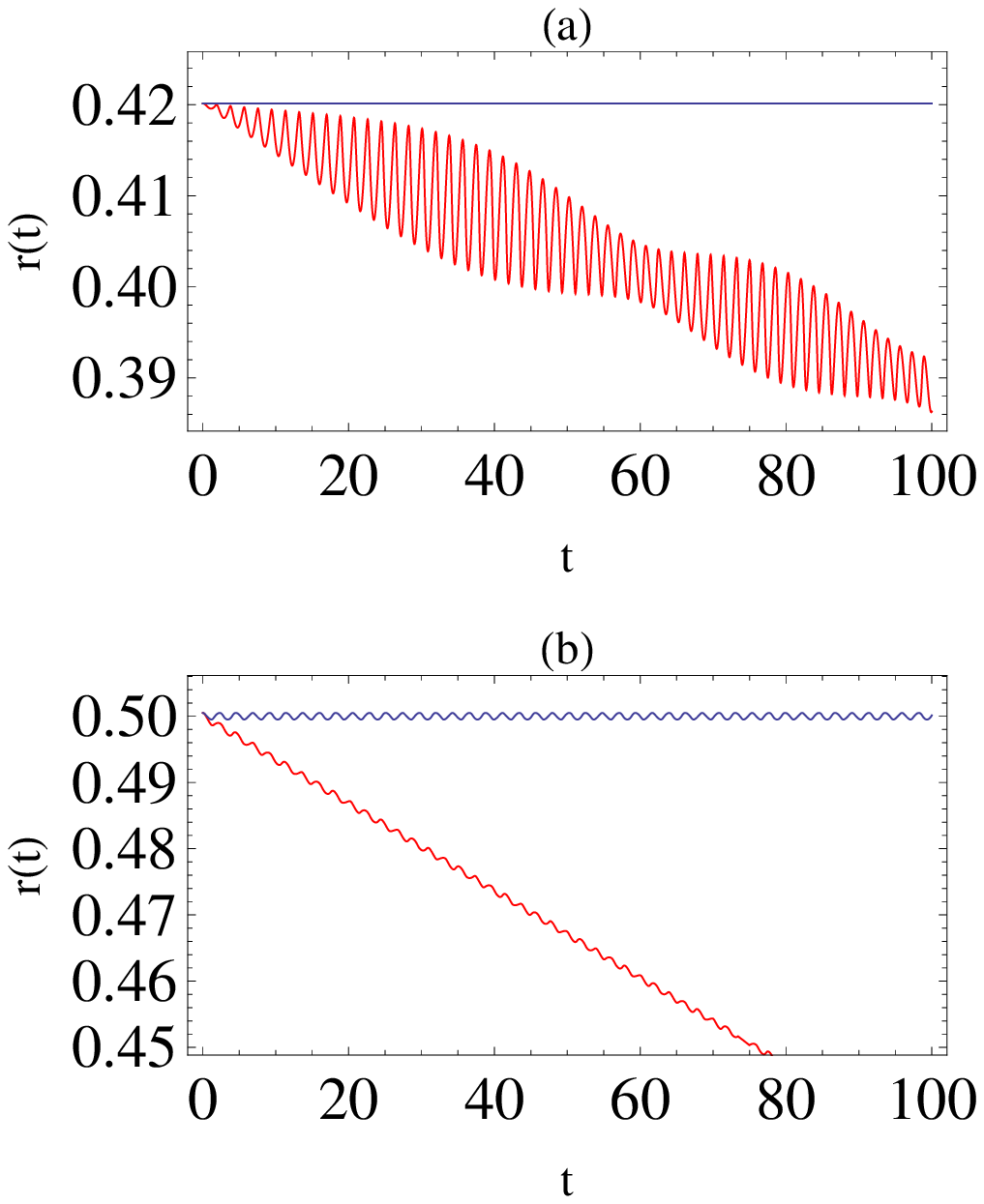}
    \caption{This plot shows the evolution of a bound Eulerian orbit under the dissipative force of Eq.~(\ref{thisSF}). The evolution
    of the orbits have been computed with 
    respect to normal time, $t$, and not with 
    respect to Mino-type time, $\lambda$.
    Diagram (a) 
    corresponds to a geodesic 
    (when the external force is absent) 
    circular orbit that initially
    is at resonance, that is, $\Omega_r(0)=2\Omega_\theta(0)$.
    The blue curve shows the
    evolution of the radial
    coordinate when the external force is zero. Since the orbit is circular
    there is no change on $r(t)$ then.
    The red curve shows the evolution
    when the self-force is turned on.
    Due to resonance, radial oscillations are excited, while the orbit drifts to lower $r$ values. 
    Diagram (b) shows the corresponding evolution of an orbit when there is no resonance. More specifically
    this orbit was chosen to have initially
    $\Omega_r(0)/(2 \Omega_\theta(0))=0.6$. If the self-force is absent (blue curve)
    the orbit oscillates slightly 
    due to the small initial eccentricity
    introduced in the initial conditions. 
    When the self-force 
    is turned on (red curve) the radius of the orbit drifts to lower values
    without any apparent increase 
    in its eccentricity. The different average 
    slope of the two red curves is due to
    the fact that the initial conditions 
    are completely different.}
    \label{fig:resandnonres}
\end{figure}

We have actually investigated such oscillations by
performing numerical integrations of spherical orbits
under an artificial  dissipative self-force, of quite
arbitrary form. More specifically we have used a
self-force of the form
\begin{equation}\label{thisSF}
{\bf F}^{(SF)}= - \epsilon a \frac{1-\eta^2}{r} {\bf v},
\end{equation}
where $\bf v$ is the velocity on oblate spheroidal coordinates (see Appendix \ref{App:1}). The
form of the self-force has been constructed 
so as to lead to a loss of energy and angular
momentum, while its strength is enhanced at 
lower $r$ values where the field is stronger, 
and depends on the $\eta$ coordinate in a 
reflection-symmetric way. 
The components of the self-force on spheroidal coordinates are:
\begin{eqnarray}
F_r^{(SF)}&=&-\epsilon a
\sqrt{\frac{r^2+a^2 \eta^2}{r^2+a^2}}\frac{(1-\eta^2) \dot{r}}{r}
\\
F_{\eta}^{(SF)}&=& -\epsilon a 
\sqrt{\frac{r^2+a^2 \eta^2}{1-\eta^2}}\frac{(1-\eta^2) \dot{\eta}}{r}\\
F_{\phi}^{(SF)}&=& -\epsilon a\sqrt{(1-\eta^2)(r^2+a^2)}
\frac{(1-\eta^2)\dot{\phi}}{r},
\end{eqnarray}
(c.f.~Appendix \ref{App:1}).
We have numerically integrated the time evolution
of an Eulerian orbit under the action of the self-force
given in the previous paragraph (apart of the gravitational force). The strength parameter
$\epsilon$ was adjusted to such a low value,
$(1/1000)$,
that the orbit does not deviate significantly
from the corresponding geodesic orbit for 
a time period equivalent to a few times
the maximum period of all orbital frequencies.
The initial conditions of the orbit was prepared
to obtain a spherical orbit when the self-force
was absent. Furthermore different initial 
conditions were constructed so that the
orbit was either at resonance ($\Omega_r=2
\Omega_\theta$), or not.
What we have observed in our numerical experiments is
that when the almost spherical orbit does not
satisfy the resonance condition, the effect of
the above self-force is simply a continuous
drift of the radius of the orbit towards
closer (lower $r$ values) spherical orbits, 
without any apparent increase in its eccentricity
(see Figure \ref{fig:resandnonres}(b)).
However, if the initial spherical orbit
meets the resonance condition, there is a noticeable
increase in its eccentricity (see
Figure \ref{fig:resandnonres}(a)), while the 
average $r$ coordinate of the orbit
drifts to lower values due to the 
dissipative self-force.
The increase of eccentricity though is not
monotonic. At some point, it starts
decreasing, like a beating effect.
Obviously the resonance condition 
is then lost and the amplitude of 
$r$-oscillations 
starts decreasing. It should be noted here that
although the initial conditions we considered
were describing a perfect spherical geodesic orbit,
when the self-force were absent, 
the appearance of an extra self-force 
is destroying  its integrability,
and some kind of initial eccentricity 
was then indirectly induced in the orbit. 
This tiny eccentricity
could either increase (due to resonance) 
or remain small during the evolution
of the orbit when the orbital 
characteristics keep it out of resonance.

\begin{figure}[b]
    \includegraphics[width=\linewidth]{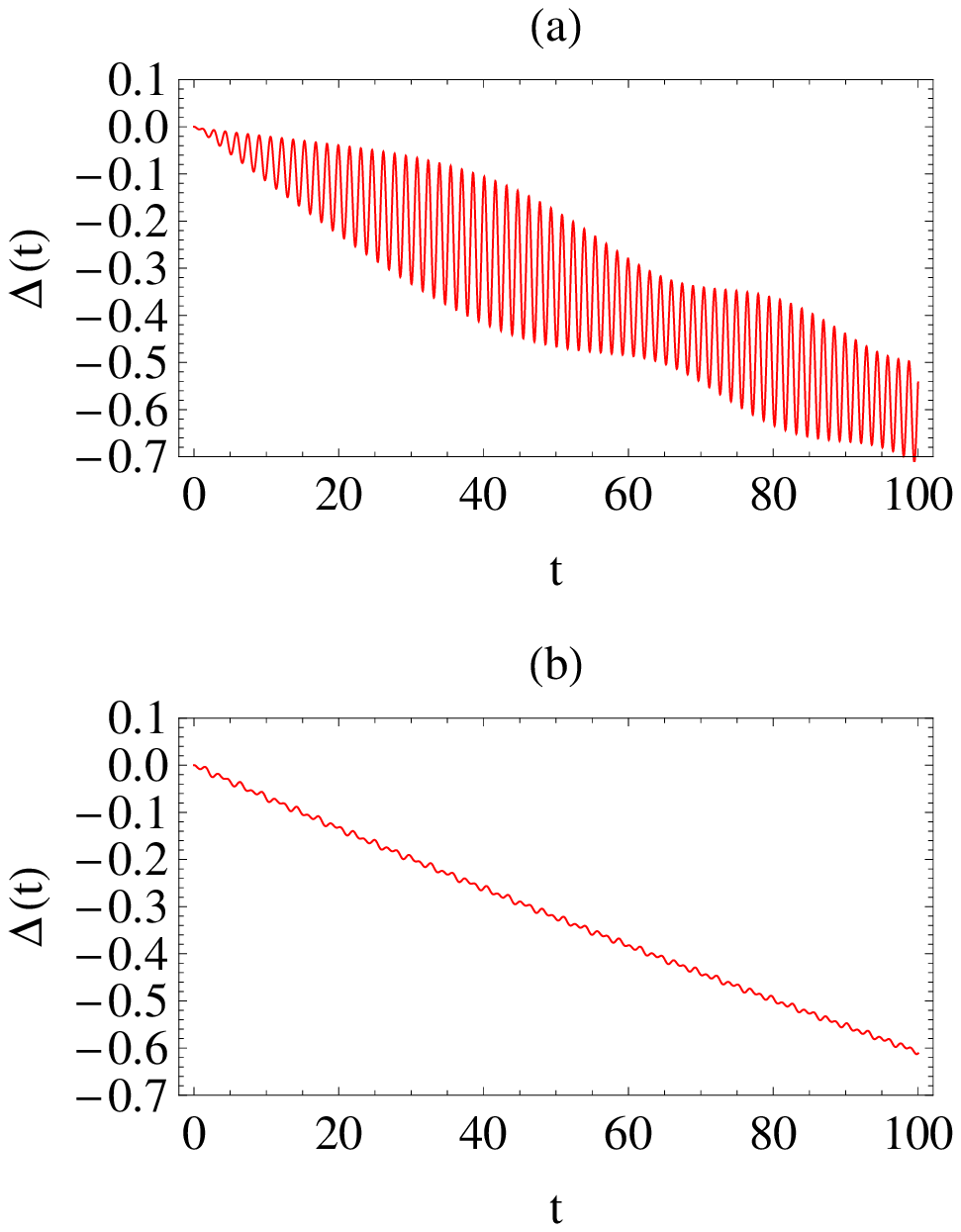}
    \caption{The two diagrams show the evolution
    of the driven harmonic oscillator described in Eq.~(\ref{model}). The frequency of the oscillatory part of the driving force is $k \omega_0$, where $\omega_0$ is
    the initial frequency of the oscillator.
    In plot (a), $k=1$. Thus  the amplitude of the oscillator increases due to resonance. Later on the resonance condition is lost due to frequency shift and the amplitude starts decreasing. In the second plot (b), $k=0.6$.  Now there is no resonance, and the oscillator simply drifts to lower values (the $e_3$ term in Eq.~(\ref{model}) is positive) 
    without any apparent amplitude increase.}
    \label{fig:model}
\end{figure}

In order to show the relevance of our mathematical model, described by Eq.~(\ref{model}), 
of a simple driven and drifting harmonic oscillator,
we have numerically solved Eq.~(\ref{model})
with parameters: $\omega_0=4$, $\epsilon_1=0.021$, $\epsilon_2=0.01$, $\epsilon_3=0.011$, $\epsilon_4=0.04$,
$\phi_1=\phi_2=0$, and $k$ either 1, which signifies resonance, or 0.6, which describes a non-resonance
condition. 
The evolution of $\Delta$, with
initial conditions $\Delta(0)={\dot\Delta}(0)=0$,
is shown in Figure \ref{fig:model}. The
evolutionary behavior of $\Delta$ looks like
what we got in the resonant (or non-resonant) forced 
Eulerian orbit. We tried to excite parametric resonance in our mathematical model, as well,
by choosing $k=1/2$, but we found that
we do need extremely high,  
non-physical, values of $\epsilon_2$
to achieve this goal. Therefore the 
oscillating part $A_\eta(\lambda)$ in
Eq.~(\ref{eqm2}) seems rather unimportant.

Before ending this  section
we should note once again the 
differences between the two times: the Mino-type time parameter $\lambda$ and the
normal time parameter $t$. With respect to
$\lambda$ both the $r$ oscillations and the $\eta$
oscillations are periodic with frequency
$\omega_r$ and $\omega_\eta$ respectively.
The corresponding oscillations with
respect to $t$-time are not in general
periodic. The fundamental frequencies 
are in this case $\Omega_r$ and $\Omega_\theta$.
The two types of frequencies are not equal,
but the ratio between them is the same, that is
\begin{equation}
    \frac{\Omega_r}{\Omega_\theta}=
    \frac{\omega_r}{\omega_\eta},
\end{equation}
since the $\omega$'s frequencies 
(with respect to $\lambda$) are simply the
$Y_i$ components used in the construction of the 
fundamental frequencies (c.f.~Eqs.~(\ref{Yr},
\ref{Y8}, \ref{Yf}) in  Appendix 
\ref{App:5}).

 \section{Conclusion - Discussion}
 \label{sec:7}
 
 In this article we have studied 
 thoroughly a Newtonian 
 gravitational field that shares a lot of 
 similarities with the relativistic gravitational
 field of a Kerr black hole. The fundamental
 property of both problems that makes them
 look similar is the fact that both are 
 integrable and separable  problems, 
 characterized by three constants of motion, 
 the physical meaning of which is completely
 analogous in the two cases. These
 three integrals of motion are the energy, 
 the $z$-component of the angular momentum, 
 and the Carter constant, a quantity
 that is quadratic with respect to momenta
 which could be viewed as a continuous 
 transformation of the square of the total 
 angular momentum of the corresponding
 spherically symmetric fields (monopole Newtonian and
 Schwarzschild) when the extra parameter 
 $a$ is introduced so as to destroy 
 the spherical symmetry while retaining 
 the integrability in the two problems. 
 It is quite intriguing that
 this  $a$ parameter 
 has a completely  different origin in these
 two problems: while it represents the spin 
 parameter in Kerr case, it is simply a
 length in the Euler problem, that defines an
 imaginary distance between the two half masses
 governing the axi-symmetric gravitational field.
 Of course the different physical frame 
 assumed when one studies each problem
 (the relativistic frame in the former one and 
 the Newtonian in the latter one) introduces
 restrictions in considering the two problems as
 completely analogous. For example the horizons 
 in Kerr are absolutely absent in the Euler problem
 since they are of completely relativistic origin.
 Also the consideration of space-time as a dynamic
 entity in the Kerr field leads to the dragging
 of frames, which has no analogue in the Newtonian
 frozen space of the Euler problem.
 
 We have attempted to give a complete list of
 properties of the two problems that are 
 qualitatively (and sometimes quantitatively)
 comparable, taking into account their 
 intrinsic differences. In some parts 
 of our comparison we  found a way to bridge
 these differences; the frame-dragging effect has 
 been artificially neutralized, by considering 
 the rotational frequency of a test body with 
 respect to a ZAMO observer 
 as a physical tool to construct 
 equivalent prograde and retrograde 
 orbits in Kerr, so that the physical characteristics of each such pair
 could be compared,
 on  equal footing, with the  omni-directional
 orbits in the Euler potential. With respect to
 other fundamental differences, like the existence 
 of horizons in Kerr, we have avoided to 
 approach such regions by
 considering only orbits that are bound in a fixed
 region of the gravitational field in both
 problems without reaching either the plunging
 region of the Kerr field, or the interior
 allowed region of the Euler field. Schematically,
 we have named the orbits that either move 
 exclusively in the interior region or in a
 region that has been produced by a merge of the
 two regions of the Euler field, ``plunging'' or
 ``effectively plunging'', as well). 
 
 The similarities between the two problems 
 are summarized in the following list: (i) The 
 separability of the two problems leads to
 two potentials $V_r(r)$ and $V_\theta(\theta)$
 that describe the orbital dynamics. Both share a lot of common properties:
 $V_\theta$'s have exactly the same form in both problems,
 while $V_r$'s are described by 4th order polynomials
 in which the physical constants of motion are
 introduced in similar but not exactly equal
 manner (part of the difference is due to
 the frame-dragging dynamic property of 
 space-time itself which is incorporated in Kerr's potential).
 (ii) Due to separability, the wave properties 
 in both fields have analogous characteristics, especially the scalar case.
 (iii) The multipolar structure of the Euler
 field is exactly that of Euler, if the
 current-moments of Kerr are neglected. This
 makes the two problems behave exactly like each 
 other at least at large distances.
 (iv) The bound orbits are precessing orbits that
 move in and out radially. (v) There is an ISCO
 for both problems. This is of great astrophysical interest since one could describe disks of matter
 with a finite internal radius.
 (vi) Although the $a=0$ case of the Euler problem
 is the monopole gravitational field leading to
 simple closed elliptical orbits, whatever 
 the initial  conditions are, the $a \to 0$
 Eulerian problem
 has an ISCO and bound orbits that are not closed
 in the strong field region, which is also true
 in the Schwarzschild case.
 (vi) The bound orbits  are characterized 
 by a set of 
 three fundamental frequencies in both problems.
 The expressions for the frequencies 
 could be written 
 in similar forms, in terms of 
 elliptical integrals,
 for both problems. Especially in the strong field 
 regime analogous characteristics arise. Thus
 in a specific region of orbital characteristics,
 both problems have pairs of distinct orbits with the same 
 set of fundamental frequencies.
 
 A simple first exploitation of this unique
 analogy between the two problems is to
 consider the Boyer-Lindquist coordinates
 that are usually used to describe the Kerr 
 metric not as the analogous of 
 spherical coordinates
 (used in the Schwarzschild case) but rather
 as some kind  of oblate spheroidal coordinates
 that are suitably adjusted by the spin parameter
 $a$ of the Kerr black hole, which is directly
 related with the quadrupole moment of the 
 black hole itself. This different view-point
 elucidates the difference in radial proper distance  across the equator and across the axis of symmetry of the Kerr metric.
 
Also, taking into account all the above 
similarities, one could 
use the Newtonian problem as a
simple mechanical model to discuss and 
clarify some subtle issues regarding 
the dynamics of 
orbits in a Kerr black-hole field. 
In this article we have reexamined 
the old argument
according to which the ``circular'' orbits 
evolve into circular orbits in Kerr, under
the action of a self-force arising from
the radiation reaction of EMRIs. 
The evolution of a ``spherical'' (as we 
call it here) orbit in the Euler problem 
under a generic dissipative self-force
has been investigated. We have analyzed
the dynamic evolution of radial oscillations
under the influence of such an external force
and ended up into a simple harmonic oscillator toy-model
with a drifting and oscillating frequency
driven by a force that is partly linear with 
respect to time and partly oscillating. Both 
oscillating parts in this mechanical model
have the frequency of $\eta^2$, that is twice 
the $\Omega_\theta$ frequency and higher 
even harmonics.
According to our toy-model the radial oscillations
will grow, mainly due to simple resonance when the resonance condition $\Omega_r=2 k\Omega_\theta$ (with $k$ integer)
is met. As explained in Section \ref{sec:6}
the parametric resonance (corresponding to 
$k=1/2,1,3/2,\ldots$), although in principle
capable to increase
exponentially the eccentricity of the orbit, it
is quite sensitive to the resonance condition; 
consequently it is rather difficult to arise
when all frequencies are drifting. On the other hand 
the simple resonance condition (which is met when
$k=1,2,\ldots$) will make the eccentricity grow
linearly with time or it will cause a beating effect
when the resonance condition is turned on and off due to frequency drift.
This analysis could be used to analyze and explain the adiabatic evolution of a spherical orbit either in Kerr or 
in Euler under any type of perturbative dissipative external force. While in Kerr the fundamental
resonance condition has not been found to
hold for any kind of spherical orbits, this is not true
for the Euler problem. The small quantitative
differences between the two problems render the 
Euler problem suitable to test our model.
Having the analytical tools to seek an orbit
in the frequency parameter space with suitable
frequencies in resonance, it was quite easy to find such an orbit
with $\Omega_r=2 \Omega_\theta$ initially. Then we 
numerically evolved it under a small external
(non-gravitational) dissipative force of very
simple form. We actually confirmed that
our mathematical model captures the exact 
qualitative behavior of the orbital
eccentricity.

We believe that the analogy presented in this 
paper could be further used to study the adiabatic
passage of an orbit around a black hole through 
a resonance due to gravitational radiation.
The Euler analogue is a perfect tool, 
we believe, to
thoroughly study such a delicate issue
that might have implications in the corresponding
signal either in the  Kerr case (an exact 
integrable system), or
that of a modified slightly non-integrable system
(e.g.~a perturbed Kerr black
hole) by employing, accordingly, a suitable 
Newtonian analogue of a modified Euler 
gravitating system that is constructed to be slightly non-integrable. We plan to address these issues
in a following paper.

\section*{Acknowledgements}

TA would like to thank a lot of people with whom 
he had long discussions,  
during the past years, about this problem.
Katerina Chatziioannou, and later George Pappas 
were the first persons who performed the
first computations regarding some of
the similarities relating the orbital 
characteristics in Euler 
and in Kerr. Also Charalampos Markakis
who shared with us his ideas with respect to generalizations
of the notion of Carter constant. 
Later Kostas Glampedakis
collaborated with TA to investigate the similarities 
at the level of wave properties. Technical advice from Maarten van de Meent's who has 
acquired a deep intuition on Kerr's 
resonances was of great help as well. We offer our thanks also to 
Georgios Lukes-Gerakopoulos for sharing his insightful
ideas regarding chaotic behavior of orbits in
non-integrable systems with us. AE would like to thank 
the research funding program I.K.Y. (MIS-5003404)
for its support, which was of great importance for the completion of this research project.


\appendix

\section{Oblate Spheroidal coordinates}
\label{App:1}

The Lagrangian in oblate spheroidal coordinates $(r,\eta,\phi)$ is:
\begin{eqnarray}
    L&=&\frac{1}{2}(r^2+a^2\eta^2)\left(\frac{\dot{r}^2}{r^2+a^2}+\frac{\dot{\eta}^2}{1-\eta^2}\right)\nonumber \\
    &+&\frac{1}{2}(r^2+a^2)(1-\eta^2)\dot{\phi}^2+\frac{M r}{r^2+a^2\eta^2}
\end{eqnarray}
where $r=a\xi$ ($\xi$ is the radial oblate spheroidal coordinate). An overdot denotes differentiation with respect to time coordinate $t$.

From Euler-Lagrange equations, we derive the equations of motion:
\begin{eqnarray}
    \ddot{r}&=&\frac{r}{r^2+a^2\eta^2}
    \left(-\frac{a^2\dot{r}^2(1-\eta^2)}{r^2+a^2}+
    \frac{\dot{\eta}^2(r^2+a^2)}{1-\eta^2}\right)
    -\frac{2a^2\eta\dot{\eta}\dot{r}}{r^2+a^2\eta^2} \nonumber \\
    &&+\frac{r(r^2+a^2)(1-\eta^2)\dot{\phi}^2}{r^2+a^2\eta^2}
    -\frac{M(r^2-a^2\eta^2)(r^2+a^2)}{(r^2+a^2\eta^2)^3},\\
    \ddot{\eta}&=&-\frac{\eta}{r^2+a^2\eta^2}
    \left(-\frac{a^2\dot{r}^2(1-\eta^2)}{r^2+a^2}+
    \frac{\dot{\eta}^2(r^2+a^2)}{1-\eta^2}\right)
    -\frac{2r\dot{r}\dot{\eta}}{r^2+a^2} \nonumber \\
    &&-\frac{\eta(r^2+a^2)(1-\eta^2)\dot{\phi}^2}{a^2+r^2\eta^2}
    -\frac{2Ma^2r\eta(1-\eta^2)}{(r^2+a^2\eta^2)^3},
    \\
    \ddot{\phi}&=&\left(-\frac{2r\dot{r}}{r^2+a^2}+\frac{2\eta\dot{\eta}}{1-\eta^2}\right)\dot{\phi}.
\end{eqnarray}
We used the above equations for integrating numerically the orbits and not  Eqs.~(\ref{quadr1})-(\ref{quadr3}). When the latter ones where used in a numerical integration scheme,
Eqs.~(\ref{quadr1})-(\ref{quadr3}) accumulate error at the turning points  due to the square roots. Moreover the signs of the $r$ and $\theta$ velocities have to be changed every time the orbit passes through a turning point.

The oblate spheroidal unit vectors are \cite{Nig}:
\begin{align*}
    \hat{\bf r}&=r\sqrt{
    \frac{1-\eta^2}{r^2+a^2\eta^2}} \cos{\phi}
    \;\hat{\bf i}+
    r\sqrt{\frac{1-\eta^2}{r^2+a^2\eta^2}}
    \sin{\phi}\;\hat{\bf j}+
    \eta\sqrt{\frac{r^2+a^2}{r^2+a^2\eta^2}}
    \;\hat{\bf k}\\
    \boldsymbol{\hat{\eta}}&=
    -\eta\sqrt{\frac{r^2+a^2}{r^2+a^2\eta^2}}
    \cos{\phi}\;\hat{\bf i}-
    \eta\sqrt{\frac{r^2+a^2}{r^2+a^2\eta^2}}
    \sin{\phi}\;\hat{\bf j}+
    r\sqrt{\frac{1-\eta^2}{r^2+a^2\eta^2}}
    \;\hat{\bf k}\\
    \boldsymbol{\hat{\phi}}&=
    -\sin{\phi}\;\hat{\bf i}+
    \cos{\phi}\;\hat{\bf j},
\end{align*}
where $\left(\hat{\bf i},\hat{\bf j},\hat{\bf k}\right)$ are the Cartesian unit vectors. 
Finally the position vector $\bf r$ is:
\begin{equation}
\boldsymbol{r}=
r\sqrt{\frac{r^2+a^2}{r^2+a^2\eta^2}}
\;\hat{\bf r}-
a^2\eta\sqrt{\frac{1-\eta^2}{r^2+a^2\eta^2}}
\boldsymbol{\hat{\eta}},
\end{equation}
while the velocity vector ${\bf v}={d\bf r}/{dt}$ expressed in terms of oblate spheroidal coordinates is:
\begin{equation}
    {\bf v}=\dot{r}
    \sqrt{\frac{r^2+a^2\eta^2}{r^2+a^2}}
    \;\hat{\bf r}+
    \dot{\eta}\sqrt{\frac{r^2+a^2\eta^2}{1-\eta^2}}\;\boldsymbol{\hat{\eta}}+
    \dot{\phi}\sqrt{(r^2+a^2)(1-\eta^2)}
    \;\boldsymbol{\hat{\phi}}.
\end{equation}

\section{Parameterization of orbits}
\label{App:2}

The potentials $V_r(r)$ of Eq.~(\ref{Vr}) and $V_\theta$ of Eq.~(\ref{Vth}), that govern the bound motion, could be rewritten as:
\begin{equation}\label{Vrroot}
    V_r(r)=2E(r-r_1)(r-r_2)(r-r_3)(r-r_4),
\end{equation}
\begin{equation}\label{Vthroot}
    V_{\theta}(\theta)=\frac{-2a^2E}{1-\cos^2{\theta}}(z_--\cos^2{\theta})(z_+-\cos^2{\theta}).
\end{equation}
Note the $E$ here is the initial Eulerian energy
before, its substitution by the corresponding relativistic analogue.
The radial potential has either four real roots with order $r_4\leq r_3\leq r_2\leq r_1$ or two real roots with order $r_2\leq r_1$ and two complex conjugate roots $r_3,r_4$. The roots of longitudinal potential are $\pm\sqrt{z_-},\pm\sqrt{z_+}$  which satisfy the inequalities $z_-\leq 1$ and $z_+>1$. Normal bound orbits have $r_2\leq r\leq r_1$ and $-\sqrt{z_-}\leq\cos{\theta}\leq\sqrt{z_-}=\cos\theta_{\min}$. We have excluded from our study bound orbits with $r_4\leq r\leq r_3$ (considering them plunging orbits), while $z_+$ does not correspond to any physical $\theta$ value. 

At numerical calculation we have used the orbital parameters $\{e,p,\theta_{\min}\}$ as a useful parametrization of bound orbits. Where $e=(r_1-r_2)/(r_1+r_2)$ is the eccentricity, $p=2r_1r_2/(r_1+r_2)$ is the  semi-latus rectum and $\theta_{\min}$ is the lowest polar angle along the orbit. The turning points of a normal bound orbit become:
\begin{equation}
    r_1= \frac{p}{1-e} \quad,\quad r_2= \frac{ p}{1+e} \quad,\quad z_-=\cos^2{\theta_{\min}}.
\end{equation}

The rest of the roots of the potentials (\ref{Vrroot}) and (\ref{Vthroot}) can be computed from the form of the polynomials and the corresponding constants of motion $\{E,L_z,Q\}$. Analyzing (\ref{Vrroot}) and equating to (\ref{Vr}) we obtain the following equations:
\begin{align*}
    -2E(r_1+r_2+r_3+r_4)&=2M,\\
    2Er_1r_2r_3r_4&=-Qa^2,
\end{align*}
which end up to the following expressions:
\begin{equation}
r_3=\frac{A+\sqrt{A^2-4 B}}{2}\quad,\quad r_4=\frac{B}{r_3} , \label{r34}
\end{equation}
where 
\begin{equation}
    A=-\frac{M}{E}-(r_1+r_2)
\end{equation}
and 
\begin{equation}
    B=\frac{a^2 Q}{-2 E r_1 r_2}.
\end{equation} 
Also, equating (\ref{Vth}) and (\ref{Vthroot}) we find:
\begin{equation*}
    -2a^2Ez_-z_+=Q
\end{equation*}
which gives:
\begin{equation}\label{z+}
    z_{+}=\frac{Q}{-2a^2 E z_{-}}.
\end{equation}

Next we need to express the constants of motion $\{E, L_z, Q\}$, showing up in the expressions above, in terms of the orbital parameters $\{p, e, \theta_{\min}\}$. For the Kerr space-time, similar expressions have been given by Schmidt in Appendix B of \cite{Schmidt}. 
We use the condition of turning points ($d\theta/dt=0$) at $\theta=\theta_{\min}$ to express $Q$ as a function of $(E,L_z,z_-)$:
\begin{equation}\label{Qcomp}
    Q=z_{-}\left[
    -2a^2E+\frac{L_z^2}{1-z_{-}}
    \right],
\end{equation}
and  rewrite the radial potential as:
\begin{equation}
    V_r(r)=2Ef(r)-L^2_zg(r)+d(r),
\end{equation}
where the functions:
\begin{align*}
f(r)&=r^4+a^2(1+z_{-})r^2+a^4z_{-},\\
g(r)&=\frac{r^2+a^2z_{-}}{1-z_{-}},\\
d(r)&=2Mr^3+2Ma^2r.
\end{align*}
Furthermore we impose $dr/dt=0$ at $r_1$ and $r_2$. (For circular orbits with $r_1=r_2=r_0$ we should solve simultaneously the equations $dr/dt=0$ and $d^2r/dt^2=0$ at $r_0$.) The energy $E$ and angular momentum $L_z$ are 
then given by:
\begin{align}\label{Ecomp}
E&=-\frac{\kappa}{2\rho},\\
\label{Lzcomp}
L_z&=\pm (\frac{\tau}{\rho})^{1/2},
\end{align}
where we the determinants $\rho, \kappa, \tau$
are defined as:
\begin{eqnarray}\label{dets1}
\rho&=f_1g_2-f_2g_1,\\\label{dets2}
\kappa&=d_1g_2-d_2g_1,\\\label{dets3}
\tau&=f_1d_2-f_2d_1.
\end{eqnarray}
In the above expressions for the determinants the subscripts 1, 2 have the following meaning:
\begin{enumerate}
\item[i.] for eccentric orbits ($e\neq 0$):
\begin{eqnarray} \label{fgd0}
(f_1,g_1,d_1)&=(f(r_1),g(r_1),d(r_1)),\\
(f_2,g_2,d_2)&=(f(r_2),g(r_2),d(r_2)),
\end{eqnarray}
\item[ii.] for circular orbits ($e=0$ and $r_1=r_2=r_0$):
\begin{eqnarray} \label{fgdn0}
(f_1,g_1,d_1)&=(f(r_0),g(r_0),d(r_0)),\\
(f_2,g_2,d_2)&=(f'(r_0),g'(r_0),d'(r_0)).
\end{eqnarray}
\end{enumerate}

In total, starting from the three orbital parameters $\{e,p,\theta_{\min}\}$: (a) 
we construct the determinants 
$\rho, \kappa, \tau$, using
Eqs.~(\ref{dets1}-\ref{dets3}), (b) from them we
compute the constants of motion $E, L_z$
and $Q$ through Eqs.~(\ref{Ecomp}, \ref{Lzcomp},
 \ref{Qcomp}), and (c) we finally obtain
the four radial roots of $V_r$ (from Eq.~(\ref{r34})) and
the second root, $z_+$, of $V_\theta$
(from Eq.~(\ref{z+})).

\section{Separatrix}
\label{App:3}

The separatrix describes all orbits 
that are essentially circular although their
eccentricity is not necessarily zero.
It is defined as the set of orbits with
$r_2=r_3$. Due to this double root
the orbit spends infinite time
to approach this root, therefore it 
evolves into an eternally circular
orbit.

In this Appendix we parametrize
these orbits by two parameters
$e$ (the eccentricity) and $x=r_4/r_3$.
From these two parameters we will show that
one could construct the rest orbital 
parameters $(p,z_{-})$, as well as
the constants of motion $E, L_z, Q$.
We will show also that the new 
parameter $x$ is intimately related to the inclination of the orbit. Thus for $x=0$
the orbit is equatorial ($\theta_{\min}=\pi/2$),
while for $x=1$ we get the maximally
inclined orbit ($\theta_{\min}={\min}$).

Let's start from the radial potential (\ref{Vr}):
\begin{equation}\label{vr_eqn}
V_r(r)=2Er^4+2Mr^3+(2Ea^2-Q-L_z^2)r^2+2Ma^2r-Qa^2
\end{equation}
which is a polynomial of degree four:
\begin{equation}
\begin{split}\label{pol4_eqn}
P_4(r)= 2E &(r-r_1)(r-r_2)(r-r_3)(r-r_4)\\
= 2E &[r^4-(r_1+r_2+r_3+r_4)r^3\\
&+(r_1r_2+r_1r_3+r_1r_4+r_2r_3+r_2r_4+r_3r_4)r^2\\
&-(r_1r_2r_3+r_1r_2r_4+r_1r_3r_4+r_2r_3r_4)r+r_1r_2r_3r_4],
\end{split}
\end{equation}
where the roots $r_1, r_2, r_3, r_4$ are given in Appendix \ref{App:2}. The motion of a particle is restricted between $r_2\leq r\leq r_1$ (Section \ref{sec:3.2}).
On separatrix,  $r_2=r_3=p/(1+e)$. The smallest root $r_4$, always lies within the interval $[0,r_3]$ thus $r_4=x r_3=x p/(1+e)$,
where $x \in [0,1]$.
Equating the coefficients of the polynomials (\ref{vr_eqn}), and (\ref{pol4_eqn}) and 
introducing the above parametrization for $r_1, r_2, r_3, r_4$ we derive the following set of equations:
\begin{align}
2M&=-2EMp\frac{(3-e)+x(1-e)}{1-e^2},\\
2Ea^2-Q-L_z^2&=2E\frac{M^2p^2}{(1+e)^2(1-e)}((3+e)+x(3-e)),\\
2Ma^2&=-2E\frac{M^3p^3}{(1+e)^3(1-e)}((1+e)+x(3+e)),\\
-Qa^2&=\frac{2EM^4p^4x}{(1-e)(1+e)^3}.
\end{align}
Solving this system of equations we can express the constants of motion and the semi-latus rectrum as functions with respect to the eccentricity $e$ and the parameter $x$ only:
\begin{equation}\label{pex}
p(e,x)=\frac{a}{M}(1+e)\sqrt{\frac{3-e+x(1-e)}{1+e+x(3+e)}},
\end{equation}

\begin{equation}\label{Eex}
E(e,x)=-\frac{(1-e)M}{a}\sqrt{\frac{1+e+x(3+e)}{(3-e+x(1-e))^3}},
\end{equation}

\begin{equation}\label{Qex}
Q(e,x)=2Ma(1+e)x\sqrt{\frac{3-e+x(1-e)}{(1+e+x(3+e))^3}},
\end{equation}

\begin{eqnarray}\label{Lzex}
L_z^2(e,x)=16Ma\sqrt{\frac{1+e+x(3+e)}{(3-e+x(1-e))^3}}\times
\frac{(1+x)^2[1+x+e(1-x)]}{(1+e+x(3+e))^2}.
\end{eqnarray}
The $z_{-}=\cos^2{\theta_{\min}}$ is the lower root of the quadratic equation $V_\theta(\theta)=0$. That is:
\begin{equation}\label{z_eqn}
    z_{-}=\frac{C-\sqrt{C^2+8a^2 E Q}}{-4a^2 E},
\end{equation}
where $C=Q+L_z^2-2a^2 E$. Replacing (\ref{pex})-(\ref{Lzex}) in (\ref{z_eqn}), we obtain: 
\begin{equation}
    \begin{split}
        z_-(e,x)&=\frac{3-e+x(1-e)}{2(1-e)(1+e+x(3+e)} \times \\
        &\left[3+e+x(3-e) -\sqrt{(3+e+x(3-e))^2-4x(1-e^2)}\right].
    \end{split}
\end{equation}
For a fixed eccentricity $e$, $z_{-}(e,x)$ is a monotonically increasing function of $x$. It takes its greatest value at $x=1$, while for $x=0$, it is $z_{-}(e,0)=0$ (equatorial orbits), as mentioned earlier. This parameter $x$ is  not  very practical
for bound orbits in Kerr, though, since the maximally inclined orbits in Kerr do not correspond to $x=1$.

\section{Fundamental Frequencies}
\label{App:4}

In this Appendix we give analytic expressions
that one could use to calculate the 
fundamental frequencies of normal bound 
orbits in the Euler field.
We will exploit the action-angle variables formalism \cite{Arnold}. We denote the constants of motion as: $F_{i}=(H=E,L_z,Q)$. The canonical momenta are: $p_{\phi}=L_z$, $p_r=\sqrt{V_r(r)}/(r^2+a^2)$ and  $p_{\theta}=\sqrt{V_{\theta}(\theta)}$, with the potentials  given in Eqs.~(\ref{Vr}) and (\ref{Vth}).
The definition  of the action variables,
of Eq.~(\ref{J}), give:
\begin{align}
J_r&=\frac{1}{2\pi}\oint\frac{\sqrt{V_r(r)}}{r^2+a^2}dr\label{Jr},\\
J_{\theta}&=\frac{1}{2\pi}\oint\sqrt{V_{\theta}}d\theta \label{Jth},\\
J_{\phi}&=\frac{1}{2\pi}\oint p_{\phi}d\phi=L_z\label{Jph}.
\end{align}
In order to derive the corresponding frequencies
\[
\Omega_{i}(\boldsymbol{J})=\frac{\partial{H}(\boldsymbol{J})}{\partial{J}_{i}},
\]
we should first express the Hamiltonian with respect to the action variables $H(\boldsymbol{J})$, which can't be done analytically. The integrals (\ref{Jr})-(\ref{Jph}) of action variables, cannot be 
explicitly inverted. 
However, we can calculate the frequencies from the inverse derivatives $\partial{J_i}/\partial{F_{j}}$, combined with the chain rule. The non trivial partial derivatives are: 
\begin{align}
\frac{\partial{J}_r}{\partial{H}}&=\frac{Y}{\pi}\label{1}\\
\frac{\partial{J}_r}{\partial{L_z}}&=-\frac{Z}{\pi}\label{2}\\
\frac{\partial{J}_r}{\partial{Q}}&=-\frac{X}{2\pi}\label{3}\\
\frac{\partial{J}_{\theta}}{\partial{H}}&=\frac{2a^2\sqrt{z_{+}}}{\pi \beta}(K(k)-E(k))\label{4}\\
\frac{\partial{J}_{\theta}}{\partial{L_z}}&=\frac{2L_z}{\pi \beta\sqrt{z_{+}}}(K(k)-\Pi(z_{-},k))\label{5}\\
\frac{\partial{J}_{\theta}}{\partial{Q}}&=\frac{1}{\pi \beta\sqrt{z_{+}}}K(k)\label{6}
\end{align}
with $K(k)$, $E(k)$ and $\Pi(z_{-},k)$ being the 1st, 2nd and 3rd complete elliptic integrals
that are given in Eqs.~(\ref{ElK}, \ref{ElE}, \ref{ElP}),
while the quantities $Y$, $Z$ and $X$ are the radial integrals:
\begin{align}
Y&=\int^{r_2}_{r_1}\frac{r^2}{\sqrt{V_r}}dr,\label{Y}\\
Z&=\int^{r_2}_{r_1}\frac{L_zr^2}{(r^2+a^2),\sqrt{V_r}}dr\label{Z}\\
X&=\int^{r_2}_{r_1}\frac{dr}{\sqrt{V_r}}.\label{X}
\end{align}
Finally the two extra quantities $\beta,k$
shown above are defined as $\beta^2=-2a^2E$ and $k^2=z_-/z_+$.
Now we can inverse the derivatives (\ref{1})-(\ref{6}), using the chain rule
\begin{equation}
\frac{\partial{F_{i}}}{\partial{J_{j}}}\frac{\partial{J_{j}}}{\partial{F_{k}}}=\delta_{k}^{i}.
\end{equation}
By setting $F_i=H$ we obtain the system of equations:
\begin{align*}
    \frac{\partial H}{\partial J_r}\frac{\partial J_r}{\partial H}+\frac{\partial H}{\partial J_{\theta}}\frac{\partial J_{\theta}}{\partial H}&=1\\
    \frac{\partial H}{\partial J_r}\frac{\partial J_r}{\partial L_z}+\frac{\partial H}{\partial J_{\theta}}\frac{\partial J_ {\theta}}{\partial L_z}+\frac{\partial H}{\partial J_{\phi}}\frac{\partial J_{\phi}}{\partial L_z}&=0\\
    \frac{\partial H}{\partial J_r}\frac{\partial J_r}{\partial Q}+\frac{\partial H}{\partial J_{\theta}}\frac{\partial J_{\theta}}{\partial Q}&=0.
\end{align*}
which solved with respect to $\partial H/\partial J_i$ provides us with the desired
 frequencies:
\begin{align}
\Omega_r&=\frac{\pi K(k)}{a^2z_{+}[K(k)-E(k)]X+YK(k)},\\
\Omega_{\theta}&=\frac{\pi\beta\sqrt{z_{+}}X/2}{a^2z_{+}[K(k)-E(k)]X+YK(k)},\\
\Omega_{\phi}&=\frac{ZK(k)+XL_z[\Pi(z_{-},k)-K(k)]}{a^2z_{+}[K(k)-E(k)]X+YK(k)}.
\end{align}

Although the denominators of the integrals (\ref{Y}, \ref{X}) vanish when the orbit passes through a turning point, {\it Mathematica} is capable to campute the above frequencies.
We should note that the above expressions for the fundamental frequencies are identical to those
for Kerr orbits (c.f.~\cite{Schmidt}), except of
the actual form of  $V_r$ and the form of the integral $Z$.

 \section{Alternative analytical expressions for the frequencies}
 \label{App:5}

We should emphasize the fact  that the expressions for the fundamental frequencies derived in Appendix \ref{App:4},
directly from the formalism of action-angle variables  turn into indefinite expressions when $r_2 \to r_3$ (separatrix). Then  one needs to resort in approximating analytical expressions for this region, which is quite challenging especially for generic orbits. 
Another major problem is that the $r$ and $\theta$ oscillations are not periodic in coordinate time $t$, since the corresponding equations (c.f.~Eqs. (\ref{quadr1}, \ref{quadr2})) are coupled through the quantity $r^2+a^2 \cos^2\theta$. Following
\cite{Hiki}, in order to decouple the radial and polar motion, we introduce a new time variable,
\begin{equation}
\label{dtdMino}
    d\lambda=\frac{dt}{r^2+a^2 \cos^2\theta},
\end{equation}
by analogy with the Mino time, 
which is widely used in the study 
of geodesic orbits in Kerr.
Then the corresponding variables become 
strictly periodic with respect to $\lambda$. 
It is preferable to derive new analytical expressions for the frequencies exploiting the elliptic integrals and the new time variable (\ref{dtdMino}) as Fujita and Hikida did for Kerr.

The geodesic equations in the new time variable $\lambda$ become:
\begin{equation}\label{V2}
\begin{split}
\left(\frac{dr}{d\lambda}\right)^2
&=V_r(r),\\
\left(\frac{d\cos{\theta}}{d\lambda}\right)^2
&={\tilde V}_{\theta}(\cos{\theta}),\\
\frac{d\phi}{d\lambda}
&=\Phi_r(r)+\Phi_{\theta}(\cos{\theta})-a E,\\
\frac{dt}{d\lambda}
&=r^2+a^2\cos^2{\theta}.
\end{split}
\end{equation}
where the new functions $\tilde{V}_{\theta},\Phi_r(r), \Phi_{\theta}$ that are introduced above are:
 \begin{equation}
 \begin{split}
{\tilde V}_{\theta}(\cos{\theta})&=V_{\theta}(\theta) (1-\cos^2\theta)=Q-(Q-2a^2E+L_z^2)\cos^2{\theta}-2a^2E\cos^4{\theta},\\
\Phi_r(r)&=a\frac{E(r^2+a^2)-a L_z}{r^2+a^2},\\
\Phi_{\theta}(\cos{\theta})&=\frac{L_z}{1-\cos^2{\theta}}.
\end{split}
\end{equation}
The frequency of the $r$-motion and $\theta$-motion, with respect to  $\lambda$, will be denoted $Y_r$ and $Y_{\theta}$, respectively.
Similarly,  one can define the  azimuthal $\lambda$-frequency  $Y_{\phi}$ and the frequency $\Gamma$ of the coordinate time $t$ with respect to $\lambda$. Following the procedure of  \cite{Hiki}, we derive all $\lambda$ frequencies for the Euler problem. 

The radial (\ref{Vr}) and polar (\ref{Vthroot}) potentials are polynomials of order four for both Euler and Kerr, so $Y_r$ and $Y_{\theta}$ of Euler are exactly the same with that of Kerr,
when written in terms of the roots of the corresponding polynomial. We will write the final expressions and not reproduce all the calculations here (the various new quantities introduced here 
will be analytically presented at the end). The process is exactly the same with that for the Kerr field:
\begin{eqnarray}\label{Yr}
Y_r=\frac{\pi\sqrt{-2E(r_1-r_3)(r_2-r_4)}}{2K(k_r)}, \\
\label{Y8}
Y_{\theta}=\frac{\pi L_z \sqrt{\epsilon_0z_{+}}}{2K(k_{\theta})},
\end{eqnarray}
while 
the $Y_{\phi}$ and $\Gamma$ for Kerr are defined by Eqs.~(7), (8) of \cite{Hiki}. When they are translated into
the Euler case they yield the following form:
\begin{align}
\Gamma&=Y_{t^{(r)}}+Y_{t^{(\theta)}},\\
Y_{\phi}&=Y_{\phi^{(r)}}+Y_{\phi^{(\theta)}}-a E,
\end{align}
where:
\begin{align*}
    Y_{t^{(r)}}&=\langle r^2 \rangle_{\lambda}\\
Y_{t^{(\theta)}}&=\langle a^2\cos^2{\theta}\rangle_{\lambda}\\
Y_{\phi^{(r)}}&=\langle\Phi_r(r)\rangle_{\lambda}\\
Y_{\phi^{(\theta)}}&=\langle\Phi_{\theta}(\cos{\theta})\rangle_{\lambda}
\end{align*}
and $\langle f(x) \rangle_{\lambda}$ denotes the average over  $\lambda$. 
$\Phi_r$ could be recast in the following form
\begin{equation}
    \Phi_r(r)=\frac{a}{r_{+}-r_{-}}
    \left(-\frac{a L_z}{r-r_+}+\frac{a L_z}{r-r_-}\right)+aE,
\end{equation}
and using the expressions and integrals from Appendix A of \cite{Hiki},  we obtain the following expressions for the rest frequencies:

\begin{equation}\label{Yf}
\begin{split}
Y_{\phi}=\frac{2Y_{\theta}}{\pi\sqrt{\epsilon_0z_{+}}}\Pi(z_{-},k_{\theta})+\frac{2a Y_r}{\pi(r_{+}-r_{-})\sqrt{-2E(r_1-r_3)(r_2-r_4)}}\times \\
\left(\frac{-a L_z}{r_3-r_{+}}\left[K(k_r)-\frac{r_2-r_3}{r_2-r_{+}}\Pi(h_{+},k_r)\right] \right.\\
+\left.\frac{a L_z}{r_3-r_{-}}\left[K(k_r)-\frac{r_2-r_3}{r_2-r_{-}}\Pi(h_{-},k_r)\right]\right),
\end{split}
\end{equation}

\begin{equation}\label{Yg}
\begin{split}
\Gamma=&\frac{2Y_{\theta}a^2z_{+}}{\pi L_z\sqrt{\epsilon_0z_{+}}}
[K(k_{\theta})-E(k_{\theta})]\\
&+\frac{Y_r}{\pi\sqrt{-2E(r_1-r_3)(r_2-r_4)}}
\left[(r_3(r_1+r_2+r_3)-r_1r_2)K(k_r)\right.\\
&+\left.(r_2-r_3)(r_1+r_2+r_3+r_4)\Pi(h_r,k_r)+(r_1-r_3)(r_2-r_4)E(k_r)\right].
\end{split}
\end{equation}
The various quantities that appear on the formulae
for the frequencies  above are defined as:
\begin{eqnarray}
\epsilon_0=&\frac{-2a^2E}{L_z^2}, \\
k_r=&\sqrt{\frac{r_1-r_2}{r_1-r_3}
\frac{r_3-r_4}{r_2-r_4}}, \\ k_{\theta}=&\sqrt{\frac{z_{-}}{z_{+}}}, \\ r_{\pm}=&\pm ia, \\ h_{\pm}=&\frac{(r_1-r_2)(r_3-r_{\pm})}{(r_1-r_3)(r_2-r_{\pm})}, \\
h_r=&\frac{r_1-r_2}{r_1-r_3},
\end{eqnarray}
while he expressions $K(k), E(k)$ and $\Pi(n,k)$ are the complete elliptic integrals of first, second and third kind respectively (note that
\cite{Hiki} has different conventions for the 
elliptic integrals than the ones used in this article).

The fundamental frequancies in coordinate time $t$ are finally given by:
\begin{align}\label{W}
\begin{split}
    \Omega_r&=\frac{Y_r}{\Gamma}\\ \Omega_{\theta}&=\frac{Y_{\theta}}{\Gamma}\\ \Omega_{\phi}&=\frac{Y_{\phi}}{\Gamma}.
\end{split}
\end{align}

On the plots of the text all fundamental frequencies of the Euler problem were
calculated based on the analytical expressions (\ref{W}). These formulae are precise and easy to calculate by means of {\it Mathematica} even for the region close to separatrix that is when $r_2 \to r_3$. 

Especially at the separatrix the expressions for the frequencies are greatly simplified since the ratios between some elliptic integrals vanish, that is,
\begin{equation}
    \frac{E(k_r)}{K(k_r)} \xrightarrow[r_3 \to r_2]{} 0, 
    \end{equation}
and 
\begin{equation}
        (r_2-r_3)\frac{\Pi(h_{r,+,-},k_r)}{K(k_r)}
     \xrightarrow[r_3 \to r_2]{} 0.
\end{equation}
The final expressions for the frequencies
at the separatrix are still given by Eqs.~(\ref{W}), but now the various components $Y_i$'s and $\Gamma$ are
much simpler:
\begin{align}\label{Ws}
\begin{split}
    Y_r&=0 \\
    Y_{\theta}&=
    \frac{\pi L_z \sqrt{\epsilon_0 z_+}}
    {2 K(k_\theta)}\\ 
    Y_{\phi}&=
    L_z \left(
    \frac{\Pi(z_-,k_\theta)}{K(k_\theta)}-
    \frac{a^2}{r_2^2+a^2} 
    \right)\\
    \Gamma&=
    a^2 z_+ \left(1 -
    \frac{E(k_\theta)}{K(k_\theta)} 
    \right) +r_2^2.
\end{split}
\end{align}


\end{document}